\documentclass[ reprint, superscriptaddress, amsmath,amssymb,prx, longbibliography]{revtex4-2}

\usepackage[normalem]{ulem}
\usepackage[dvipsnames]{xcolor}

\newcommand{\ric}[1]{\textcolor{black}{#1}}
\newcommand{\ricreview}[1]{\textcolor{black}{#1}}
\newcommand{\ricrevieww}[1]{\textcolor{black}{#1}}

\usepackage{amsmath}
\usepackage{amsfonts}
\usepackage{pgfplots}
\DeclareUnicodeCharacter{2212}{−}
\usepgfplotslibrary{groupplots,dateplot}
\usetikzlibrary{patterns,shapes.arrows}
\usepackage{bbm}
\usepackage{cancel}
\usepackage{braket}
\usepackage{dsfont}
\usepackage{enumerate}
\usepackage{bm}
\usepackage{url}

\begin{document}


\title{State-dependent mobility edge in kinetically constrained models}
\author{Manthan Badbaria}
\email{mbadbaria@umass.edu}
\affiliation{Department of Physics, Indian Institute of Technology (Banaras Hindu University), Varanasi 221005, India}
\author{Nicola Pancotti}
\affiliation{AWS Center for Quantum Computing, Pasadena, CA 91125, USA}
\affiliation{California Institute of Technology, Pasadena, CA, USA}
\author{Rajeev Singh}
\affiliation{Department of Physics, Indian Institute of Technology (Banaras Hindu University), Varanasi 221005, India}
\author{Jamir Marino}
\affiliation{Institut f\"{u}r Physik, Johannes Gutenberg-Universit\"{a}t Mainz, D-55099 Mainz, Germany}
\author{Riccardo J. Valencia-Tortora}
\email{riccardo.valencia95@gmail.com}
\affiliation{Institut f\"{u}r Physik, Johannes Gutenberg-Universit\"{a}t Mainz, D-55099 Mainz, Germany}
\affiliation{Pritzker School of Molecular Engineering, University of Chicago, Chicago, IL 60637, USA}
\date{\today}

\begin{abstract}
In this work, we show that the kinetically constrained quantum East model lies between a quantum scarred and a many-body localized system featuring an unconventional type of mobility edge in the spectrum. 
We name this scenario \textit{state-dependent} mobility edge: 
while the system does not exhibit a sharp separation in energy between thermal and non-thermal eigenstates, the abundance of non-thermal eigenstates
results in slow entanglement growth for \textit{many} initial states, such as product states, below a finite energy density.
We characterize the state-dependent mobility edge by looking at the complexity of classically simulating dynamics using tensor network for system sizes well beyond those accessible via exact diagonalization.
Focusing on initial product states, we observe a qualitative change in the dynamics of the bond dimension needed as a function of their energy density. Specifically, the bond dimension typically grows \textit{polynomially} in time up to a certain energy density, where we locate the state-dependent mobility edge, enabling simulations for long times. Above this energy density, the bond dimension typically grows \textit{exponentially} making the simulation practically unfeasible beyond short times, as generally expected in interacting theories.
We correlate the polynomial growth of the bond dimension to the presence of many non-thermal eigenstates around that energy density, a subset of which we compute via tensor network. \ricreview{The outreach of our findings  encompasses  quantum sampling problems and   the efficient simulation of quantum circuits beyond Clifford families.}
\end{abstract}

\maketitle

In the era of noisy intermediate scale quantum  (NISQ) devices~\cite{Preskill2018}, 
a natural task that could display quantum advantage over classical computers is the simulation of quantum many-body dynamics~\cite{Feynman1982,Daley2022,Miessen2022,Kim2023}. 
The simulation of quantum systems is typically expected to be hard for the same reason quantum computers are believed to be powerful, namely entanglement. However, displaying provable quantum advantage is a challenging task in itself due to the noisy nature of current quantum computers~\cite{PRXQuantum.5.010308,PhysRevResearch.6.013326,liao2023simulation,anand2023classical,Begui2024,rudolph2023classical}. Indeed, while quantum computers are developed so are also classical algorithm, making the quantum advantage effectively a moving target.
Even more surprisingly, also assuming the ability to implement large-scale error-correcting protocols and possess a fault-tolerant quantum computer, there are strongly interacting quantum systems whose simulation is not guaranteed to be more efficient using a quantum computer over a classical one. In fact, there exist systems whose properties make them amenable to being efficiently simulated on a classical computer.
Examples include Hamiltonians exhibiting non-thermal behavior at finite energy density, e.g. area-law excited eigenstates, such as quantum many-body scarred systems~\cite{Bernien2017,Turner2018,PhysRevLett.122.220603,Serbyn2021,PhysRevB.98.155134,PhysRevB.98.155134,PhysRevB.99.161101,PhysRevX.11.021021,Moudgalya2022,chandran2023quantum, PhysRevX.13.031013} and many-body localized systems (MBL) systems~\cite{doi:10.1146/annurev-conmatphys-031214-014726, RevModPhys.91.021001}, or mechanisms hindering the propagation of quantum correlations, such as dynamical confinement~\cite{kormos2017real,PhysRevB.102.041118,PhysRevX.10.021041,PhysRevLett.122.150601,JavierValenciaTortora2020,10.21468/SciPostPhys.12.2.061}. As these systems are characterized by a slow growth of entanglement, typically considered a measure of `classical complexity,' they are amenable to be efficiently simulated using classical algorithms (e.g., via tensor networks) with modest computational resources up to long times~\cite{PhysRevLett.109.017202,PhysRevLett.110.260601,ehrenberg2022simulation,PhysRevB.105.224203,sierant2024many}.  The presence of exceptions to the naively expected quantum advantage in computing dynamics raises an intriguing possibility: 
can non-thermal features be diagnosed and characterized by examining how efficiently dynamics can be computed using classical algorithms?\\

To this aim, we focus on one-dimensional non-integrable systems. 
A key point of our approach is the choice of an appropriate candidate for efficient classical simulations susceptible to the complexity due to entanglement. Tensor-network methods fulfill such requirement as
the resources required—the bond dimension—typically scale exponentially with time in far-from-equilibrium dynamics. This occurs since finite-energy volume-law eigenstates of the Hamiltonian participate predominantly in the dynamics~\cite{PhysRevLett.111.127205}. However, the presence of non-thermal eigenstates of the Hamiltonian could significantly reduce complexity, enabling efficient simulations for `long' times~\cite{PhysRevLett.109.017202,PhysRevLett.110.260601,ehrenberg2022simulation}.
Having selected a classical method, we need to define complexity. 
We categorize a computational task as `hard' (or `easy') if the required computational resources scale (sub-)exponentially. 
We distinguish complexity along the space and time domain. For the time domain, we adopt the bond dimension $\chi$ of the matrix product state representation of the evolved states as a measure of complexity.  We choose $\chi$ as it encodes the actual computational time needed in performing operations, such as dynamics, in tensor network methods. Moreover, $\chi$ is closely linked to entanglement, widely regarded as an indicator of `hardness' in representing quantum states on a classical computer. 
If $\chi$ grows (sub-)exponentially in time the task is said to be (`easy') `hard' in the time domain. 
Instead, for the complexity along the space domain, we look at the degree of separability of the evolved state, which directly reflects on the dimension of the potentially accessible Hilbert space. Specifically, we distinguish whether the state describing the whole system can or cannot be written as product states of (potentially entangled) states describing smaller disconnected subsystems. In case it can, we name the state as separable and the space-complexity is said to be `easy.' Otherwise, we say that is inseparable and space-complexity is `hard.' If the state is separable, the accessible Hilbert space is effectively reducible in smaller disconnected Hilbert spaces each describing different subsystems, as the degrees of freedom defined on such subsystems are not entangled. Instead, if the state is inseparable, the accessible Hilbert space is strictly irreducible. Combining space and time complexity in simulating dynamics of initial product states, we then aim to infer the properties of non-thermal eigenstates of the Hamiltonian.\\

For the sake of concreteness, we test our complexity-oriented proxy on the kinetically constrained quantum East model~\cite{PhysRevB.92.100305,PhysRevX.10.021051,PhysRevLett.132.223201,PhysRevB.108.104317,PhysRevLett.132.120402,PhysRevLett.132.080401,PRXQuantum.3.020346,geissler2023slow,brighi2023hilbert,PhysRevB.110.014301}.
 From a condensed matter and statistical physics perspective, it attracts interest because it exhibits both fast and slow thermalizing dynamical phases alongside with localization, despite being translationally invariant and non-integrable~\cite{PhysRevB.92.100305,PhysRevX.10.021051,PhysRevLett.132.223201,PRXQuantum.3.020346,PhysRevB.108.104317,geissler2023slow,brighi2023hilbert,klobas2023exact, bertini2024exact, de2024exact,causer2024quantum}. This behavior is markedly different from MBL systems, where localization arises due to many-body wave function interference caused by quenched disorder, while instead it arises by making transport a higher-order process in the quantum East model. Moreover, it features extreme slowdown of thermalization as well as dynamical heterogeneity similar to structural glasses~\cite{PhysRevB.92.100305,PhysRevLett.121.040603}, from which its classical counterpart is inspired~\cite{Garrahan2009,Chleboun2013,Garrahan2018,PhysRevLett.121.040603,PhysRevE.102.052132}.
From a quantum information perspective, the digital (i.e. Floquet) version hosts special points where it reduces to a Clifford circuit~\cite{PhysRevLett.132.120402}, making it a candidate for investigating deviations from the latter, and has been shown to still display localization despite lack of energy conservation~\cite{PhysRevLett.132.080401}. Additionally, its localized nature has been shown to aid in passively protecting quantum information against certain types of coupling to an external environment~\cite{PRXQuantum.3.020346}.\\
\begin{figure}[t!]
\centering
\includegraphics[width=\linewidth]{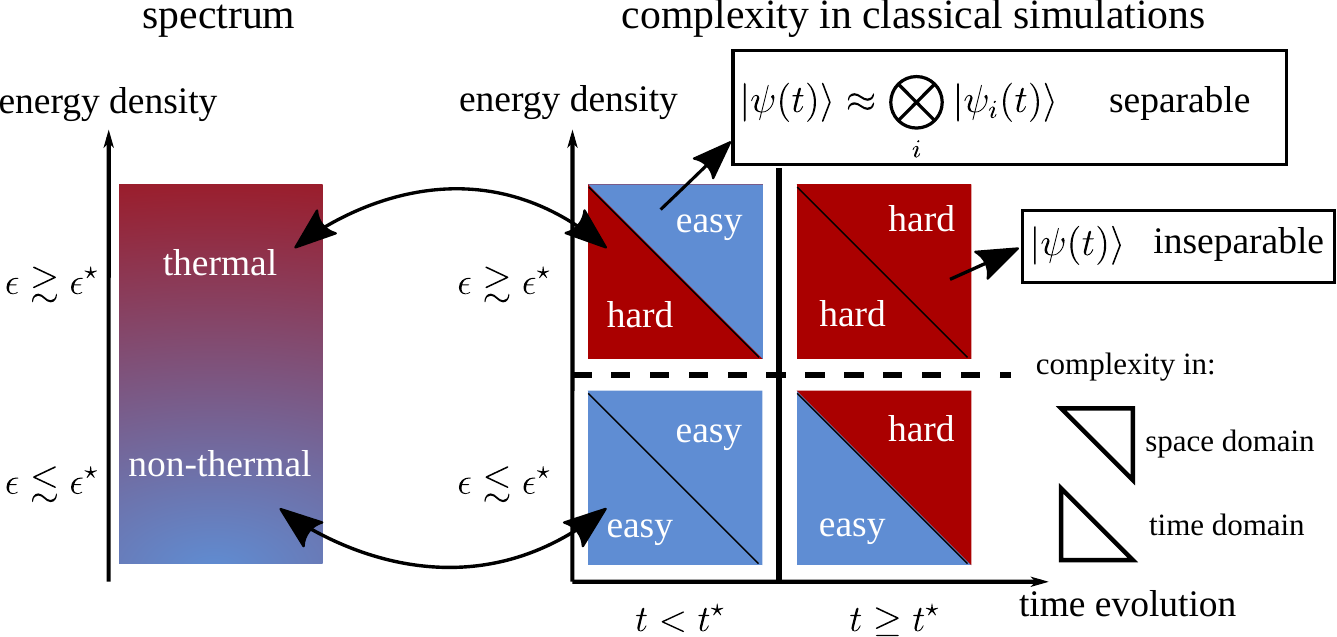}
\caption{Cartoon summarizing the complexity phase diagram of simulating dynamics of initial product states in the quantum East model as a function of their energy density $\epsilon \in [0,1]$ and the physical time $t$ reached. Each box is split into two triangles which encode the complexity either along the space domain (upper triangle) or along the time domain (lower triangle). A simulation is (easy) hard in the time domain if the computational resources needed, i.e. the bond dimension of the tensor network, grows (sub-)exponentially in time. A simulation is easy (hard) in the space domain if the evolved state is (in)separable. Simulations of \ricreview{a large class of} initial product states are typically easy in the time domain if the energy density is below a certain energy density $\epsilon^\star$, dependent on the parameters of the Hamiltonian, indicating the existence of a \textit{state-dependent} mobility edge in the spectrum (cf. Sec.~\ref{sec_mob_edge_energy}). 
The sharp differentiation between active and inactive regions, typical of glassy-like systems, makes the space-complexity undergo a transition from easy to hard as a function of time $t$: for $t< t^\star$ the entanglement between the different active regions remains negligible and the state can be written as a tensor product of (potentially entangled) states describing the different active regions, i.e. it is separable; for $t>t^\star$, the entanglement between the different active regions is no longer negligible and the state is inseparable. 
In the separable case, the accessible Hilbert space $\mathcal{H}$ appears as if it is reducible in smaller disconnected subspaces $\mathcal{H}_i$, each describing the $i$-th active region, similarly to \textit{fragmented} systems~\cite{PhysRevX.10.011047,PhysRevX.12.011050,Moudgalya2022}.} 

\label{fig_1_sketch}
\end{figure}
\begin{figure*}[t!]
    \centering
    \includegraphics[width=\linewidth]{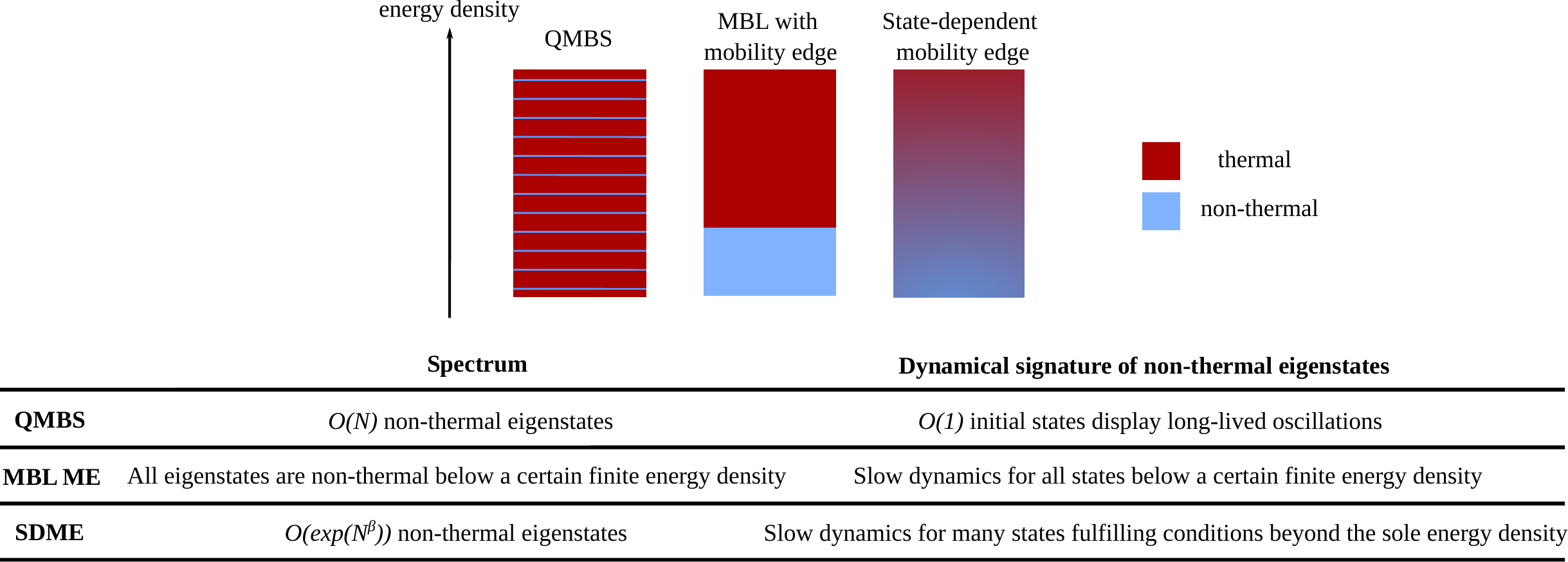}
    \caption{\ricrevieww{Comparison between quantum-many body scars (QMBS), many-body localization with mobility edge (MBL ME), and state-dependent mobility edge (SDME).}}
    \label{fig_comparison_QMBS_MBL_SDME}
\end{figure*}

In our work, we combine a quantum information and statistical physicsoriented approach to characterize the quantum East model at finite energy density. Specifically, we aim to inspect spectral properties by using the complexity of simulating dynamics of initial product states via tensor network, going beyond the small system sizes accessible via exact diagonalization. By doing so, we can identify parameter regimes where \textit{many} initial states are easily simulable. Then, by closely inspecting the properties of such states, we discover and compute, via DMRG-X~\cite{PhysRevLett.116.247204}, a family of non-thermal finite-energy density eigenstates of the Hamiltonian mostly responsible for the observed slow dynamics.\\

Interestingly, the easily simulable states are mostly clustered in their energies, and so it is tempting to identify a sort of mobility edge in the spectrum -- a separation of a thermal region and a localized one in the spectrum of the Hamiltonian-- challenging the common belief that a mobility edge is exclusive to disordered systems~\cite{PhysRevB.92.064203,PhysRevB.91.081103,Naldesi2016}.
However, small-scale exact diagonalization calculations suggest that the spectrum of the quantum East model does not exhibit a distinct separation between non-thermal eigenstates and thermal ones~\cite{PhysRevX.10.021051} (see Appendix~\ref{appendix_full_spectrum}).
Instead, its spectrum displays features more reminiscent of quantum many-body scarred (QMBS) systems, even though the non-thermal eigenstates are not evenly spaced in energy, are exponentially many in the system size, and are nearly product states~\cite{PhysRevX.10.021051} (see Appendix~\ref{appendix_full_spectrum}). This contrasts with QMBS systems, where the\ricreview{re are a polynomial number in the system size of }non-thermal eigenstates evenly spaced in energy~\cite{Moudgalya2022}. 
All these ingredients open up an intriguing possibility: the existence of a \textit{state-dependent} mobility edge. In other words, the system behaves akin to having a many-body mobility edge for a large class of initial states, such as product states. To draw an analogy, our scenario resembles QMBS systems, where the impact of the non-thermal eigenstates significantly influences the dynamics of some initial states. However, differently from QMBS systems, the exponential abundance of non-thermal eigenstates suggests that the non-thermal behavior becomes prevalent across a wide range of initial states, particularly among product states. Under this lens, we expect, and numerically confirm, that simulating dynamics of initial product states can be carried out efficiently via tensor network up to a finite energy density, while instead being computationally hard above it. \ricrevieww{For reference,  in Fig.~\ref{fig_comparison_QMBS_MBL_SDME}, we summarize the differences between quantum-many body scars, MBL with a mobility edge, and the state-dependent mobility edge we introduce.}\\

As a byproduct of our analysis, the existence of parameters and many initial states whose dynamics can be carried out efficiently suggests that kinetically constrained inspired circuits can be easily simulable on classical computers, alongside Clifford, MBL~\cite{PhysRevB.98.134204} and fractonic ~\cite{PhysRevX.9.021003,PhysRevX.10.011047} inspired.
Additionally, identifying regimes where dynamics of product states are classically challenging holds practical relevance since they are the easily preparable initial states in current NISQ devices~\cite{PhysRevLett.132.223201, PhysRevLett.111.215305,PhysRevA.98.021804,PhysRevA.93.040701,PhysRevA.90.011603,PhysRevA.97.011603,PhysRevA.99.060101,PhysRevLett.118.063606,Lorenzo2017}.

\subsection{Summary of results}
In Sec.~\ref{sec_dyn_proxy} and Sec.~\ref{sec_model}  we introduce our measure of complexity and the model, respectively. In Sec.~\ref{sec_state_dependent} we discuss our results,  
summarized in Fig.~\ref{fig_1_sketch}, which are:
\begin{enumerate}[(i)]
\item While time-complexity is typically hard in the delocalized phase, as expected for generic non-integrable models, we observe both easy and hard regimes in the localized phase, indicating the existence of a \textit{state-dependent} mobility edge (see Sec.~\ref{sec_mob_edge_energy});
\item The sharp differentiation between active and inactive regions, typical of kinetically constrained models and glassy systems~\cite{Garrahan2018,PhysRevB.92.100305,lan2018quantum,zadnik2023slow,PhysRevX.10.021051}, makes the initial distribution of excitations matter in dictating time-complexity (see Sec.~\ref{sec_mob_edge_structure}) and space-complexity (see Sec.~\ref{sec_fragmented_ergodic_transition}). In the latter, we observe a transition from easy to hard space-complexity as a function of time. Specifically, in the easy regime, the state is approximately separable and its dynamics can be faithfully computed by the dynamics of suitably chosen non-overlapping subsystems. Instead, in the hard regime the state is inseparable;
\item We connect the points above by computing a novel family of non-thermal localized eigenstates at finite energy density via tensor network (see Sec.~\ref{sec_mob_edge_unifying_picture}). We use such states to \ricrevieww{identify an exponentially large, in the system size, corner of the Hilbert space which can be efficiently simulated via tensor networks, on which simple product states have large weight in overlap (see Sec.~\ref{sec_unifying_picture}). This} explains the extremely slow dynamics observed for a large class of initial states.
\end{enumerate}

The significance of our results lies in being in stark contrast to what expected in generic non-integrable models, where simulating dynamics using tensor networks is computationally challenging at any finite energy density due to the linear growth of entanglement~\cite{PhysRevLett.111.127205}. A known exception to this are MBL systems, where entanglement grows slowly~\cite{PhysRevLett.109.017202,PhysRevLett.110.260601,sierant2024many} and our analysis could be potentially applied and give nontrivial results. However, our analysis would be computationally more demanding in MBL systems due to the need for disorder averaging. Also, it would face additional challenges from potential avalanches as the system size increases, which could destabilize the MBL phase~\cite{vsuntajs2020quantum,Abanin2021,PhysRevB.105.224203,sierant2024many,PhysRevB.95.155129}, making finite size scaling a delicate procedure. In contrast, our system avoids these issues because there is no disorder and the atypical features arise not from interferences, as in MBL, but from the kinetic constraints, making the overall physical picture more transparent.\\

Finally, in Sec.~\ref{sec_conclusions} we list possible fruitful directions in the context of \ricreview{higher dimensional systems and} random quantum circuits~\cite{Fisher2023}.

\section{Complexity oriented proxy of non-thermal eigenstates \label{sec_dyn_proxy}}

To overcome the limitations faced by exact diagonalization, here we propose a complexity-oriented proxy based on tensor-networks for detecting non-thermal eigenstates. Our approach is based on the observation that typical thermalizing systems display an exponential growth of the bond dimension during dynamics, as opposed to systems displaying non-thermal behavior (e.g. MBL systems) where instead the bond dimension grows polynomially in time~\cite{PhysRevLett.109.017202,PhysRevLett.110.260601}.
An intermediate scenario is constituted by systems with a mobility edge, i.e. systems displaying non-thermal eigenstates in a certain energy window, and thermal ones in the others. Specifically, in such system we envision that the complexity of simulating the dynamics depends on the energy of the specific state at hand, potentially allowing a distinction between the two regions.\\

 To this end, we 
consider the dynamics of initial product states. 
Specifically, we first select initial states with a small energy variance so that the eigenstates of the Hamiltonian participating in the dynamics are mostly within a small energy window. Then, we use the way the bond dimension $\chi$ grows in time as a measure of time-complexity. In typical non-integrable systems, we expect an exponential growth of $\chi$ after a quench, as the bipartite entanglement entropy $S \propto \log \chi$ grows linearly in time~\cite{PhysRevLett.111.127205}. Instead, a sub-exponential growth of $\chi$ could indicate the existence of non-thermal eigenstates in the spectrum. In the following, we label it as an (easy) hard task if the bond dimension needed scales (sub-)exponentially in time. 
 In such a manner, we can potentially locate a mobility edge within the spectrum at fixed parameters of the Hamiltonian.\\

 Before continuing, we also mention Krylov complexity used as a measure for characterizing how operators explore the available space~\cite{PhysRevX.9.041017,rabinovici2022krylov}, which has been recently applied also to the quantum East model here discussed~\cite{menzler2024krylov}, and dynamics of operators using tensor networks~\cite{PhysRevB.96.174201}. However, 
 computing the dynamics of operators generally involves democratically the full spectrum of the Hamiltonian, and so lacks the desired energy resolution needed for detecting a mobility edge.


\section{\label{sec_model}Model}

We study the quantum East model~\cite{PhysRevB.92.100305,PhysRevX.10.021051} in open boundary conditions with Hamiltonian 
\begin{equation}
\label{eq:Hamiltonian}
    \hat{H}=-\frac{1}{2}\sum^{N}_{j=0}\hat{n}_{j}\left(e^{-s}\hat{\sigma}^{x}_{j+1} - 1 \right),
\end{equation}
where $\hat{\sigma}^{\alpha}_{j}$ is the Pauli-$\alpha$ matrix on site $j$; $\hat{n}_{j}=(1-\hat{\sigma}^{z}_{j} )/2$ is the projector onto the state $\ket{1}$ in the local $z$ basis. The term $\hat{n}_j$ in Eq.~\eqref{eq:Hamiltonian} is the kinetic constraint, which translates to a nontrivial action of the Hamiltonian solely to the right (`East') of a previously excited ($\ket{1}$) site. Consequently, the Hamiltonian acts trivially on empty strings without any excited sites to its left, making the location and occupation of the first excited spin a conserved quantity. Thus, the Hilbert space splits into $N$ dynamically disconnected sectors indexed with the position of the first occupied site $\ket{1}$; i.e. the $k$-th sector has $(k-1)$ zeros preceding $\ket{1}$ on the $k$-th site.
This feature is in stark contrast with systems undergoing Hilbert space fragmentation, where the number of disconnected subspaces is $\mathcal{O}(\exp(N))$~\cite{PhysRevX.10.011047,PhysRevX.12.011050,Moudgalya2022} (cf. Sec.~\ref{sec_comparison_HSF} for a comparison with Hilbert space fragmentation).\\

Because of the trivial action of Hamiltonian on empty sites, the results do not depend on the sector considered in the thermodynamic limit. 
Thus, we fix $k=1$ throughout our work, without loss of generality. Once the `East symmetry' sector is fixed, any product state can be dynamically accessed by any other. In other words, the sector is irreducible,  as the Hamiltonian does not possess any other nontrivial (excluding the energy) conserved quantity. 

\subsection{Localization}
Despite being non-integrable and translational invariant, numerical evidence mostly based on exact diagonalization indicates that the quantum East model displays a dynamical transition separating a fast and slow thermalizing phase as a result of the competition of the kinetic term, controlled by $e^{-s}$, and the potential one $\propto \sum_j \hat{n}_j$~\cite{PhysRevB.92.100305,PhysRevX.10.021051}. Intuitively, when the kinetic term dominates ($s\lesssim 0$), excitations propagate ballistically, making the details of the initial state rapidly lost, while instead when it is small $s\gtrsim 0$, excitations propagate slowly, making the details of the initial state \textit{potentially} matter up to long times. 
In Ref.~\cite{PhysRevX.10.021051} it was shown that such dynamical transition is linked to the delocalization-localization transition occurring in the ground state. When $s<0$ the ground state is delocalized; namely, the wave function is spread along the lattice with homogeneous probability and amplitude of finding an occupied site. Instead, the ground state is localized for $s>0$, namely the corresponding wave
functions contain nontrivial excitations only on a small region of the lattice around the first excitation fixing the East symmetry, while it is approximately in the vacuum
state everywhere else. In other words, for $s>0$ the probability of finding an occupied site in the ground state decays exponentially as~\cite{PhysRevX.10.021051}
\begin{equation}
    \braket{\hat{n}_j} \sim e^{-j/\xi},
\end{equation}
where $\xi$ is the localization length, parametrically small in $s$, beyond which the ground state can be approximated as a product state of empty sites. In turn, the localized ground state can be used, together with empty strings, as a basis of area-law states arbitrarily close to true eigenstates of the Hamiltonian, partially explaining the possible slow thermalization in the localized phase~\cite{PhysRevX.10.021051,PRXQuantum.3.020346}. However, such construction is mostly limited to low-energies and so there is still a lack of a complete understanding of the dynamical phase transition at finite-energy density. Here, we aim to investigate such dynamical phase transition more closely, focusing on whether the system hosts a mobility edge despite being disorder-free. However, from exact diagonalization calculations, a sharp separation of a thermal and non-thermal region looks unlikely (see Appendix~\ref{appendix_full_spectrum})~\cite{PhysRevX.10.021051}.  Nonetheless, since many non-thermal eigenstates have a large overlap with product states (see Appendix~\ref{appendix_full_spectrum}), there is the possibility that the system behaves akin to having a many-body mobility edge for a large class of initial states, such as product states, and displays a `state-dependent mobility edge.' 
We will address such open questions using the complexity-oriented proxy introduced in Sec.~\ref{sec_dyn_proxy}. 

\subsection{Comparison with Hilbert space fragmentation \label{sec_comparison_HSF}}
Before presenting our results, we briefly summarize the differences between the quantum East model and systems displaying Hilbert space fragmentation. We do so by briefly summarizing the main properties of fragmented systems, while we refer to, e.g., Ref.~\cite{Moudgalya2022} for a more comprehensive and complete discussion.\\

Given a Hamiltonian $\hat{H}$, the Hilbert space $\mathcal{H}$ on which it acts can be generally decomposed into dynamically disconnected subspaces $\{\mathcal{H}_n\}$, referred to as Krylov subspaces, as
\begin{equation}
    \mathcal{H} = \bigoplus_{n=1}^Q \mathcal{H}_n,\qquad \mathcal{H}_n = \text{span}_t\{e^{-i\hat{H}t}|\psi_n\rangle\} ,
\end{equation}
where $Q$ is the number of Krylov subspaces and
\begin{equation}
\label{eq_spanning}
    \text{span}_t\{e^{-i\hat{H}t}|\psi_n\rangle\} \equiv \text{span}\{|\psi_n\rangle, \hat{H}|\psi_n\rangle, \hat{H}^2 |\psi_n\rangle ,\dots \}
\end{equation}
denotes the subspace spanned by the time evolution of the state $|\psi_n\rangle$. The states $|\psi_n\rangle$ are chosen so that they are not eigenstates of the Hamiltonian and their Krylov subspaces are distinct. More concretely, $|\psi_n\rangle$ are typically chosen to be product states, as they are the ones more easily accessible experimentally, although recent works are investigating the case where they are entangled states~\cite{PhysRevX.12.011050,PhysRevResearch.5.043239,brighi2023hilbert}. In doing such decomposition, a key question concerns how many Krylov subspaces the system displays, as well as the properties of each of them. Concerning the number $Q$ of Krylov subspaces, we can distinguish two main scenarios depending on how $Q$ scales with the system size $N$.	 Specifically, we could have 
\begin{equation}
    Q = \left\{
    \begin{split}
    &\mathcal{O}(N^a), \quad \text{with } a\geq 0\\
    &\mathcal{O}(\exp(N))
    \end{split}\right.
\end{equation}
The first scenario occurs for systems exhibiting no or `conventional' abelian or non-abelian symmetries, where with `conventional' we mean that they can be written as the sum of local terms (e.g. $U(1)$ symmetry), or products of one-site unitary (e.g. $\mathbb{Z}_2$ symmetry). In this case, the different Krylov subspaces are labeled by the quantum numbers associated with such symmetries. Instead, when $Q =\mathcal{O}(\exp(N))$ the system is said to be fragmented, and it occurs when the system displays non-local conserved quantities~\cite{PhysRevX.12.011050} (e.g. the presence of bit strings which are invariant under the action of the Hamiltonian) associated with non-conventional symmetries. Once the different subspaces are labeled, a relevant question concerns investigating their properties, such as whether the system thermalizes or not within each Krylov subspace~\cite{PhysRevB.100.214313,PhysRevB.103.134207,Moudgalya2021}.\\

Given these definitions, it is evident that the quantum East model does not display Hilbert space fragmentation, since it simply conserves  the energy and the location of the first excited site. Given that in a system of size $N$ there are $N$ different ways in which we can locate the first excited site, we can label $N$ different Krylov subspaces. Each of the Krylov subspaces cannot be further reduced, namely, it is ergodic, as seeding any product state belonging to such subspace in Eq.~\eqref{eq_spanning} allows us to explore the whole subspace. However, the remarkable aspect is that despite having the possibility to explore an exponentially large Hilbert space, the way such exploration occurs undergoes a dynamical phase transition from a regime where exploration occurs quickly to a regime where it happens extremely slowly, impeding thermalization for exponentially long times in the system size.
Additionally, we envision the scenario where the efficiency with which the Hilbert space is explored depends not only on the Hamiltonian but also on the initial state seeded $|\psi_n\rangle$. For instance, there could be, at fixed Hamiltonian parameters, a product state $|\psi_n\rangle$ for which the connectivity effectively results higher than others, allowing to explore a larger portion of the accessible Hilbert space faster, which could be enough to make the system thermalize for that specific initial state~\cite{han2024exponentially}. This point of view is equivalent to the original question we posed, about the existence of a \textit{state-dependent} mobility edge in the model at hand, and further justifies why we look at the dynamics of initial product states.

\section{\label{sec_state_dependent}State-dependent mobility edge}

\begin{figure*}[t!]
  \centering
  \includegraphics[width=\linewidth]{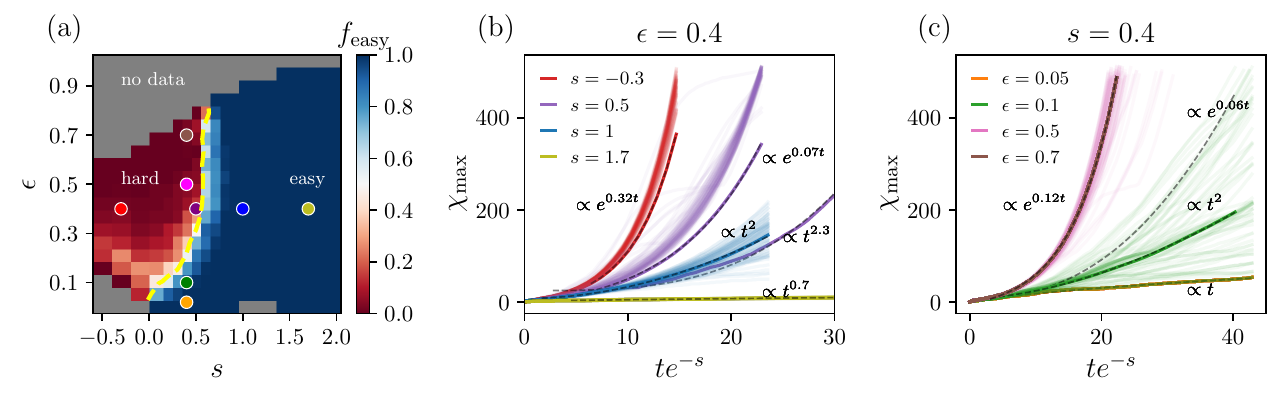}
  \caption{
  (a) Fraction of sampled initial product states (cf. Eq.~\eqref{eq_initial_state}) for which time-complexity is easy $f_\text{easy} \in [0,1]$ as a function of the normalized energy density $\epsilon=(\langle H \rangle-E_\text{min})/(E_\text{max}-E_\text{min})$ and $s$ \ricreview{in a system of size $N=30$}. 
  The grey region indicates the lack of data, as the chosen initial states (cf Eq.~\eqref{eq_initial_state}) do not span the whole energy spectrum. The yellow line serves as guidance to indicate where $f_\text{easy} = 0.5$. 
  Deep in the localized ($s>1$) and delocalized phase ($s<0$), the behavior displays typicality, namely either $f_\text{easy} \approx 0$ (typically hard) or $f_\text{easy} \approx 1$ (typically easy) weakly depending on $\epsilon$. Near the transition point on the localized side ($0 \lesssim s \lesssim 0.5$) $f_\text{easy}$ highly depends on $\epsilon$.
  (b-c) Dynamics of $\chi_\text{max}(t)$ for all the sampled states for different values of $s$ and $\epsilon$, marked in (a). The dashed line represents a fit for a representative state. 
  Deep in the easy and hard regime the growth of $\chi_\text{max}(t)$ weakly depends on the parameters of the initial states. Instead, in the regions separating typically hard or easy we observe different behaviors for states with the same $\epsilon$, signaling a role played by other features of the initial state, such as its spatial structure.
  }
  \label{fig:Fig1}
\end{figure*}

In this section, we investigate the time-complexity and space-complexity of evolving initial product states as a function of their properties, i.e. energy density and spatial structure, and the Hamiltonian parameters. Based on this, we discover a family of localized eigenstates with finite energy density responsible for the observed behavior for many initial states considered, corroborating the sensitivity of our complexity-oriented proxy to non-thermal eigenstates of the Hamiltonian. 

\subsection{Initial states \label{sec_initial_state}}
As we are interested in investigating the existence of a \textit{state-dependent} mobility edge, we need to specify the class of states of our interest. 
To our ends, we investigate the dynamics of initial product states in the computational basis 
\begin{equation}
\label{eq_initial_state}
|\psi\rangle = |1\rangle \bigotimes_{k=1}^N|0/1\rangle_k,
\end{equation}
where we keep the first site fixed to $|1\rangle$, making the dynamics occur in the largest irreducible `East' symmetry sector of our model, while $|0/1\rangle$ could be either $|0\rangle$ or $|1\rangle$. As discussed in Sec.~\ref{sec_dyn_proxy}, a quench protocol could be used to probe information about the eigenstates when the energy variance of the initial state is small. Product states as the one in Eq.~\eqref{eq_initial_state}, with $M$ excitations ($|1\rangle$), have 
\begin{equation}
\label{eq_energy_and_variance}
\begin{split}
\langle \psi|\hat{H} |\psi\rangle = \frac{M}{2},&\quad \langle \psi | \hat{H}^2|\psi\rangle = \frac{e^{-2s}}{4}M + \frac{M^2}{4},\\
\frac{\sqrt{\Delta H}}{\langle\psi| \hat{H} |\psi\rangle} &= \frac{e^{-s}\sqrt{M}}{M} \sim \frac{1}{\sqrt{M}},
\end{split}
\end{equation} 
where $\Delta H \equiv \langle \psi| \hat{H}^2|\psi\rangle - \langle\psi| \hat{H} |\psi\rangle^2$.
Since we are interested in making statements at finite energy density (i.e. we want $\langle \hat{H}\rangle/N$ finite for $N\to \infty$), we set $M=m N$, with $m\in[0,1]$ the density of excitations. In such a manner, the energy density is finite while the relative fluctuation around the mean goes to zero in the thermodynamic limit. Thus, such states are good candidates for analyzing the spectrum of the quantum East model.
Notice that the average energy depends solely on $m$ and not on their location, allowing us to isolate the impact of the spatial structure in the dynamics keeping the energy fixed.

\subsection{Details on the numerical methods \label{sec_sampling_states}}
For a fixed value of $s$ and \ricreview{$M$}, we sample up to $100$ random product states to mitigate sample biases \ricreview{(we discuss the impact of the finite number of sampled initial states in Appendix~\ref{appendix_scaling_analysis_initial_state})}, and we simulate their dynamics using the Time Evolving Block Decimation algorithm~\cite{PhysRevLett.91.147902}.  We keep the Schmidt singular values larger than $10^{-14}$ and we set the timestep $\Delta t=10^{-3}$. We stop the simulation when either the maximum bond dimension reaches $512$ or the time reached is enough to compute the quantities of interest. For each simulation, we investigate how the max bond dimension $\chi_\text{max}(t) = \max_{j\in[1,N-1]}\chi_j(t)$ grows in time by fitting either a polynomial ($\propto t^{\alpha}$) or exponential ($\propto e^{rt}$) function, where $\alpha$ and $r$ are positive constants, depending on which one better approximates the data \ricreview{(cf. Appendix~\ref{appendix_details_fit} for details on the fitting procedure)}. 
Then, we link such behavior to the spectral properties of the Hamiltonian, as detailed in Sec.~\ref{sec_dyn_proxy}. 
We highlight that trotterizing the continuous-time dynamics induces undesired errors tied to the finite time step. However, we do not expect the quantity of interest (i.e., the way the bond dimension grows) to be qualitatively affected by such errors, provided that we use a small enough time step.
We present results \ricreview{mainly for} system size $N=30$, \ricreview{and we discuss how results change as $N$ increases upon performing a detailed scaling analysis in the system size $N$.} 
All the results were obtained using the python package quimb~\cite{Gray2018}.

\subsection{Role of energy density \label{sec_mob_edge_energy}}
Here we focus on the impact of the energy density of the initial states in Eq.~\eqref{eq_initial_state} in dictating the time-complexity of simulating the dynamics.
While time-complexity is typically hard in the delocalized phase ($s<0$), as expected for generic non-integrable models, we remarkably observe both easy and hard regimes in the localized phase ($s>0$) also at finite energy density, indicating the existence of a \textit{state-dependent} mobility edge.\\

To extract the behavior in the thermodynamic limit, we measure the energy with respect to the ground state and most excited state energies, namely the normalized energy density $\epsilon = (\langle \hat{H} \rangle - E_\text{min}) / (E_\text{max} - E_\text{min}) \in [0,1]$, where $E_\text{min}$ and $E_\text{max}$ are the energy of the ground state and the most excited state (which can be computed via DMRG minimizing the energy of $-\hat{H}$), respectively. 
 Despite the advantage of initializing product states (cf. Eq.~\eqref{eq_initial_state}) in isolating the interplay of their properties, as we will discuss, they have a drawback: they do not always allow an efficient sampling over $\epsilon$. Specifically, as $s$ decreases, it is not possible to sample from the extremes of the spectrum, 
as the ground state and the most excited states are `far' from the singly occupied state and the completely filled state, respectively. Nonetheless, as we are mostly interested in the central region of the spectrum, associated with high-temperature, such limitation does not play a major role in our results. Finally, we highlight that the many-body spectrum is not symmetric around $\epsilon=0.5$ ($\hat{H} \neq -\hat{H}$ up to a unitary transformation) and so there are no reasons to expect a symmetric mobility edge. \\

In Fig.~\ref{fig:Fig1}(a), we show the fraction $f_\text{easy} \in [0,1]$ of states for which time-complexity is easy, i.e. $\chi_\text{max}(t)$ grows sub-exponentially in time,  at a given $\epsilon$ and $s$. Specifically, $f_\text{easy}$ is the average, over the sampled initial states at a given energy density $\epsilon$, of a binomial variable which takes the value $0$ if the time-complexity is hard, or $1$ if it is easy. \ricreview{For the sake of clarity, we have grouped the states (sampled as a function of the number of excitations $M$) in a small energy window around any given $\epsilon$.}
As expected, in the delocalized phase ($s<0$) the system mostly displays exponential growth of the bond dimension ($f_\text{easy}\approx 0$) as we move towards the middle of the spectrum $\epsilon$ (cf. Fig.~\ref{fig:Fig1}a), reflecting the thermal nature of the whole spectrum.  On the other hand, for $s>0$ we observe regions where $f_\text{easy}$ is large not only near the extreme of the spectrum but also at finite energy density $\epsilon$ (cf. Fig.~\ref{fig:Fig1}(b)). 
For $0 \lesssim s \lesssim 0.5$ (cf. Fig.~\ref{fig:Fig1}(c)), we observe as $\epsilon$ increases an `inversion' of $f_\text{easy}$, namely for small $\epsilon$ the dynamics is typically `simple' to be simulated ($f_\text{easy}\approx 1$), while moving towards the center of the spectrum dynamics is typically hard to be simulated ($f_\text{easy}\approx 0$).\\ 

\ricrevieww{We address how such a picture changes upon increasing the system size $N$. Specifically, we consider some values of $s \in (0,0.5]$ and we perform the same sampling procedure discussed above up to $N=120$. From the data gathered, the energy density at which the easy-to-hard inversion occurs decreases up to the considered maximal value of $N$ (see Fig.~\ref{fig_random_scaling_analysis}), and it is hard to infer whether it reaches a plateaux. Such behavior could be either a symptom of the thermal nature of the spectrum at any finite energy density for $N\to\infty$; or it could be signaling that other features of the initial state, beyond its energy, are playing a key role in dictating the complexity of simulating its dynamics. We will quantitatively support this latter statement by identifying a 
corner of the Hilbert space $\mathcal{H}_\text{easy}$ that can be easily simulated via tensor networks (see Sec.~\ref{sec_unifying_picture}). We will do so by investigating the role of the distribution of the initial excitations in Secs.~\ref{sec_mob_edge_structure} and~\ref{sec_fragmented_ergodic_transition}, which will lead to the discovery of a large number of localized eigenstates at finite energy density in Sec.~\ref{sec_mob_edge_unifying_picture}. Here, we anticipate the features of $\mathcal{H}_\text{easy}$.\\
\\
Product states with large weight in overlap on $\mathcal{H}_\text{easy}$ have the following features: 
\begin{enumerate}[(i)]
    \item \textit{not too large} clusters of excitations (consecutive $|1\rangle$);
    \item \textit{far enough} clusters of excitations.
\end{enumerate}
We refer to Sec.~\ref{sec_mob_edge_unifying_picture} for the details about how \textit{large} and how \textit{far} cluster of excitations should typically be. Here instead, we stick to a more intuitive picture. Conditions (i-ii) are a direct consequence of the kinetic-constrained nature of the system, which gives a sharp differentiation between active ($|1\rangle$) and inactive ($|0\rangle$) regions. Intuitively, (i) implies low activity within each cluster, while instead (ii) between different clusters, as we prove in Secs.~\ref{sec_mob_edge_structure}~\ref{sec_fragmented_ergodic_transition}.\\
\begin{figure}[t!]
\centering
\includegraphics[width=\linewidth]{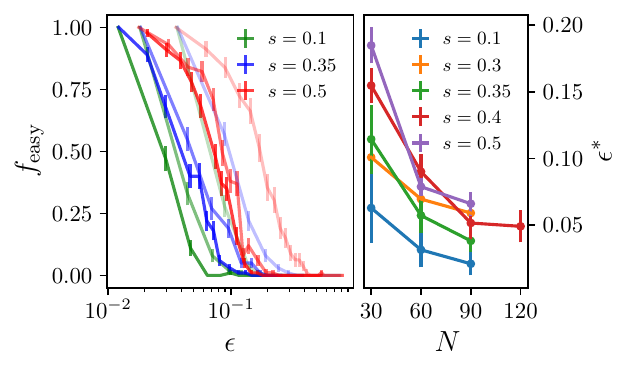}
\caption{Left panel: Fraction of randomly sampled initial product states for which time-complexity is easy ($f_\text{easy}$) as a function of the energy density $\epsilon$ at fixed $s$ and different system sizes $N=\{15,30,60,90,120\}$ (from light to dark curves). The bars correspond to the statistical uncertainty due to the finite number of initial states sampled. Deep in the localized phase (e.g. $s=1$), the results are not affected \ricrevieww{up to the largest} $N$ \ricrevieww{considered} and simulations are always easy (not shown). Instead, the closer $s$ is to the localization point ($s= 0$), the more $f_\text{easy}$ is affected by $N$. \ricrevieww{This arises from factors beyond energy, such as the spatial distribution of excitations, playing a key role in the dynamics as $s$ decreases. Therefore, these factors should be considered when sampling initial states (see the main text for the required ingredients).} 
Right panel: $\epsilon^*$, where $f_\text{easy}(\epsilon^*)=0.5$, as a function of $N$. The error bars come from the statistical uncertainty in $f_\text{easy}$ which makes multiple $\epsilon$ compatible, within a standard deviation, with our definition of $f(\epsilon^*)=0.5$. 
}
\label{fig_random_scaling_analysis}
\end{figure}
From (i-ii) we can also guess the characteristics of $\mathcal{H}_\text{easy}$ in relation to the energy density $\epsilon$ and $s$, explaining what has been observed so far.  Intuitively, as $s$ is positive and increases, the kinetic constraint is stronger, and criteria (i-ii) become looser. On the other hand, due to the one-to-one correspondence between energy density $\epsilon$ and the density of excitations $M/N$, it is obvious that (i-ii) are more easily satisfied the smaller is $\epsilon$. In principle, it is possible that above a certain $\epsilon$, (i-ii) cannot be fulfilled. To appreciate the consequences of this, let us split $\mathcal{H}_\text{easy}$ based on the energy density $\epsilon$ of the states belonging to it. Specifically, we write it as}
\begin{equation}
    \ricrevieww{\mathcal{H}_\text{easy} = \bigcup_{\epsilon} \mathcal{H}_\text{easy}(\epsilon,s)},
\end{equation}
\ricrevieww{where $\mathcal{H}_\text{easy}(\epsilon,s)$ contains the easily simulable states having energy density $\epsilon$ at a given $s$. From (i) and (ii) it follows that the size of each $\mathcal{H}_\text{easy}(\epsilon,s)$ is determined by combinatorial arguments, which can be approximated by exponentials. Thus, the dimension $\text{dim}(\mathcal{H}_\text{easy}(\epsilon,s)) = \mathcal{O}\left(e^{\beta(\epsilon,s) }\right)$, where $0 \leq \beta(\epsilon,s) \leq 1$. For any $s<0$, the system is delocalized and $\beta(\epsilon>0,s<0) = 0$, since only low-energy initial states are easily simulable (as generic in interacting theories). For $s>0$, $\beta(\epsilon,s>0)>0$ up to a not too large $\epsilon$, as otherwise (i-ii) cannot be satisfied. In the extreme case of $s = \infty$, $\beta(\epsilon,s=\infty)=1$ for any $\epsilon$, as the chain is disconnected and so dynamics is trivially easy for any initial state. Excluding the trivial $s = \infty$ case, the easily simulable subspace is still a zero fraction of the whole Hilbert space of dimension $\mathcal{D}=2^N$, since $\mathcal{D}_\text{easy}/\mathcal{D} \to 0$ for $N \to \infty$.}\\
\\
\ricrevieww{Since $\mathcal{D}_\text{easy}/\mathcal{D} \to 0$, the probability of extracting a product state that has a large overlap with $\mathcal{H}_\text{easy}$ from a uniform probability distribution, goes to $0$ as $N$ increases. This explains the observation in Fig.~\ref{fig_random_scaling_analysis}, where we observe the hard-to-easy transition moving towards smaller values of $\epsilon$ as $N$ increases. However, the simultaneous presence of easy and hard regions is still observed at sizable values of $N$, cf. Fig.~\ref{fig:Fig1}(a), since states with the features (i) and (ii) can remain abundant below a certain energy density $\epsilon^*$. The value of $s$    plays also a role:  for  large values of $s$,   $\mathcal{H}_\text{easy}$ grows bigger, since the system is more localized and  entanglement   suppressed.  For instance, at $s=1$, we find that  for $N=60$ (not shown) all the simulations are  still easy.\\
\\
The presence of easy-simulable many-body states below a certain energy density, is   the behavior observed in a canonical MBL system hosting a mobility edge~\cite{PhysRevB.92.064203,PhysRevB.91.081103,Naldesi2016}, with the caveat that in our model we have to take into account also the structure of many-body states, besides  energy density -- hence the name state-dependent mobility edge. We refer to Fig.~\ref{fig_comparison_QMBS_MBL_SDME} for a summary of the differences between the typical mobility edge and a state-dependent mobility edge. The rest of the paper is devoted to guide the reader towards a quantitative  identification of the easily simulable corner of the Hilbert space $\mathcal{H}_\text{easy}$, which, as we have seen, can still affect the physics of the model up to large values of $N$ (for instance, in real-time dynamics).}
\ricrevieww{We highlight that taking into account features of the initial state is reminiscent of what occurs in quantum many-body scarred systems, where non-thermal behavior is observed in the dynamics of some initial states~\cite{Bernien2017,Turner2018,PhysRevLett.122.220603,Serbyn2021,PhysRevB.98.155134,PhysRevB.98.155134,PhysRevB.99.161101,PhysRevX.11.021021,Moudgalya2022,chandran2023quantum, PhysRevX.13.031013}. Still, in the latter, such anomalous dynamics occur for $\mathcal{O}(1)$ initial states, while in our case for $\mathcal{O}(e^{N^\beta})$ states.}\\ 

\begin{figure*}[t!]
\centering
\includegraphics[width=0.33\linewidth]{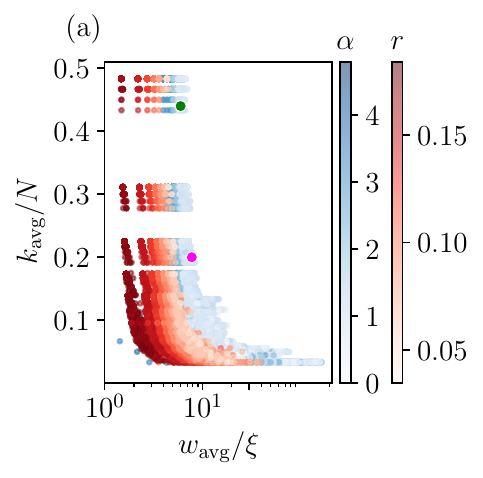}
\includegraphics[width=0.63\linewidth]{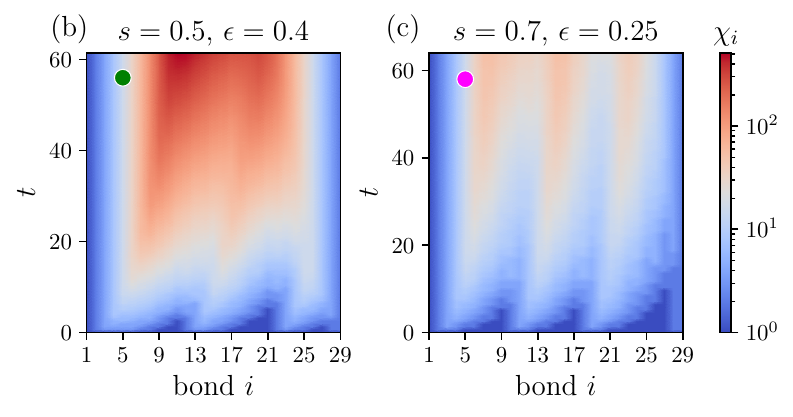}
\caption{
(a) The same data as presented in Fig.~\ref{fig:Fig1} labeling the states based on the initial average size of clusters of consecutive excitations ($|1\rangle$) normalized by $N$ ($k_\text{avg}/N$) and average distance between clusters $w_\text{avg}$ in units of $\xi$ ($w_\text{avg}/\xi$), with $\xi$ the localization length of the ground state at the respective value of $s$. The red dots correspond to hard time-complexity ($\chi_\text{max}(t) \propto e^{rt}$) with exponential rate $r$. The blue dots correspond to easy time-complexity ($\chi_\text{max}(t) \propto t^\alpha$) with power-law exponent $\alpha$. Simulations are typically hard if either $k_\text{avg}/N$ increases and $w_\text{avg}/\xi$ decreases. (b-c) dynamics of the bond dimension $\chi_i$ along each cut for two representative states. In (b), clusters of excitations have started to entangle beyond a transient time. Whereas in (c), clusters of excitations haven't entangled yet due to the presence of large islands of consecutive $|0\rangle$.
}
\label{fig_dependence_on_spatial_distribution}
\end{figure*}

\subsection{\label{sec_mob_edge_structure}Role of spatial structure}
\ricrevieww{In order to identify the easily simulable corner of the Hilbert space, we address} the role of the initial state structure in the time-complexity. In doing so, we show how the size of excited regions and their distance plays a crucial role in dictating the time-complexity of simulating their dynamics~\cite{PhysRevB.92.100305,PhysRevX.10.021051}. Additionally, we show that the localization length of the ground state serves as a useful length scale in the system for understanding finite-energy density phenomena. \ricrevieww{}\\

Long-time dynamics of slowly thermalizing systems can depend on features of the initial state beyond the conserved quantities of the system. In kinetically constrained models (KCMs), the spatial structure of the initial state plays a key role up to extremely long times, in a manner reminiscent of glassy systems~\cite{PhysRevLett.121.040603,lan2018quantum,zadnik2023slow,PhysRevX.10.021051,PhysRevB.92.100305}.
Such a dependence stems from the sharp differentiation between active and inactive regions in KCMs. Typically, as in the quantum East model, active regions are constituted by $|1\rangle$, while inactive regions by $|0\rangle$. As a result, a natural parameter to capture the spatial structure is the initial size $w$ of inactive regions (number of consecutive $|0\rangle$) as it controls when active regions will entangle~\cite{PhysRevB.92.100305}. However, this parameter does not contain information regarding the size of the active regions. 
To capture such information, we look at the initial size of active regions $k$ (number of consecutive $|1\rangle$), which is also in a one-to-one correspondence with their average energy (cf. Eq.~\eqref{eq_energy_and_variance}) for initial product states in the computational basis.\\

As we are dealing with initial states with multiple active regions, we trade $k$ and $w$ with their averages over the system $k_\text{avg} $ and $w_\text{avg}$, respectively. \ricreview{Operatively, $k_\text{avg}$ is equal to the total number of $|1\rangle$'s divided by the number of distinct active regions. Instead, $w_\text{avg}$ is the total number of $|0\rangle$'s divided by the number of inactive regions. For example, given the state $|11000101\rangle$, we have:
\begin{equation}
\begin{split}
    k_\text{avg} &= \frac{2+1+1}{3} = \frac{5}{3},\\
    w_\text{avg} &= \frac{3+1}{2} = 2.
\end{split}
\end{equation}
} Inspecting the time-complexity as a function of these two parameters, we observe that they are not so effective in predicting the time-complexity of simulating the states at hand (see Appendix~\ref{Appendix_importance_w}).
However, by including the localization length $\xi$ of the ground state at the corresponding $s$ (we restrict to $s>0$), \ricreview{which we extract by fitting the occupation number $\langle \hat{n}_i\rangle \sim e^{-j/\xi}$ on the ground state obtained via DMRG,} we observe that the predictive power improves (see Fig.~\ref{fig_dependence_on_spatial_distribution}(a)). 
The localization length $\xi$ serves as a good length scale since active regions expand exponentially slowly and remain within few $\xi$ up to long times in the localized phase. 
As $k_\text{avg}$ increases the simulations are typically harder, while instead, the opposite occurs as excitations are farther apart. Such behavior could be understood as the interplay of dynamics within each cluster of excitations and between different ones. 
Intuitively, if clusters are far from each other on average ($w_\text{avg}/\xi\gg 1$), the time-complexity is mostly dictated by the dynamics within each cluster, since the propagating fronts generate little entanglement.
In turn, the larger the cluster is, and so is its energy, the more it is typically hard to simulate, in agreement with Fig.~\ref{fig:Fig1}(a).\\

A possible analogy to explain such behavior is given by looking at the system as a collection of subsystems with a certain temperature, directly linked to the number of excitations, separated by completely inactive regions at zero temperature (the empty state $|00\dots 0\rangle$ is the true ground state of the quantum East model): if the inert regions are too extended, the hottest source dominates the hardness in simulating the dynamics.
Such a picture is supported by inspecting more closely how the bond dimension, or equivalently entanglement, at each possible bipartition, evolves in time (cf. Fig.~\ref{fig_dependence_on_spatial_distribution}(b,c)). In Fig.~\ref{fig_dependence_on_spatial_distribution}(b) islands of excitations have started to appreciably entangle beyond a transient time. Instead, in 
Fig.~\ref{fig_dependence_on_spatial_distribution}(c) clusters of excitations are not strongly entangled due to the presence of large inactive regions between them.\\

The distinction between inactive and active regions opens up the possibility of understanding the dynamics of generic product states in terms of concatenated and weakly entangled clusters in the localized phase. We further investigate this observation in the next sections.

\subsection{Transition in space-complexity \label{sec_fragmented_ergodic_transition}}
We have observed that the structure of the initial state plays a leading role in dictating the time-complexity in the localized phase. Such a dependence stems from the sharp differentiation between active and inactive regions in kinetically constrained models (KCMs). Here, we go a step further showing that such distinction also leads to a crossover in the space-complexity from a fragmented regime, where the state is separable, to a fully ergodic one, where the state is non-separable.\\

Heterogeneity is a hallmark of classical glassy systems, which manifests also in their quantum counterpart as observed in other studies~\cite{PhysRevLett.121.040603,zadnik2023slow}. The quantum East model makes no exception, as it is evident from the dynamics of various observables, such as occupation number and entanglement entropy. 
This feature opens up the possibility of distinguishing two timescales: one where dynamics occurs \textit{mostly} within each cluster of excitations (intra-cluster), and another when it also \textit{appreciably} involves different clusters of excitations (inter-cluster). 
Specifically, we could define a time $t^\star$ up to which the system is approximately separable, since the entanglement between the different clusters is negligible, and the whole dynamics is encoded in the dynamics of each cluster separately. Such separability in the evolved state could be formalized using the Lieb-Robinson bound~\cite{lieb1972finite}: quantum correlations propagate at most ballistically with exponentially small corrections in systems with a finite local Hilbert space and short-range interactions. In KCMs, empty regions are completely inactive, and the finite velocity of propagation of entanglement in the system (upon neglecting the exponentially small correction in the distance) implies the separability just mentioned up to time $t^\star$ where the active regions appreciably merge. 
As a consequence, an initial product state (cf. Eq.~\eqref{eq_initial_state}) with $\mathcal{N}$ cluster evolves as
\begin{equation}
\label{eq_dynamics_product_states}
    |\psi(t)\rangle \approx  \bigotimes_{n=1}^{\mathcal{N}}|\psi_n(t)\rangle \quad t \lesssim t^\star
\end{equation}
where $|\psi_n(t)\rangle$ is the time-evolved state describing the $n$-th cluster. In other words, the evolved state is given by a product state of each cluster, i.e. it is separable (up to exponentially small corrections coming from the propagating fronts).
The transition from separability to fully ergodic is accompanied by a change in the dimension of the accessible Hilbert space $\mathcal{D}$. Specifically,
\begin{equation}
\mathcal{D}(t)  
= \left\{
    \begin{split}
    \sim \mathcal{N} \times 2^{N/\mathcal{N}}& \qquad t \lesssim t^\star,\\
    2^N & \qquad t \gtrsim t^\star,\\
    \end{split}
   \right.
\end{equation}
which in words translates, keeping $N/\mathcal{N}$ fixed, to a transition in the dimension of the accessible Hilbert space from \textit{polynomial} to \textit{exponential} in the system size $N$. Such transition could be interpreted as a transition between a fragmented regime~\cite{PhysRevX.10.011047,PhysRevX.12.011050,Moudgalya2022} to a fully ergodic one. 
However, we once more highlight that such a transition is present in the quantum East model upon neglecting the exponential small corrections coming from the Lieb-Robinson bound. Instead, in exactly fragmented systems the different Hilbert space sectors are completely disconnected, and the time $t^\star$ is formally infinite.\\

\begin{figure}[t!]
\centering
\includegraphics[width=\linewidth]{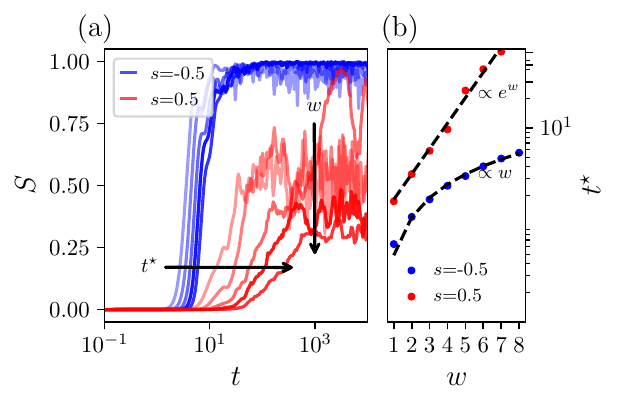}
\caption{Dynamics of the entanglement entropy $S$, computed on the last bond, upon initializing the state $|111\rangle \otimes |0\rangle^w \otimes |1\rangle$ with $w\in[1,8]$ in the delocalized ($s<0$) and localized phase ($s>0$). 
(b) Time $t^\star$ at which the two clusters appreciably entangled, i.e. when $S>\Delta$,  as a function of $w$. We set $\Delta=10^{-2}$ for the sake of concreteness (see text for a more compelling discussion in terms of Lieb-Robinson bound). In the delocalized phase ($s<0$), $t^\star$ grows linearly in $w$ as excitations propagate ballistically. Instead, $t^\star$ grows exponentially in $w$ in the localized phase ($s>0$), due to the extreme slowdown in dynamics. The results were obtained via exact diagonalization.}
\label{fig_merge_times}
\end{figure}
As the Lieb-Robinson bound constitutes an upper bound, it is in principle possible to observe slower than ballistic propagation. This is the case for the quantum East model, where the propagation could be exponentially slow in the localized phase~\cite{PhysRevB.92.100305}. As a consequence, the dependence of the time $t^\star$ with the distance $w$, defined as the number of empty sites $|0\rangle$ between two clusters, strongly depends on whether the system is localized or delocalized. To test this, we compute the evolution of the entanglement entropy between two clusters, of size $3$ and $1$ respectively, at a distance $w$, i.e. $|\psi(t=0)\rangle = |111\rangle \otimes |0\rangle^w \otimes |1\rangle$. Specifically, we look at the entanglement entropy $S$ on the last bond, and we define a threshold $\Delta$ so that we consider the two clusters not entangled if $S<\Delta$. 
The specific value of $\Delta$ is arbitrary and chosen just to show the dependence of $t^\star$ on $w$ (cf. Fig.~\ref{fig_merge_times}).
In the delocalized phase, $t^\star$ grows linearly in $w$ as excitations propagate ballistically. Instead, $t^\star$ grows exponentially in $w$ in the localized phase, due to the extreme slowdown in dynamics (cf. Fig.~\ref{fig_merge_times}(b)). 
Since $t^\star$ could be very large in the system size, our observation could prove valuable in different directions:
\begin{enumerate}[(i)]
\item For time $t<t^\star$ the dynamics of the whole system can be efficiently simulated as a collection of its subsystems making a negligible error~\cite{Haah2021,AnthonyChen2023}. 
\item Separability up to time $t^\star$ justifies the investigation of smaller system sizes in order to grasp the behavior or larger ones. Specifically, from Eq.~\eqref{eq_dynamics_product_states}, the dynamics of a product state given by $\mathcal{N}$ cluster is completely encoded looking at the dynamics of $\mathcal{N}$ single-cluster states up to time $t^\star$. 
\end{enumerate}
Based on (ii), we now investigate the time-complexity in simulating the time evolution of single-cluster states, namely kink states. We will show that such investigation not only will confirm the observed time-complexity in the previous sections, but also allow us to identify a class of area-law states responsible for such behavior. This will provide evidence of the sensitivity of our complexity-oriented proxy to non-thermal eigenstates.

\begin{figure}[t!]
\centering
\includegraphics[width=1\linewidth]{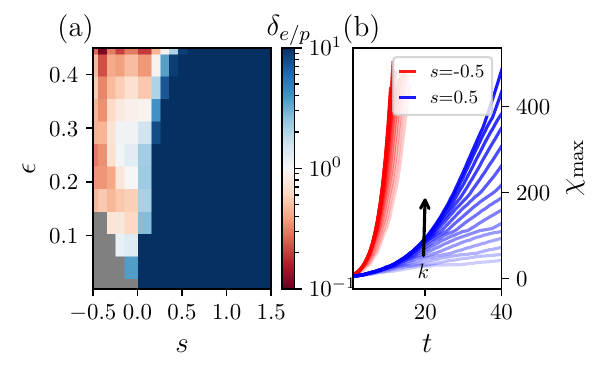}
\caption{Time-complexity upon initializing kink states  $|\bm{k}\rangle = \otimes_{j=1}^k|1\rangle \otimes_{j=k+1}^N \otimes |0\rangle$ with $k\in [1,N/2]$ with $N=30$. Polynomial and exponential fits are performed to determine the complexity of simulating these states. (a) Relative error $\delta_{e/p}$ between the exponential and polynomial fits as a function of energy density $\epsilon$ and $s$. For $\delta_{e/p}<1$, the state is hard to simulate. Conversely, for $\delta_{e/p}>1$, the state is easy to simulate. We see a complexity trend similar to the one observed in Fig.~\ref{fig:Fig1}(a), signaling that most of the complexity of simulating product states could be understood in terms of the evolution of kink states. (b) Dynamics of $\chi_\text{max}$ for kink states with varying $k$, in the delocalized ($s=-0.5$) regime and in the localized ($s=0.5$) regime. In the delocalized regime, states are hard to simulate irrespective of $k$. Instead, in the localized regime, $k$ has a huge impact on the time-complexity.}
\label{fig_complexity_kink}
\end{figure}
\subsection{Localization at finite energy-density \label{sec_mob_edge_unifying_picture}}
Following the conclusions of the previous section, we investigate the time-complexity of simulating the dynamics of the kink states
\begin{equation}
\label{eq_kink_state}
|\bm{k}\rangle = |1\rangle^k \otimes |00\dots 0\rangle.
\end{equation}
As always for product states in the computational basis, we have a one-to-one \ricreview{correspondence} between $k$ and the energy density $\epsilon$. Since we now have a single state for each point in the parameter space $\epsilon$ vs $s$, we show in Fig.~\ref{fig_complexity_kink} the relative error $\delta_{p/e}$ between the polynomial and the exponential fit performed on the time evolution of the max bond dimension $\chi_\text{max}(t)$. If $\delta_{p/e}>1$, the exponential fit better approximates the data, while instead vice versa if $\delta_{p/e}<1$. Time-complexity of simulating kink states follows a similar trend as the one observed in Fig.~\ref{fig:Fig1}(a). This signals that, as expected, the hardest part to simulate is given by regions densely excited, or in other terms the `hottest' regions.\\

To better understand our observations, we more closely investigate the eigenstates of the Hamiltonian. In particular, from Exact Diagonalization calculations, we discover that the kink states in Eq.~\eqref{eq_kink_state} have a large overlap with a limited number of eigenstates of the Hamiltonian deep in the localized phase. Such eigenstates have a density profile similar to a kink state with an additional localized tail on the right edge. For this reason, we term them \textit{localized kink states}. Due to the limited system sizes accessible via Exact Diagonalization, we further characterize such states resorting to DMRG-X, a variation of the standard DMRG, which allows finding area-law eigenstates which are `near' (in overlap) to the initial state seeded to the algorithm~\cite{PhysRevLett.116.247204}. In our case, we seed the kink states (cf. Eq.~\eqref{eq_kink_state}) and let the algorithm find the best approximating eigenstate, setting a maximal bond dimension of $100$. For each of these states, the algorithm could either fail to converge, which we interpret as the absence of a localized kink state, or converge.\\
\begin{figure}[t!]
\centering
\includegraphics[width=\linewidth]{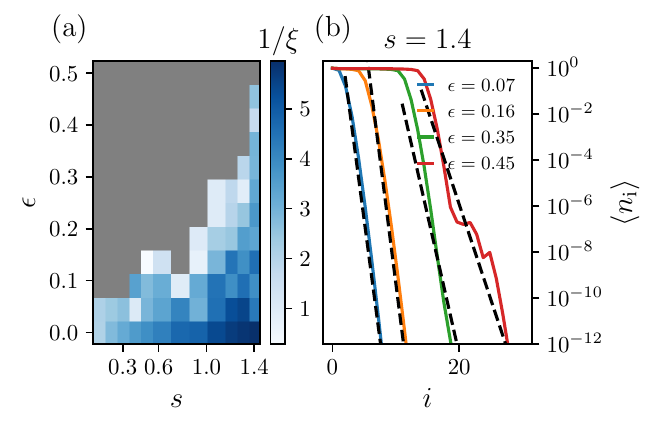}
\caption{(a) Inverse of the localization length $\xi$ of localized kink eigenstates (variance of energy $<10^{-5}$) obtained via DMRG-X~\cite{PhysRevLett.116.247204}, as a function of their energy density $\epsilon = (\langle H\rangle - E_\text{min})/(E_\text{max}-E_\text{min})$ in a system of size $N=30$. The algorithm converges not only at zero energy density ($\epsilon=0$), corresponding to the ground state, but also at finite energy $\epsilon>0$, indicating the existence of localized kink-states along the spectrum for $s>0$. (b) Density profile of some localized kink states found via DMRG-X for different $\epsilon$ at fixed $s=1.4$. The dashed lines are the exponential fit used to extract $\xi$.}
\label{fig_localized_kink_states}
\end{figure}

In Fig.~\ref{fig_localized_kink_states} we show the localization length $\xi$ of the states for which the algorithm was able to converge (we set as convergence criteria a variance of the Hamiltonian $<10^{-5}$). We observe that the algorithm can find a state with $\epsilon=0$, faithfully reproducing the results from DMRG, as it corresponds to the ground state of the Hamiltonian. Additionally, the algorithm can find other localized states at finite energy density as $s$ increases, corresponding to the localized kink states above introduced. We observe that $\xi$ is parametrically small in $s$, while instead it is parametrically large in $\epsilon$. In Appendix~\ref{appendix_dmrgx} we provide further details on the results obtained via DMRG-X.
This further provides evidence of the effectiveness of our complexity-oriented quantity for detecting the presence of thermal and non-thermal eigenstates.

\subsection{A unifying picture \label{sec_unifying_picture}}
The localized kink states found could be used, together with empty strings, as building blocks for defining a large class of states close to eigenstates of the Hamiltonian for larger system sizes. To show this, let us name $|\widetilde{k}^N\rangle$ the localized kink-states, with $k$ the number of excitations in the kink (excluding the localized tail) and $N$ the system size.
First of all, we observe that the state 
\begin{equation}
|\Psi_k^L\rangle \equiv |\widetilde{k}^N\rangle \otimes |0\rangle^{\otimes (L-N)},
\end{equation}
with support on $L>N$ sites, has energy variance 
\begin{equation}
\label{eq_superk_variance}
\langle \Psi_k^L|\Delta \hat{H}|\Psi_k^L\rangle \sim e^{-s} \langle  \widetilde{k}^N|\hat{n}_N|\widetilde{k}^N\rangle \sim e^{-s} e^{-(N-k)/\xi}
\end{equation}
which depend solely on the occupation on the last site of the state $|\widetilde{k}^N\rangle$, since the only contribution to the variance comes from the boundary term between $|\widetilde{k}^N\rangle$ and the subsequent string of empty sites. Since the variance in Eq.~\eqref{eq_superk_variance} is independent on $L$, it is small if $(N-k) \gg \xi$, i.e. $|\widetilde{k}^N\rangle$ is localized. Since $\xi$ does not scale with $N$, such a condition is satisfied if $k/N \ll 1$ for large $N$, which can be fulfilled for $k < c N$ with $c \ll 1$ a small constant and $k < N^p$ with $p<1$. Recalling the correspondence between the number of excitations and the average energy, the former case corresponds to states with an extensive energy, while the latter to states with a sub-extensive one. The bound in $k$ imposed is quite strict in practice and could be relaxed due to the exponential dependence of the variance on $(N-k)/\xi$. Specifically, by setting a target variance $\varepsilon$ in Eq.~\eqref{eq_superk_variance}, it would be enough to have $(N-k) \sim \xi \log(1/\varepsilon)$.\\

Having proven that the localized kink states found at finite $N$ are quasi-eigenstates of larger systems, we can go a step further. Specifically, we could use them as a basis for states in a system of size $L>N$ \ricreview{by concatenating them one after the other in real space}. Allowing $k \in [0,k_\text{max}]$, where $|\widetilde{0}\rangle \equiv |0\rangle^{\otimes N}$ and $k_\text{max}$ corresponds to the maximum number of excitations for which the system display localization \ricreview{for a given $s$}, we can define $(k_\text{max}+1)^{L/N}$ states, \ricreview{which can be exponentially many in the system size $L$}. Let us call $|\Psi^L\rangle \in \{|\widetilde{0}^N\rangle,|\widetilde{1}^N\rangle,\dots ,|\widetilde{k}_\text{max}^N\rangle\}^{\otimes L/N}$. Since $|\widetilde{0}^N\rangle$ blocks do not contribute to the variance, the only contribution comes from the states $|\widetilde{k}^N\rangle$, giving 
\begin{equation}
\langle \Psi^L|\Delta \hat{H}|\Psi^L\rangle \sim e^{-s}\sum_{j=1}^\mathcal{M} e^{-(N-k_j)/\xi} < \mathcal{M}e^{-(N-k_\text{max})/\xi},
\end{equation}
where $\mathcal{M} = L/N$ is the number of states $|\widetilde{k}\rangle$ with $k>1$, and $k_j$ is the size of the $j$-th concatenated localized kink state. We highlight that $\mathcal{M}$ is potentially unbounded in the thermodynamic limit, making the variance inevitably large. The energy is given by
\begin{equation}
\langle \Psi^L | \hat{H}|\Psi^L\rangle \sim \sum_{j=1}^\mathcal{M} \langle \widetilde{k}_j|\hat{H}|\widetilde{k}_j\rangle \sim \sum_{j=1}^\mathcal{M} k_j.
\end{equation}	
Given these results, we now aim to find a trade-off between small variance and finite energy density. To this end, let us embed the maximum number of localized states possible, namely $\mathcal{M}=L/N$, and consider that the kink states embedded have $k\sim c N$, with $c\ll 1$. Let us consider the case where $N=L^\alpha$, with $\alpha<1$, to which corresponds $\mathcal{M}=L^{1-\alpha}$ and $k \sim c L^{\alpha}$. For this parametrization, we have $\langle \Psi^L| \hat{H}|\Psi^L\rangle \sim cL$ and $\langle \Psi^L| \Delta \hat{H}|\Psi^L\rangle < L^{1-\alpha} e^{-L^\alpha/\xi}$, implying that we can have states with finite energy density states and small variance, which are thus slowly evolving.\\
\begin{figure}[t!]
    \centering
    \includegraphics[width=\linewidth]{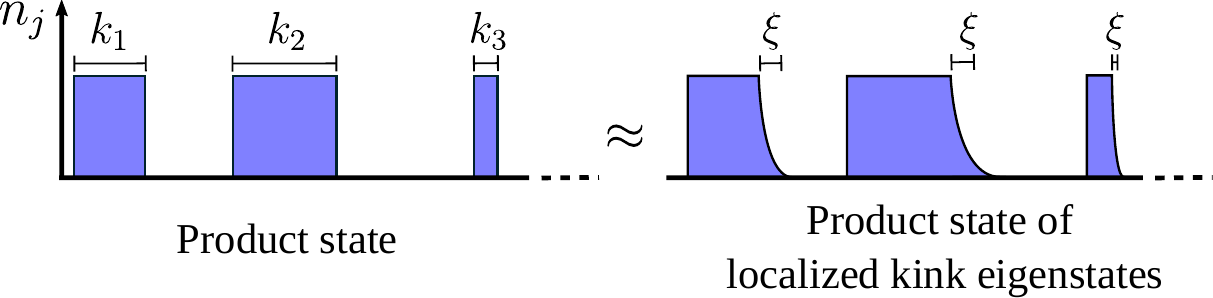}
    \caption{\ricrevieww{An initial product state (left) can be well approximated by a product state of localized kink states (right) in the localized phase. In this case, the dynamics of the former are slow and efficiently simulable up to long times (at least exponentially long in the distance between the kink states, following the approximate separability discussed in Sec.~\ref{sec_fragmented_ergodic_transition}).}}
    \label{fig_sketch_overlap}
\end{figure}

\ricrevieww{
The existence of localized kink eigenstates, which include the ground state~\cite{PhysRevX.10.021051},  allows us to explain the ease of simulating the dynamics of a large class of initial states deep in the localized phase. Specifically, product states in the computational basis can be potentially very close (in overlap) to concatenated localized kink states. If so, such a product state would have small energy variance and be slowly evolving, making its simulation via tensor network efficient. Operatively, for a given $s$, a state given by: 
\begin{enumerate}[(i)]
    \item  concatenated clusters of excitations of size $k \leq k_\text{max}(s)$;
    \item  separated by empty regions larger than the localization length $\xi$ (where $\xi \sim 1$ as computed in Fig.~\ref{fig_localized_kink_states});
\end{enumerate}
fulfill such requirements  (Fig.~\ref{fig_sketch_overlap} offers a cartoon).
 As a remark, such conditions are the ones mentioned in Sec.~\ref{sec_mob_edge_energy}, here rigorously derived.}\\

\ricrevieww{The number of states which fulfill such conditions are exponentially many in the system size $L$, as its size goes like $(k_\text{max}(s)+1)^{L/N}$, where $N$ is the size of the localized kink state. Such subspace constitutes the easily simulable corner $\mathcal{H}_\text{easy}$ of the Hilbert space mentioned in Sec.~\ref{sec_mob_edge_energy}, which even though exponentially large in $L$, is still a small fraction of the whole Hilbert space of dimension $2^L$ as $L$ increases. 
This explains the observation in Sec.~\ref{sec_mob_edge_energy}, where the energy density (or in other words the density of excitations) is not so informative on its own as the system size increases: randomly sampling initial product states does not efficiently target the desired subspace spanned by localized kink states. However, as we have now elucidated the structure of such subspace, it becomes possible to guide the sampling procedure to target the desired subspace. Also, from (i-ii) it follows that $\mathcal{H}_\text{easy}$ contains states up to a certain energy density $\epsilon^\star$ (density of excitations), since above this threshold, (i-ii) cannot be fulfilled. Above such energy density, excluding the existence of other easily simulable subspaces, the time-complexity of simulations is always hard. This constraint gives rise to the notion of a state-dependent mobility edge at the core of this paper.}

\section{Discussion and Perspectives \label{sec_conclusions}}
In this work, we have characterized the kinetically constrained quantum East model by inspecting the complexity of simulating the dynamics via tensor networks. In doing so, we have distinguished complexity along the time (time-complexity) and space (space-complexity).  We have linked the ease of time-complexity to the presence of numerous non-thermal eigenstates at finite energy densities, a subset of which we have computed via DMRG-X. The abundance of non-thermal eigenstates causes the system to exhibit behavior akin to having a mobility edge when initializing \ric{certain} product states, a scenario we term the \textit{state-dependent} mobility edge. Space-complexity, on the other hand, is susceptible to the separability of the evolved state. We have shown that the sharp distinction between active and inactive regions, characteristic of kinetically constrained models, induces a transition in space-complexity during the dynamics of initial product states, as summarized by Eq.~\eqref{eq_dynamics_product_states}.\\

\ricreview{Although generic non-integrable systems are not expected to display a dynamical transition in time-complexity, it would be natural to ask whether similar behavior can be found in other interacting systems featuring non-ergodic dynamics, for instance in models with dynamical confinement~\cite{kormos2017real}, where non-thermalizing oscillations are observed for a broad set of Hamiltonian parameters. In a recent work  we have proposed how to tune between this model and the quantum East model using a Rydberg array which combines staggered drive fields and anti-blockade~\cite{PhysRevLett.132.223201}. It would be intriguing to study the fate of the complexity transition as the drive fields are varied such that Rydbergs dynamics exit the East regime to enter dynamical confinement. A long-term related question is whether this type of transition can present different universality classes and which would be the key ingredients dictating them, in terms of symmetries, dimensionality, and nature of degrees of freedom involved. While the topic of dimensionality will be shortly covered in Sec.~\ref{sec:highd}, we observe that the role of local Hilbert space size could be easily addressed through the bosonic variant discussed in Refs.~\cite{PRXQuantum.3.020346,geissler2023slow}. This is not a mere technical exercise, since it has been shown in~\cite{PRXQuantum.3.020346} that the localized phase of the bosonic East model can host non-Gaussian entangled states, therefore offering a potential connection between the complexity-transition explored in this work, and quantum metrological gain. }\\

Furthermore, the transition encoded in Eq.~\eqref{eq_dynamics_product_states} is reminiscent of the dynamical transition in the complexity of performing the so-called sampling task~\cite{aaronson2010computational,Lund2017}. The sampling problem involves the extraction of events according to the probability distribution provided by the many-body quantum state, and it is widely regarded as a leading contender in demonstrating provable quantum advantage. 
Additionally, and more relevant for us, the sampling problem has gathered attention also for defining novel kinds of dynamical phase transitions linked to an easy to hard transition during dynamics, which could already happen in free bosons systems in a lattice~\cite{PhysRevLett.121.030501}. Specifically, in Ref.~\cite{PhysRevLett.121.030501}, it was considered free bosons sparsely located along a lattice: for times smaller than the time at which particle interfere, controlled by the Lieb-Robinson bound, the task is easy as quantum particles behave as distinguishable classical ones; for larger times than the interference time the task becomes hard, as interference between particles can no longer be disregarded. This strongly resembles the transition encoded in Eq.~\eqref{eq_dynamics_product_states}. \\

\ricreview{Before concluding, we concretely address two instances where our results can find straightforward extension: higher dimensional generalizations and dynamics of random circuits with `East' gates.}

\begin{figure}[t!]
    \centering
    \includegraphics[width=0.95\linewidth]{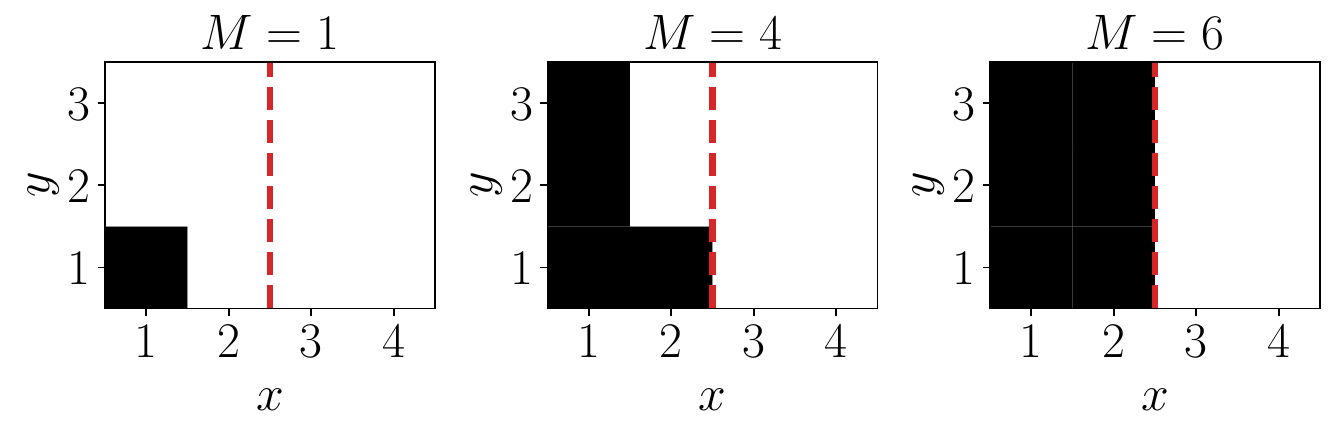}
     \includegraphics[width=\linewidth]{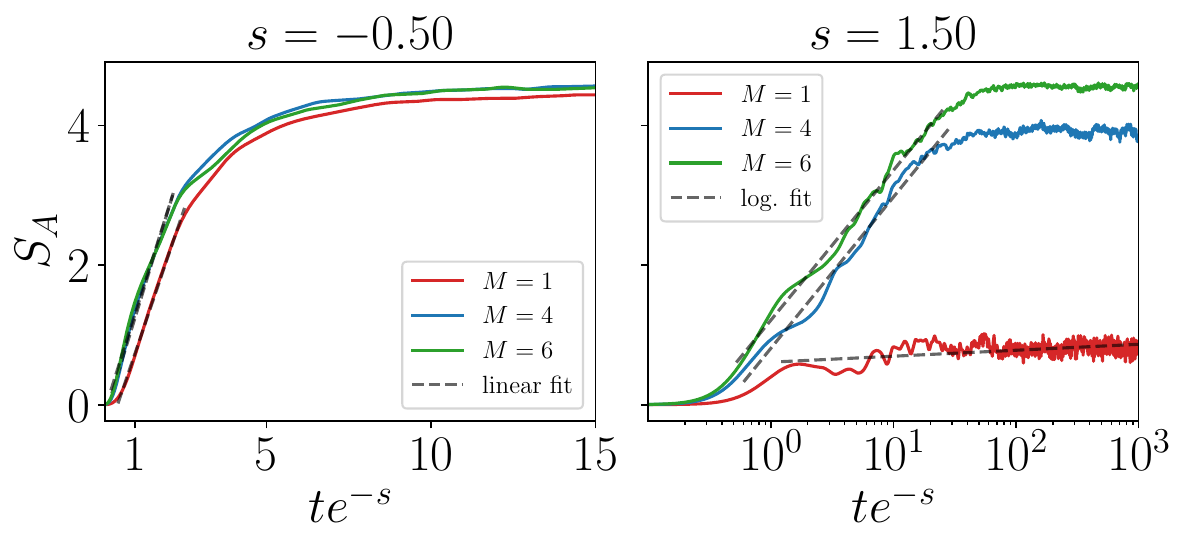}
    \caption{\ricreview{Dynamics of entanglement entropy in the 2D quantum North-East model (Eq.~\eqref{eq_HNE_model}) upon initializing 3 initial product states with a given number of excitations $M = \{1,3,6\}$ in a system of $(L_x,L_y)=(4,3)$ (in the first row we show the density profile of the initial states, with black corresponding to an excitation). The entanglement entropy is computed considering as $A = [[1,L_x/2],[1,L_y]]$, namely the bipartition along the vertical red cut depicted in the first row. For $s<0$, we observe ballistic growth as expected for a non-integrable system, before reaching a plateaux due to finite size effects. Instead, for $s>0$, entanglement grows in a logarithmic fashion. 
    } 
    }
    \label{fig_2D_NE_model}
\end{figure}
\subsection{Higher dimensions}
\label{sec:highd}
\ricreview{An intriguing natural direction involves the investigation of higher dimensional kinetically constrained models. Kinetically constrained \textit{classical} models are not very susceptible to   dimensionality~\cite{Garrahan2018}, and a natural question is whether this is true in their higher dimensional \textit{quantum} counterparts. To be   concrete, let us consider the 2D extension of the quantum East model, the quantum North-East model with Hamiltonian~\cite{PhysRevX.10.021051}
\begin{equation}
\label{eq_HNE_model}
    H_{2D} = - \frac{1}{2} \sum_{i,j} \hat{n}_{i,j} \left[e^{-s}\left( \hat{\sigma}_{i+1,j}^x + \hat{\sigma}_{i,j+1}^x\right) - 1\right],
\end{equation}
where a spin flip on site $(i,j)$ occurs if at least one of the sites below or to its left is excited ($|1\rangle$).
We consider open boundary conditions and we pin the (1,1) site to be with $n_{1,1}=1$ (similarly to the 1D version). The system still displays a delocalization-localization transition in the ground state at $s=0$~\cite{PhysRevX.10.021051}. Thus, an intriguing question is whether also the 2D case displays slow entanglement growth after quantum quenches. To this end, we consider small system sizes amenable to exact diagonalization calculation ($L_x \times L_y = 4 \times 3$ lattice). Following our procedure based on initializing product states, we consider the dynamics of product states with $M$ excitations ($|1\rangle$). Specifically, we consider $M = \{1,3,6\}$ (see Fig.~\ref{fig_2D_NE_model}), having increasing energy density with $M$ ($M=6$ lies near the middle of the spectrum, effectively corresponding to a high-temperature state). For each of them, we compute the dynamics and the entanglement entropy along the $x=L_x/2$ cut (cf. Fig.~\ref{fig_2D_NE_model}). For $s<0$, the entanglement entropy grows ballistically and quickly saturates to a finite value due to finite size effects. Instead, for $s>0$, we observe an extreme slowdown of entanglement entropy (compatible with a logarithmic growth), in a fashion analogous to the one-dimensional case. Based on our arguments, such a slowdown opens up the possibility that 2D disorder-free kinetically constrained systems are amenable to be simulated using tensor networks~\cite{10.21468/SciPostPhys.15.6.222}, alongside systems displaying slow-entanglement growth due to other mechanisms, such as confinement~\cite{tindall2024confinement,pavevsic2024constrained}. Additionally, from a statistical mechanics point of view, such a slowdown could hint at the existence of many sub-thermal eigenstates at finite energy density. However, due to the limited system sizes here considered, no conclusive data can be gathered, leaving these as intriguing directions for future investigations.}

\subsection{(Random) unitary circuits and thermal bubbles \label{sec_random_unitary_circuits}}
The results obtained in the quantum East model open up the intriguing possibility that the corresponding quantum circuit could be efficiently simulable as well. More concretely, the generator of the dynamics could be Trotterized as 
\begin{equation}
    \begin{split}
e^{-i \hat{H} T} &= \prod_{n=1}^{T/\Delta t} e^{-i \hat{H} \Delta t} \\
&\approx \prod_{n=1}^{T/\Delta t} \left( \prod_{j=1}^{N-1} e^{-i \hat{h}_{j,j+1}\Delta t}\right)
    \end{split}
\end{equation}
where the two-sites operator $e^{-i \hat{h}_{j,j+1}\Delta t}$ for the quantum East model results equal to 
\begin{equation}
\begin{split}
e^{-i \hat{h}_{j,j+1}\Delta t} = e^{i J\hat{n}_j \hat{\sigma}_{j+1}^x\Delta t/2} e^{-i \hat{n}_j \Delta t/2}
\end{split}
\end{equation}
where $J \equiv e^{-s}$. By using $\hat{n}_j^k = \hat{n}_j$ for $k>0$, and $(\hat{\sigma}_j^x)^k = 1$ for $k$ odd, we can write the evolution operator exactly as a 
\begin{equation}
\label{eq_circuit}
\begin{split}
    e^{-i \hat{h}_{j,j+1}\Delta t}=& U_j + U_{j,j+1}\\
    U_j = &1-\hat{n}_j + \hat{n}_j \cos(J\Delta t/2) e^{-i\Delta t/2}\\
    U_{j,j+1}= & i\hat{n}_j\hat{\sigma}_{j+1}^{x}\sin(J\Delta t/2) e^{-i\Delta t/2}  ,
\end{split}
\end{equation}
which can be seen as a \textit{tilted}-CNOT, i.e. a CNOT which does not necessarily perform conditioned perfect spin-flips.
This opens up the possibility that a subgroup of such a family of circuits, which does not belong to Clifford circuits (apart for fine-tuned parameters~\cite{PhysRevLett.132.120402}), could be efficiently simulated via classical algorithms. In such direction, the Floquet version of the quantum East model~\cite{PhysRevLett.132.080401} has been shown to display localization starting from a single excitation, despite the absence of energy conservation. However, analogously to what is here discussed in the continuous time version ($\Delta t\to 0$), also in the Trotterized one the initial state could play a role in dynamics. A detailed analysis of such a circuit could lead to the discovery of a novel family of \textit{tilted}-CNOTs that can be simulated efficiently, alongside Clifford, MBL inspired~\cite{PhysRevB.98.134204}, as well as fractonic random circuits~\cite{PhysRevX.9.021003,PhysRevX.10.011047}. \\
\begin{figure}[t!]
\centering
\includegraphics[width=\linewidth]{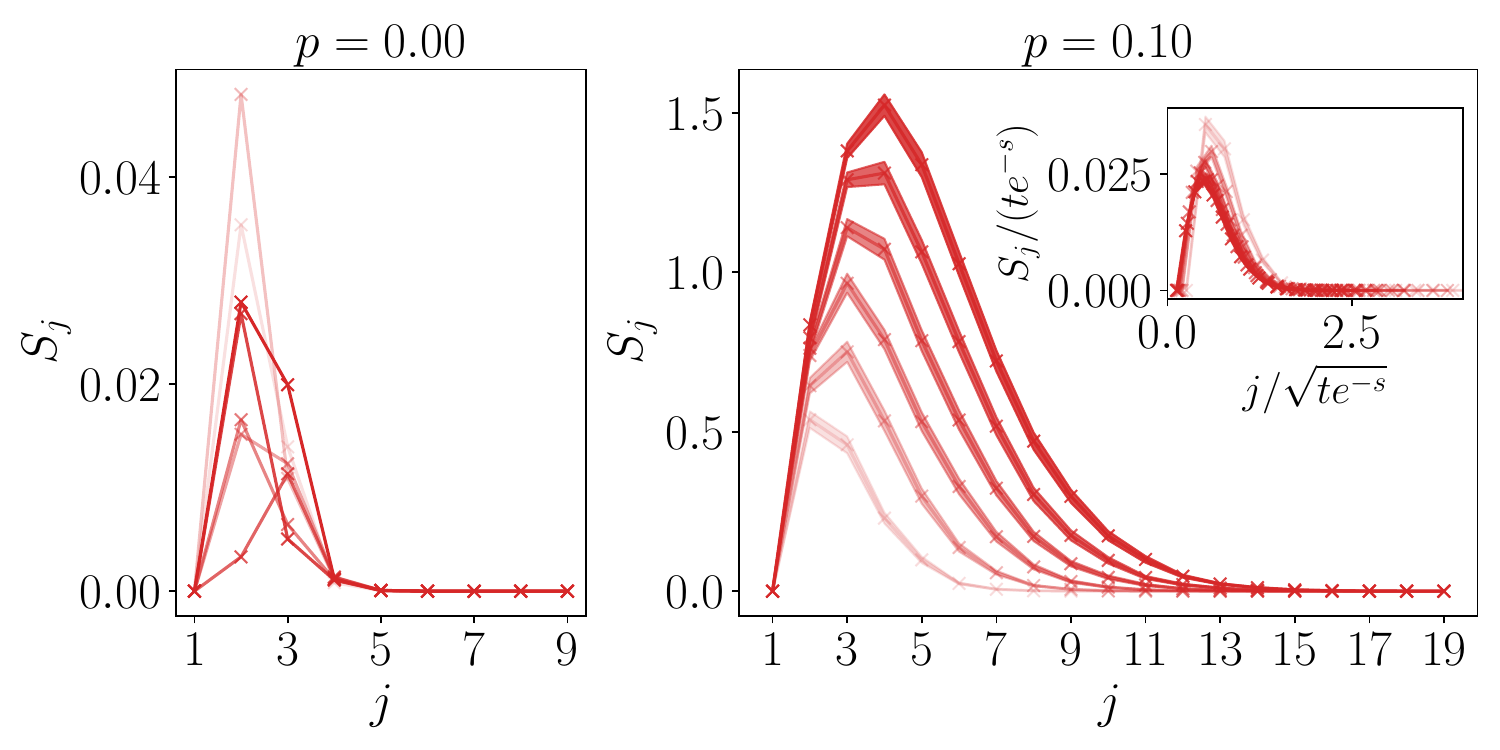}
\caption{Snapshots at different times $te^{-s}=\{15,25,...,75\}$ (from light to dark curves) of the entanglement entropy on each cut under the quantum East model deep in the localized phase ($s=1.5$) with $Z$ gates applied with probability $p$ on each site after a cycling time $\tau=0.5$. For $p=0$, entanglement remains localized near the initially seeded excitation. For $p>0$, the active region heats up, as indicated by the linear growth of $S_j$ on each $j$ (until saturating its upper bound given by an infinite temperature thermal state), and expands diffusively towards the inactive region (see inset). Results averaged over $80$ random realization for $p=0.10$ (shaded areas are statistical uncertainties). The initial state is $|1\rangle \otimes_{j=2}^N |0\rangle$. The value of $\tau$ does not qualitatively change the observed behavior.}
\label{fig_circuit_results}
\end{figure}

Going beyond a deterministic evolution, an intriguing direction could be the addition of randomness in the circuit. We can envision different ways to do that. For instance, by randomly applying gates belonging to the one listed in Eq.~\eqref{eq_circuit} with random $J$. In such a scenario, we envision that the system could display an entanglement transition between volume law to area law (if $J$ is sampled mostly from the localized side, namely $J < 1$). 
Another possibility is applying Eq.~\eqref{eq_circuit} deterministically for time $\tau$ (even by keeping $\Delta t$ small so that we are approximating the continuous time evolution), and then applying some gates to each site with probability $p$, and repeat. In such a protocol, notice that the density of random gates in the total space-time volume is finite. By doing so, we could naively envision that the addition of gates preserving the directional character together with the sharp differentiation between active (given by $|1\rangle$) and inactive (given by $|0\rangle$) regions should not spoil the main features of the quantum East model. However, we observe that such a picture is challenged also by the random application of the $Z$-gate. 
Even an infinitesimal $p$, leads to the disruption of the slow entanglement growth, making it ballistic, and locally heats up the system to infinite temperature as a result of the non-commutativity of $Z$-gates with the kinetic constraint terms. To show this, we initialize a state with a single excitation. We observe that the $Z$-gate heats the system solely in the already active regions as expected, which in turn heats the surroundings. Intriguingly, the application of $Z$ gates turns the spread of entanglement  (see  Fig.~\ref{fig_circuit_results}) from subdiffusive (for $p=0$, corresponding to the deterministic dynamics) to diffusive (for $p>0$). Such phenomenon is reminiscent of avalanches in MBL systems~\cite{PhysRevB.95.155129,PhysRevLett.121.140601,PhysRevLett.119.150602,PhysRevB.105.174205,PhysRevB.106.L020202}, and it could potentially serve as a toy model for their investigation.

\section*{Acknowledgements}
We thank Juan P. Garrahan and Shane P. Kelly for insightful discussions.  
This project has been supported by the Deutsche Forschungsgemeinschaft (DFG, German Research Foundation) through the Project-ID 429529648 -- TRR 306 QuCoLiMa (``Quantum Cooperativity of Light and Matter''); and by the Dynamics and Topology Center funded by the State of Rhineland Palatinate. Parts of this research were conducted using the Mogon supercomputer and/or advisory services offered by Johannes Gutenberg University Mainz (hpc.uni-mainz.de), which is a member of the AHRP (Alliance for High Performance Computing in Rhineland Palatinate,  www.ahrp.info) and the Gauss Alliance e.V. We gratefully acknowledge the computing time granted on the Mogon supercomputer at Johannes Gutenberg University Mainz (hpc.uni-mainz.de) through the project ``DysQCorr''. M.B. acknowledges the support and the resources provided by PARAM Shivay Facility under the National Supercomputing Mission, Government of India at the Indian Institute of Technology, Varanasi
\appendix
\section{Full spectrum of the Hamiltonian for small system size \label{appendix_full_spectrum}}

\begin{figure}[b!]
\centering
\includegraphics[width=0.8\linewidth]{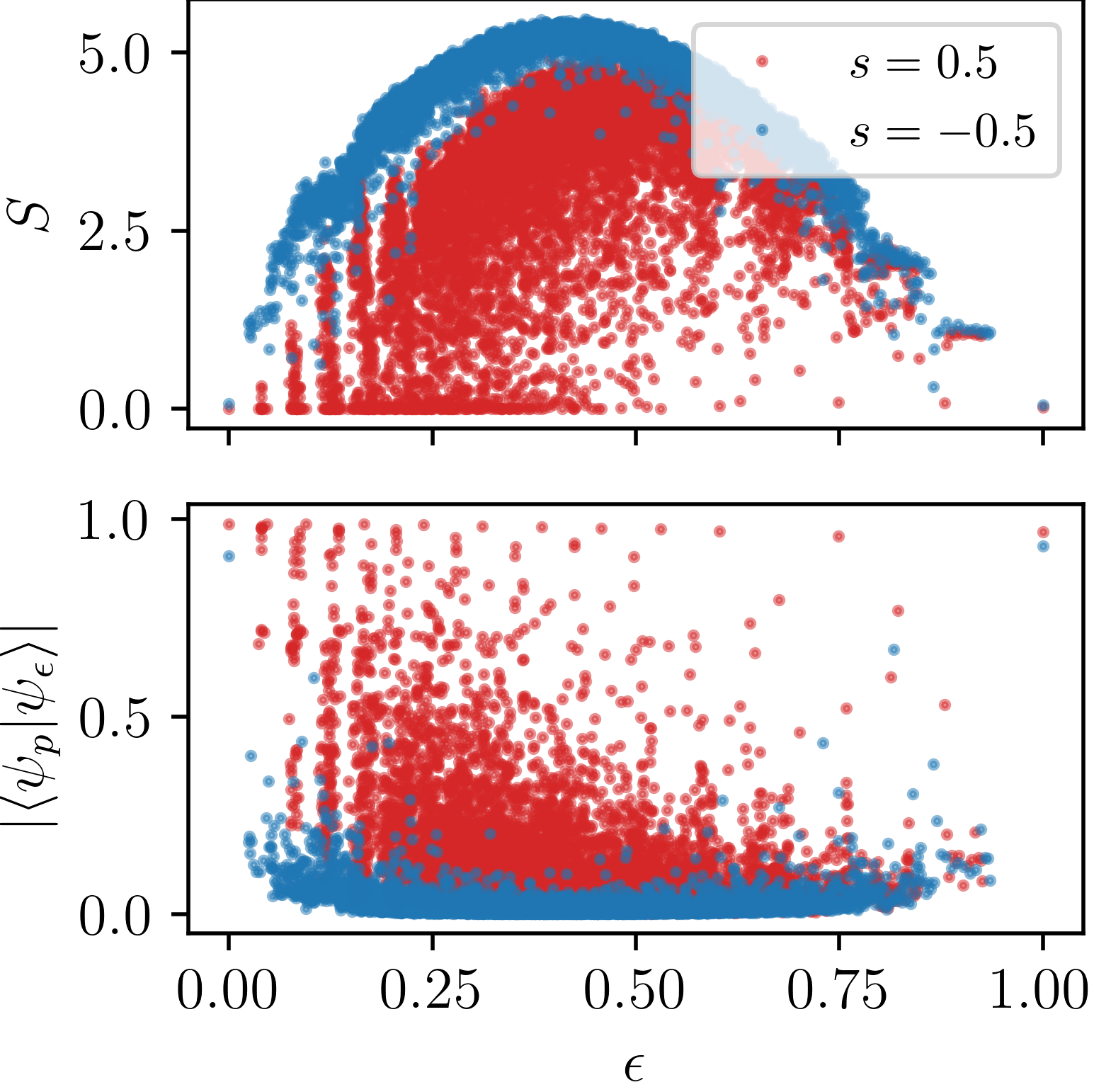}
\caption{
\textbf{Upper panel:} half-cut entanglement entropy of the eigenstates $|\psi_\epsilon\rangle$ of the quantum East model (cf. Eq.~\eqref{eq:Hamiltonian}) as a function of the normalized energy $\epsilon = (E-E_\text{min})/(E_\text{max}-E_\text{min})$. \textbf{Lower panel:} overlap between $|\psi_\epsilon\rangle$ and its best approximating product state $|\psi_p\rangle$ as a function of $\epsilon$. In the delocalized phase ($s<0$) the spectrum appears to be thermal, as typical for non-integrable systems. Instead, in the localized phase ($s>0$) there are (exponentially) many non-thermal eigenstates with
large overlap with product states. We show data for $N=13$ sites.}
\label{fig:ED_eigenstate_ovelap}
\end{figure}

Here we reproduce the results from Ref~\cite{PhysRevX.10.021051} showing the presence of an exponentially large number of non-thermal eigenstates with large overlap with product states. To do so, we compute the full spectrum of the quantum East model (cf. Eq.~\eqref{eq:Hamiltonian}) for small system sizes via exact diagonalization. We fix the first site in the $|1\rangle$ state. In the upper panel of Fig.~\ref{fig:ED_eigenstate_ovelap} we show the half-cut entanglement entropy $S$ of the eigenstates $|\psi_{\epsilon}\rangle$ in the delocalized regime ($s=-0.5$) and the localized regime ($s=0.5$) as a function of the normalized energy $\epsilon = (E_n-E_\text{min})/(E_\text{max}-E_\text{min})$, where $E_n$ is the energy of the $n$-th eigenstate, $E_\text{max}$ is the maximum energy, and $E_\text{min}$ is the ground state energy. In the delocalized ($s<0$) phase the spectrum appears to be thermal as expected, while instead in the localized regime ($s>0$) there are many non-thermal eigenstates with a large overlap with the product states (see lower-panel). The product states $\{|\psi_p\rangle\}$ were obtained by writing the eigenstates $\{|\psi_\epsilon\rangle\}$ as MPS and then truncate their bond-dimension to $1$ on each bond.
\begin{figure}[b!]
\centering
\includegraphics[width=0.8\linewidth]{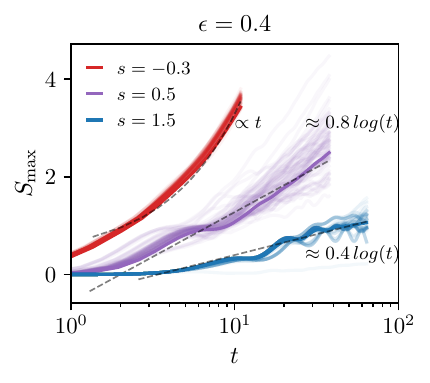}
\caption{Dynamics of the maximum entanglement entropy of $\sim 100$ product states (cf. Eq.~\eqref{eq_initial_state}) at fixed normalized energy $\epsilon=(\langle H\rangle - E_\text{min})/(E_\text{max}-E_\text{min})=0.4$ and different values of $s$. Deep in the delocalized phase ($s<0$), we observe a linear growth of $S_\text{max}$ as expected for thermal systems. Whereas in the localized phase ($s>0$), we have logarithmic growth of $S_\text{max}$ due to localization. \ricreview{Results obtained for a system size $N=30$}.}
\label{fig:fig_ent}
\end{figure}

\section{\label{appendix_entropy}Entanglement entropy dynamics}
In Fig.~\ref{fig:Fig1}(a) we have located a state-dependent mobility edge for initial product states using the way $\chi_\text{max}(t)$ grows in time, distinguishing a hard and easy time-complexity region.  Here we present the data for the dynamics of max entanglement entropy $S_\text{max}(t)$ along the system, to support the arguments presented in Sec.~\ref{sec_initial_state}.\\

For states represented in the form of MPS with bond dimension $\chi$, the maximal entanglement entropy is limited to $\log\chi$. Thus, if $S_\text{max}(t) \propto t$, the bond-dimension needed grows exponentially. Instead, if $S_\text{max}$ grows sub-linearly, e.g. logarithmically as occurs for localized systems, the needed bond dimension scales polynomially. In Fig.~\ref{fig:fig_ent}, we show the data for the growth of $S_\text{max}$ for a sample of product states at a fixed $\epsilon=0.4$ and different $s$. For states deep in the delocalized regime (e.g. $s=-0.3$), we have a linear growth of $S_{max}(t)$, as expected in the thermal regime. 
In the localized phase $s>0$, we observe $S_{max}(t) \propto \log t$, indicating non-thermal properties. 

\begin{figure}[b!]
    \centering
    \includegraphics[width=0.8\linewidth]{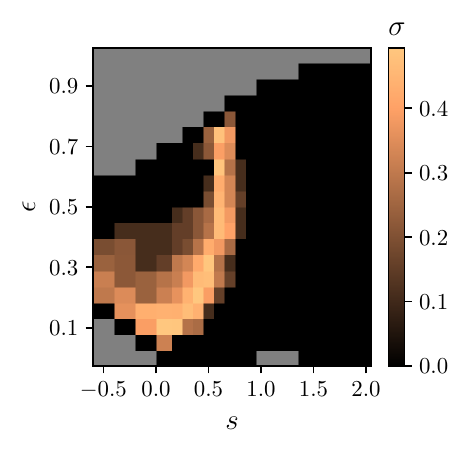}
    \caption{\ricreview{Standard deviation of the outcomes $\{ f_\text{easy}\}_{k=1}^n$ for any given value of energy density $\epsilon$ and Hamiltonian parameter $s$. The distribution is highly peaked in the typically easy and hard time-complexity regions. Instead, along the crossover, the distribution is wider, as the outcome strongly depends not only on the energy density but also on other features of the initial. Results obtained for a system size $N=30$.}}
    \label{fig_standard_deviation_total}
\end{figure}

\section{Details on the fitting procedure \label{appendix_details_fit}}
\begin{figure}[t!]
    \centering
    \includegraphics[width=0.8\linewidth]{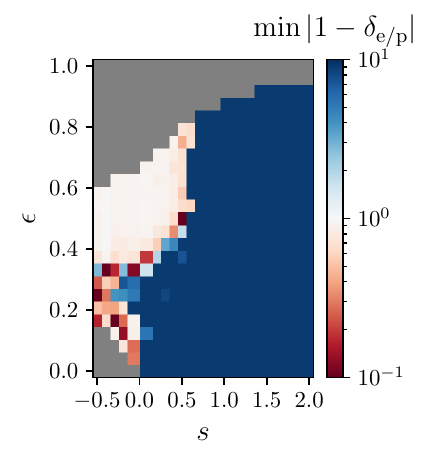}
    \caption{\ricreview{Mimimum of $|1-\delta_{e,p}|$ over the sampled initial states at any given $s$ and $\epsilon$, with $\delta_{e,p}$ the relative error between the exponential and polynomial fit (cf. Eq.~\eqref{eq_relative_error_fit}). Results obtained for a system size $N=30$.}}
    \label{fig_relative_error}
\end{figure}

\begin{figure}[t!]
    \centering
    \includegraphics[width=\linewidth]{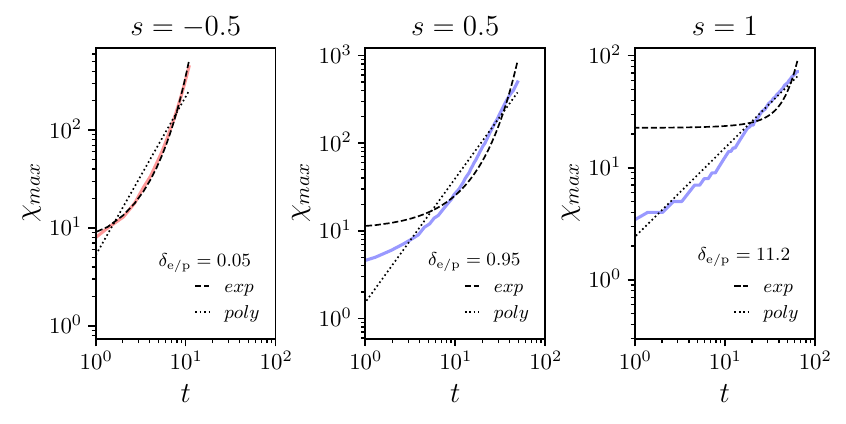}
    \caption{\ricreview{Fitting procedure for some representative states and some value of $s$ together with the corresponding relative error Eq.~\eqref{eq_relative_error_fit}. We show three panels where the two fits are easily distinguished (first, and third panel) and the one in the middle where it is hardly distinguished. Results obtained for a system size $N=30$.}}
    \label{fig_fitting_show}
\end{figure}

\ricreview{Here we provide additional details on the fitting procedure. The procedure can be summarized as follows:
given the dynamics of the maximal bond dimension $\chi_\text{max}(t)$, we fit a polynomial or an exponential curve; then, we select the best fit by comparing the error. In our simulations we observe that the errors are typically an order of magnitude different, making simple the selection of one over the other, while in the worst case scenarios the errors differ by $\sim 10\%$. To better quantify this, we compute the quantity $|1-\delta_{e,p}|$, with 
\begin{equation}
\label{eq_relative_error_fit}
    \delta_{e,p} = \frac{\text{error of exponential fit}}{\text{error polynomial fit}}.
\end{equation}
The quantity $|1-\delta_{e,p}|$ behaves as follows:
\begin{enumerate}
    \item $|1-\delta_{e,p}| \gg 1$ if the polynomial fit is better than the exponential one since $\delta_{e,p} \gg 1$;
    \item $|1-\delta_{e,p}| \approx 1$ if the exponential fit is better than the polynomial one since $\delta_{e,p} \ll 1$;
    \item $|1-\delta_{e,p}| \approx 0$ if the two fits are comparable.
\end{enumerate}
In Fig.~\ref{fig_relative_error} we plot the minimum value achieved at any given energy density $\epsilon$ and $s$, which constitutes the worst-case scenario. As anticipated, $|1-\delta_{e,p}|$ is always $\gtrsim 0.1$, making the two fits appreciably distinguishable in the majority of cases. In Fig.~\ref{fig_fitting_show} we show some representative fitting procedures, comparing exponential over polynomial fit.}

\begin{figure}[b!]
    \centering
    \includegraphics[width=0.8\linewidth]{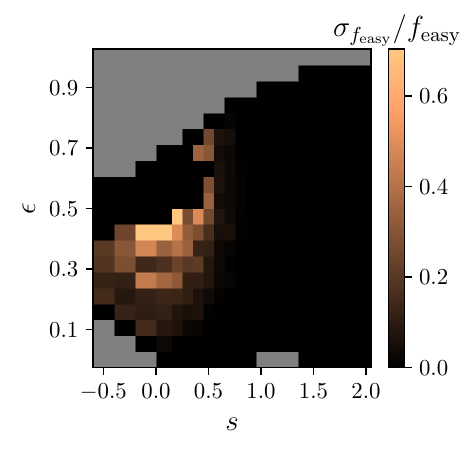}
    \caption{\ricreview{Fluctuations of the average $f_\text{easy}$ for any given value of energy density $\epsilon$ and Hamiltonian parameter $s$. The error is within a few percent in the whole space parameter, except for the typically hard regions where $f_\text{easy} \to 0$, making the ratio artificially huge. In the regions where $f_\text{easy}=0$ exactly, we set by hand the ratio equal to $0$ being singular ($\sigma_\text{easy}=f_\text{easy}=0$).  Results obtained for a system size $N=30$.}}
    \label{fig_standard_deviation_average}
\end{figure}

\section{\label{appendix_scaling_analysis_initial_state} Impact of a finite number of sampled initial states }
\begin{figure}[t!]
    \centering
    \includegraphics[width=\linewidth]{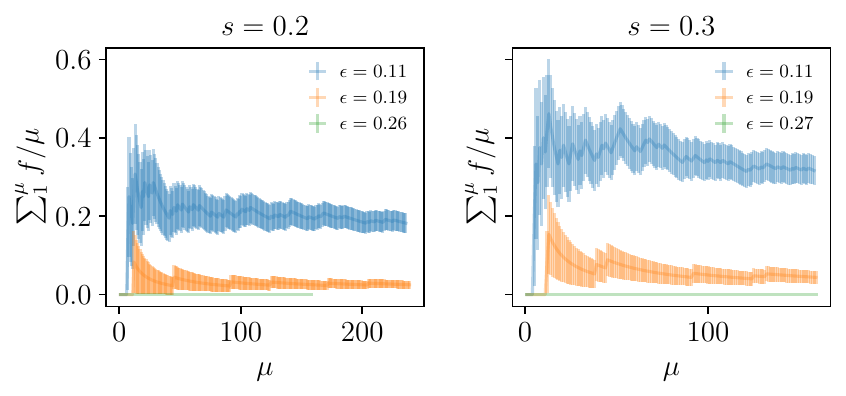}
    \caption{\ricreview{Convergence with the number of sampled states of the average outcome (cf. Eq.~\eqref{eq_running_average}). We show the worst-case scenarios, i.e. the crossover region, where even though the convergence is slower in the number of samples, we still achieve a relative error of a few percent. We highlight that the number of total sampled states is at most $100$ for each value of the number of excitations $M \in [1,N]$. However, as the initial states are not eigenstates of the Hamiltonian (and so have a non-zero energy variance, see Eq.~\eqref{eq_energy_and_variance}) we group them in a small energy window around $\epsilon$, thus increasing the number of samples at a given $\epsilon$.  Results obtained for a system size $N=30$.}}
    \label{fig_running_number_samples}
\end{figure}

\ricreview{Given a system of $N$ spin-1/2, the number of initial product states is $2^N$; this makes a thorough investigation  unfeasible for large $N$. Here, we show that by a limited sample of initial states uniformly extracted, we still can capture the features of our interest, namely the existence of a state-dependent mobility edge for $N=30$. However, we highlight that as $N$ increases, uniformly sampling product states is not a viable option, as other features (i.e. distribution of excitations) have to be taken into account as we have extensively discussed in Sec.~\ref{sec_mob_edge_energy}. Yet, up to a moderately large $N$ we observe good convergence in the sample size. Specifically, we now characterize the binomial distribution of the quantity $f_\text{easy}$ via its variance, together with the variance of its average (which is the quantity of our interest).\\ 
\\
We first look at how broad is the distribution of the outcomes obtained in our simulations. To this end, in Fig.~\ref{fig_standard_deviation_total} we show the standard deviation at any given $s$ and energy density $\epsilon$. For most space parameters, the distribution is narrow, indicating typicality. However, between the typically easy and hard simulable regions, where we locate the state-dependent mobility edge, the variance increases, indicating a dependence on other details of the initial state beyond the sole energy density. 
Despite the broader distribution, the average value of the distribution is computed with an uncertainty within few percent (see Fig.~\ref{fig_standard_deviation_average}). We highlight that the large values of the variance in the typically hard region is solely due to the fact that the quantity $f_\text{easy} \to 0$.\\
\\
As an additional test of convergence over the number of samples, we let the number of samples be a tunable parameter $\mu \in [1,n]$ (where $n$ is the total number of initial states sampled for a given $\epsilon$) and we compute
\begin{equation}
\label{eq_running_average}
    \frac{1}{\mu}\sum_{k=1}^\mu f_k(\epsilon,s),
\end{equation}
where $f_k(\epsilon,s) = 0$ or $1$. These two options correspond to hard or easy simulation of dynamics respectively,   upon initializing the $k$-th state at a given $s$ and energy density $\epsilon$. In Fig.~\ref{fig_running_number_samples} we show the worst-case scenarios, corresponding to the crossover regime, together with regimes where typicality is present, namely the single outcome is roughly enough.}

\section{\label{appendix_scaling_analysis}Finite size scaling}
\begin{figure}[t!]
  \centering
  \includegraphics[width=\linewidth]{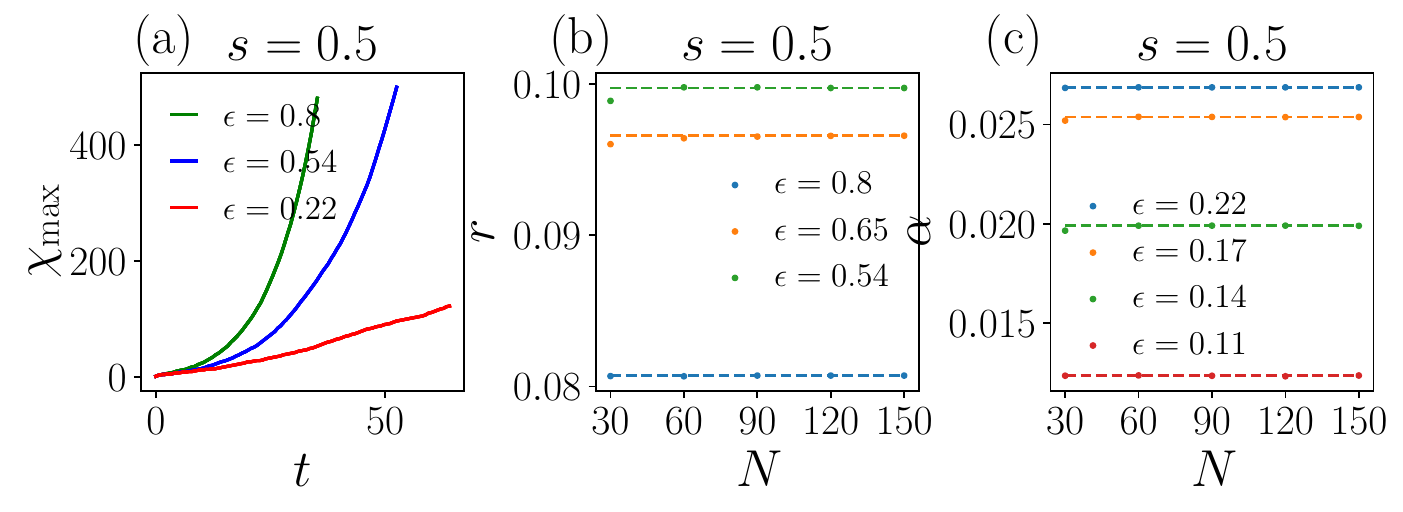}
  \caption{Dynamics of product states (cf. Eq.~\eqref{eq_states_repeated}) with similar spatial structure and same $\epsilon$ for different system sizes $N$. (a) Dynamics of $\chi_\text{max}(t)$ for different $\epsilon$ and system size $N\in[30,60,90,120,150]$. The curves for different $N$ cannot be distinguished as they overlap, proving the absence of finite size effects. (b-c) exponential rate $r$ and power-law exponent $\alpha$ as a function of $N$ for different $\epsilon$. Both $r$ and $\alpha$ are independent of the system size $N$.}
  \label{fig:PSD_fss}
\end{figure}

In the section \ref{sec_initial_state}, we uncover a mobility edge in the system for a class of product states. We studied the model on a system size of $N=30$, and the growth rate of $\chi$ was used as an indicator to classify hard or easy regimes. \ricrevieww{In  Sec.~\ref{sec_mob_edge_energy}, we have already performed a scaling analysis in $N$, by repeating the random uniform sampling procedure described in Sec.~\ref{sec_sampling_states}}.
Here, we perform a finite-size scaling analysis in the system size $N$ to show that boundary effects are irrelevant up to the times reached.\\
\\
As we have established in the main text that spatial structure plays a vital role in dictating the time-complexity, we consider a randomly sampled product state $|\psi_{N=30}\rangle$ for system size $N=30$ with energy density $\epsilon$, and we obtain states for larger $N$ by simply concatenating copies of them
\begin{equation}
\label{eq_states_repeated}
|\psi_N\rangle = |\psi_{N=30}\rangle^{ N/30}
\end{equation}
where we restrict to system sizes $N$ multiple of $30$. We highlight that these states have comparable same energy density $\epsilon$ (apart for $1/N$ corrections). We choose three representative values of $\epsilon=\{0.22,0.54,0.8\}$ at fixed $s=0.5$. In Fig.~\ref{fig:PSD_fss}(a) we show the dynamics of $\chi_\text{max}(t)$ for different values of energy density $\epsilon$. Each curve with a specific color consists of five overlapping curves for system size $N=\{30,60,90,120,150\}$. The overlap gives a clear indication that growth dynamics and hence our results are unaffected by boundary effects. As an additional test, in Fig.~\ref{fig:PSD_fss}(b-c) we show the computed exponential rates $r$ and power-law exponents $\alpha$ extracted by fitting $\chi_\text{max}(t)$ for different $\epsilon$ and $N$. Also here, there are no signatures of finite-size effects.

\section{\label{Appendix_importance_w}Role of the equilibrium localization length $\xi$ in dictating time-complexity}
\begin{figure}
  \centering
  \includegraphics[width=0.8\linewidth]{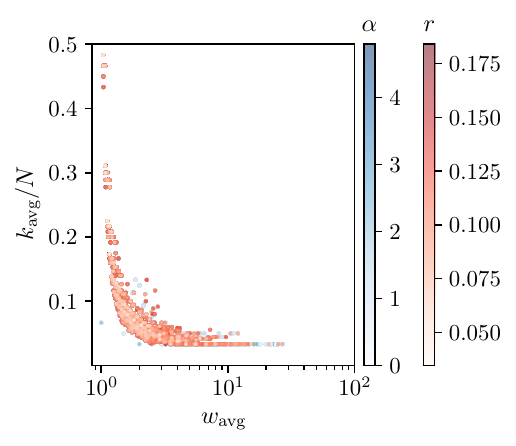}
  \caption{Same data presented in Fig.~\ref{fig_dependence_on_spatial_distribution}(a) without rescaling $w_\text{avg}$ with the ground state localization length $\xi$. The red dots correspond to hard time-complexity with $r$ the corresponding rate. The blue dots correspond to easy time-complexity with $\alpha$ the power-law. Both hard and easy simulations overlap as the effect of $\xi$ on growth dynamics is not considered, highlighting its key role also out-of-equilibrium. \ricreview{Results obtained for a system size $N=30$.}}
  \label{fig:Heterogeneity_w}
\end{figure}

In Sec.~\ref{sec_mob_edge_structure} we have discussed how the localization length $\xi$ of the ground state provides a length scale. To prove its predictive power, in Fig.~\ref{fig:Heterogeneity_w} we show the same data as in Fig.~\ref{fig_dependence_on_spatial_distribution} without rescaling the average distance between excitations $w_\text{avg}$ with $\xi$. We see that the hard and the easy simulations completely overlap and we are not able to differentiate between them. 

\section{Properties of the localized kink states found via DMRG-X \label{appendix_dmrgx}}
\begin{figure}[h!]
\centering
\includegraphics[width=0.49\linewidth]{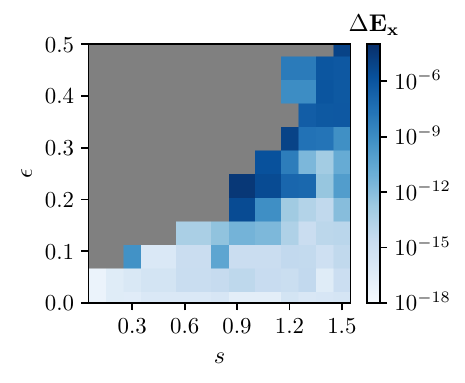}
\includegraphics[width=0.49\linewidth]{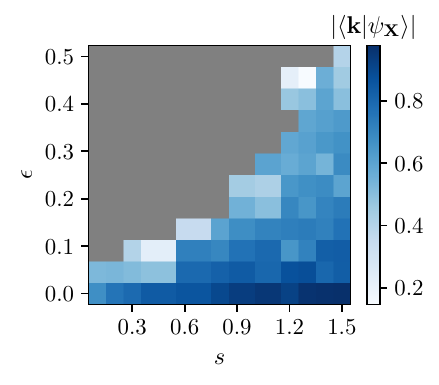}
\caption{\textbf{Left panel:} energy variance $\Delta E_\mathbf{x} = (\langle \psi_X|\hat{H}^2|\psi_X\rangle -\langle \psi_X|\hat{H}|\psi_X\rangle^2)$ computed on the states found via DMRG-X upon seeding kink states  $|\mathbf{k}\rangle=|1\rangle^{\otimes k} \otimes |0\rangle^{\otimes (N-k)}$. We consider the algorithm converged solely if $\Delta E_\mathbf{X} \leq 10^{-5}$. \textbf{Right panel:} modulus of the overlap between the states obtained via DMRG-X $|\psi_X\rangle$ and the initially seeded kink states $|\mathbf{k}\rangle=|1\rangle^{\otimes k} \otimes |0\rangle^{\otimes (N-k)}$. Results obtained for system size $N=30$.}
\label{fig_detail_DMRGX}
\end{figure}

Here we provide some additional results on the states $|\psi_X\rangle$ obtained via DMRG-X. We refer to Ref.~\cite{PhysRevLett.116.247204} for the interested reader on the technical details. In Fig.~\ref{fig_detail_DMRGX} we show the energy variance $\Delta H$ computed on $|\psi_X\rangle$ (setting a maximal bond dimension $\chi=100$) in a system of size $N=30$. We only show the data points for which  $\Delta H \leq 10^{-5}$. In Fig.~\ref{fig_detail_DMRGX} we show the overlap between the states $|\psi_X\rangle$ and the kink states $|\mathbf{k}\rangle=|1\rangle^{\otimes k} \otimes |0\rangle^{\otimes (N-k)}$ seeded in the variational approach. We observe large overlap (way larger than the typical overlap $1/\mathcal{D}$ with $\mathcal{D}=2^{N}$ the Hilbert space dimension), indicating that such states are highly relevant in dictating the dynamical features of the kink states $|\mathbf{k}\rangle$.
\newpage

\bibliography{mybibliography}

\begin{thebibliography}{100}%
\makeatletter
\providecommand \@ifxundefined [1]{%
 \@ifx{#1\undefined}
}%
\providecommand \@ifnum [1]{%
 \ifnum #1\expandafter \@firstoftwo
 \else \expandafter \@secondoftwo
 \fi
}%
\providecommand \@ifx [1]{%
 \ifx #1\expandafter \@firstoftwo
 \else \expandafter \@secondoftwo
 \fi
}%
\providecommand \natexlab [1]{#1}%
\providecommand \enquote  [1]{``#1''}%
\providecommand \bibnamefont  [1]{#1}%
\providecommand \bibfnamefont [1]{#1}%
\providecommand \citenamefont [1]{#1}%
\providecommand \href@noop [0]{\@secondoftwo}%
\providecommand \href [0]{\begingroup \@sanitize@url \@href}%
\providecommand \@href[1]{\@@startlink{#1}\@@href}%
\providecommand \@@href[1]{\endgroup#1\@@endlink}%
\providecommand \@sanitize@url [0]{\catcode `\\12\catcode `\$12\catcode
  `\&12\catcode `\#12\catcode `\^12\catcode `\_12\catcode `\%12\relax}%
\providecommand \@@startlink[1]{}%
\providecommand \@@endlink[0]{}%
\providecommand \url  [0]{\begingroup\@sanitize@url \@url }%
\providecommand \@url [1]{\endgroup\@href {#1}{\urlprefix }}%
\providecommand \urlprefix  [0]{URL }%
\providecommand \Eprint [0]{\href }%
\providecommand \doibase [0]{https://doi.org/}%
\providecommand \selectlanguage [0]{\@gobble}%
\providecommand \bibinfo  [0]{\@secondoftwo}%
\providecommand \bibfield  [0]{\@secondoftwo}%
\providecommand \translation [1]{[#1]}%
\providecommand \BibitemOpen [0]{}%
\providecommand \bibitemStop [0]{}%
\providecommand \bibitemNoStop [0]{.\EOS\space}%
\providecommand \EOS [0]{\spacefactor3000\relax}%
\providecommand \BibitemShut  [1]{\csname bibitem#1\endcsname}%
\let\auto@bib@innerbib\@empty
\bibitem [{\citenamefont {Preskill}(2018)}]{Preskill2018}%
  \BibitemOpen
  \bibfield  {author} {\bibinfo {author} {\bibfnamefont {J.}~\bibnamefont
  {Preskill}},\ }\bibfield  {title} {\bibinfo {title} {Quantum computing in the
  {NISQ} era and beyond},\ }\href {https://doi.org/10.22331/q-2018-08-06-79}
  {\bibfield  {journal} {\bibinfo  {journal} {Quantum}\ }\textbf {\bibinfo
  {volume} {2}},\ \bibinfo {pages} {79} (\bibinfo {year} {2018})}\BibitemShut
  {NoStop}%
\bibitem [{\citenamefont {Feynman}(1982)}]{Feynman1982}%
  \BibitemOpen
  \bibfield  {author} {\bibinfo {author} {\bibfnamefont {R.~P.}\ \bibnamefont
  {Feynman}},\ }\bibfield  {title} {\bibinfo {title} {Simulating physics with
  computers},\ }\href {https://doi.org/10.1007/bf02650179} {\bibfield
  {journal} {\bibinfo  {journal} {International Journal of Theoretical
  Physics}\ }\textbf {\bibinfo {volume} {21}},\ \bibinfo {pages} {467–488}
  (\bibinfo {year} {1982})}\BibitemShut {NoStop}%
\bibitem [{\citenamefont {Daley}\ \emph {et~al.}(2022)\citenamefont {Daley},
  \citenamefont {Bloch}, \citenamefont {Kokail}, \citenamefont {Flannigan},
  \citenamefont {Pearson}, \citenamefont {Troyer},\ and\ \citenamefont
  {Zoller}}]{Daley2022}%
  \BibitemOpen
  \bibfield  {author} {\bibinfo {author} {\bibfnamefont {A.~J.}\ \bibnamefont
  {Daley}}, \bibinfo {author} {\bibfnamefont {I.}~\bibnamefont {Bloch}},
  \bibinfo {author} {\bibfnamefont {C.}~\bibnamefont {Kokail}}, \bibinfo
  {author} {\bibfnamefont {S.}~\bibnamefont {Flannigan}}, \bibinfo {author}
  {\bibfnamefont {N.}~\bibnamefont {Pearson}}, \bibinfo {author} {\bibfnamefont
  {M.}~\bibnamefont {Troyer}},\ and\ \bibinfo {author} {\bibfnamefont
  {P.}~\bibnamefont {Zoller}},\ }\bibfield  {title} {\bibinfo {title}
  {Practical quantum advantage in quantum simulation},\ }\href
  {https://doi.org/10.1038/s41586-022-04940-6} {\bibfield  {journal} {\bibinfo
  {journal} {Nature}\ }\textbf {\bibinfo {volume} {607}},\ \bibinfo {pages}
  {667–676} (\bibinfo {year} {2022})}\BibitemShut {NoStop}%
\bibitem [{\citenamefont {Miessen}\ \emph {et~al.}(2022)\citenamefont
  {Miessen}, \citenamefont {Ollitrault}, \citenamefont {Tacchino},\ and\
  \citenamefont {Tavernelli}}]{Miessen2022}%
  \BibitemOpen
  \bibfield  {author} {\bibinfo {author} {\bibfnamefont {A.}~\bibnamefont
  {Miessen}}, \bibinfo {author} {\bibfnamefont {P.~J.}\ \bibnamefont
  {Ollitrault}}, \bibinfo {author} {\bibfnamefont {F.}~\bibnamefont
  {Tacchino}},\ and\ \bibinfo {author} {\bibfnamefont {I.}~\bibnamefont
  {Tavernelli}},\ }\bibfield  {title} {\bibinfo {title} {Quantum algorithms for
  quantum dynamics},\ }\href {https://doi.org/10.1038/s43588-022-00374-2}
  {\bibfield  {journal} {\bibinfo  {journal} {Nature Computational Science}\
  }\textbf {\bibinfo {volume} {3}},\ \bibinfo {pages} {25–37} (\bibinfo
  {year} {2022})}\BibitemShut {NoStop}%
\bibitem [{\citenamefont {Kim}\ \emph {et~al.}(2023)\citenamefont {Kim},
  \citenamefont {Eddins}, \citenamefont {Anand}, \citenamefont {Wei},
  \citenamefont {van~den Berg}, \citenamefont {Rosenblatt}, \citenamefont
  {Nayfeh}, \citenamefont {Wu}, \citenamefont {Zaletel}, \citenamefont
  {Temme},\ and\ \citenamefont {Kandala}}]{Kim2023}%
  \BibitemOpen
  \bibfield  {author} {\bibinfo {author} {\bibfnamefont {Y.}~\bibnamefont
  {Kim}}, \bibinfo {author} {\bibfnamefont {A.}~\bibnamefont {Eddins}},
  \bibinfo {author} {\bibfnamefont {S.}~\bibnamefont {Anand}}, \bibinfo
  {author} {\bibfnamefont {K.~X.}\ \bibnamefont {Wei}}, \bibinfo {author}
  {\bibfnamefont {E.}~\bibnamefont {van~den Berg}}, \bibinfo {author}
  {\bibfnamefont {S.}~\bibnamefont {Rosenblatt}}, \bibinfo {author}
  {\bibfnamefont {H.}~\bibnamefont {Nayfeh}}, \bibinfo {author} {\bibfnamefont
  {Y.}~\bibnamefont {Wu}}, \bibinfo {author} {\bibfnamefont {M.}~\bibnamefont
  {Zaletel}}, \bibinfo {author} {\bibfnamefont {K.}~\bibnamefont {Temme}},\
  and\ \bibinfo {author} {\bibfnamefont {A.}~\bibnamefont {Kandala}},\
  }\bibfield  {title} {\bibinfo {title} {Evidence for the utility of quantum
  computing before fault tolerance},\ }\href
  {https://doi.org/10.1038/s41586-023-06096-3} {\bibfield  {journal} {\bibinfo
  {journal} {Nature}\ }\textbf {\bibinfo {volume} {618}},\ \bibinfo {pages}
  {500–505} (\bibinfo {year} {2023})}\BibitemShut {NoStop}%
\bibitem [{\citenamefont {Tindall}\ \emph {et~al.}(2024)\citenamefont
  {Tindall}, \citenamefont {Fishman}, \citenamefont {Stoudenmire},\ and\
  \citenamefont {Sels}}]{PRXQuantum.5.010308}%
  \BibitemOpen
  \bibfield  {author} {\bibinfo {author} {\bibfnamefont {J.}~\bibnamefont
  {Tindall}}, \bibinfo {author} {\bibfnamefont {M.}~\bibnamefont {Fishman}},
  \bibinfo {author} {\bibfnamefont {E.~M.}\ \bibnamefont {Stoudenmire}},\ and\
  \bibinfo {author} {\bibfnamefont {D.}~\bibnamefont {Sels}},\ }\bibfield
  {title} {\bibinfo {title} {Efficient tensor network simulation of ibm's eagle
  kicked ising experiment},\ }\href
  {https://doi.org/10.1103/PRXQuantum.5.010308} {\bibfield  {journal} {\bibinfo
   {journal} {PRX Quantum}\ }\textbf {\bibinfo {volume} {5}},\ \bibinfo {pages}
  {010308} (\bibinfo {year} {2024})}\BibitemShut {NoStop}%
\bibitem [{\citenamefont {Patra}\ \emph {et~al.}(2024)\citenamefont {Patra},
  \citenamefont {Jahromi}, \citenamefont {Singh},\ and\ \citenamefont
  {Or\'us}}]{PhysRevResearch.6.013326}%
  \BibitemOpen
  \bibfield  {author} {\bibinfo {author} {\bibfnamefont {S.}~\bibnamefont
  {Patra}}, \bibinfo {author} {\bibfnamefont {S.~S.}\ \bibnamefont {Jahromi}},
  \bibinfo {author} {\bibfnamefont {S.}~\bibnamefont {Singh}},\ and\ \bibinfo
  {author} {\bibfnamefont {R.}~\bibnamefont {Or\'us}},\ }\bibfield  {title}
  {\bibinfo {title} {Efficient tensor network simulation of ibm's largest
  quantum processors},\ }\href
  {https://doi.org/10.1103/PhysRevResearch.6.013326} {\bibfield  {journal}
  {\bibinfo  {journal} {Phys. Rev. Res.}\ }\textbf {\bibinfo {volume} {6}},\
  \bibinfo {pages} {013326} (\bibinfo {year} {2024})}\BibitemShut {NoStop}%
\bibitem [{\citenamefont {Liao}\ \emph {et~al.}(2023)\citenamefont {Liao},
  \citenamefont {Wang}, \citenamefont {Zhou}, \citenamefont {Zhang},\ and\
  \citenamefont {Xiang}}]{liao2023simulation}%
  \BibitemOpen
  \bibfield  {author} {\bibinfo {author} {\bibfnamefont {H.-J.}\ \bibnamefont
  {Liao}}, \bibinfo {author} {\bibfnamefont {K.}~\bibnamefont {Wang}}, \bibinfo
  {author} {\bibfnamefont {Z.-S.}\ \bibnamefont {Zhou}}, \bibinfo {author}
  {\bibfnamefont {P.}~\bibnamefont {Zhang}},\ and\ \bibinfo {author}
  {\bibfnamefont {T.}~\bibnamefont {Xiang}},\ }\href@noop {} {\bibinfo {title}
  {Simulation of ibm's kicked ising experiment with projected entangled pair
  operator}} (\bibinfo {year} {2023}),\ \Eprint
  {https://arxiv.org/abs/2308.03082} {arXiv:2308.03082 [quant-ph]} \BibitemShut
  {NoStop}%
\bibitem [{\citenamefont {Anand}\ \emph {et~al.}(2023)\citenamefont {Anand},
  \citenamefont {Temme}, \citenamefont {Kandala},\ and\ \citenamefont
  {Zaletel}}]{anand2023classical}%
  \BibitemOpen
  \bibfield  {author} {\bibinfo {author} {\bibfnamefont {S.}~\bibnamefont
  {Anand}}, \bibinfo {author} {\bibfnamefont {K.}~\bibnamefont {Temme}},
  \bibinfo {author} {\bibfnamefont {A.}~\bibnamefont {Kandala}},\ and\ \bibinfo
  {author} {\bibfnamefont {M.}~\bibnamefont {Zaletel}},\ }\href@noop {}
  {\bibinfo {title} {Classical benchmarking of zero noise extrapolation beyond
  the exactly-verifiable regime}} (\bibinfo {year} {2023}),\ \Eprint
  {https://arxiv.org/abs/2306.17839} {arXiv:2306.17839 [quant-ph]} \BibitemShut
  {NoStop}%
\bibitem [{\citenamefont {Begušić}\ \emph {et~al.}(2024)\citenamefont
  {Begušić}, \citenamefont {Gray},\ and\ \citenamefont {Chan}}]{Begui2024}%
  \BibitemOpen
  \bibfield  {author} {\bibinfo {author} {\bibfnamefont {T.}~\bibnamefont
  {Begušić}}, \bibinfo {author} {\bibfnamefont {J.}~\bibnamefont {Gray}},\
  and\ \bibinfo {author} {\bibfnamefont {G.~K.-L.}\ \bibnamefont {Chan}},\
  }\bibfield  {title} {\bibinfo {title} {Fast and converged classical
  simulations of evidence for the utility of quantum computing before fault
  tolerance},\ }\bibfield  {journal} {\bibinfo  {journal} {Science Advances}\
  }\textbf {\bibinfo {volume} {10}},\ \href
  {https://doi.org/10.1126/sciadv.adk4321} {10.1126/sciadv.adk4321} (\bibinfo
  {year} {2024})\BibitemShut {NoStop}%
\bibitem [{\citenamefont {Rudolph}\ \emph {et~al.}(2023)\citenamefont
  {Rudolph}, \citenamefont {Fontana}, \citenamefont {Holmes},\ and\
  \citenamefont {Cincio}}]{rudolph2023classical}%
  \BibitemOpen
  \bibfield  {author} {\bibinfo {author} {\bibfnamefont {M.~S.}\ \bibnamefont
  {Rudolph}}, \bibinfo {author} {\bibfnamefont {E.}~\bibnamefont {Fontana}},
  \bibinfo {author} {\bibfnamefont {Z.}~\bibnamefont {Holmes}},\ and\ \bibinfo
  {author} {\bibfnamefont {L.}~\bibnamefont {Cincio}},\ }\href@noop {}
  {\bibinfo {title} {Classical surrogate simulation of quantum systems with
  lowesa}} (\bibinfo {year} {2023}),\ \Eprint
  {https://arxiv.org/abs/2308.09109} {arXiv:2308.09109 [quant-ph]} \BibitemShut
  {NoStop}%
\bibitem [{\citenamefont {Bernien}\ \emph {et~al.}(2017)\citenamefont
  {Bernien}, \citenamefont {Schwartz}, \citenamefont {Keesling}, \citenamefont
  {Levine}, \citenamefont {Omran}, \citenamefont {Pichler}, \citenamefont
  {Choi}, \citenamefont {Zibrov}, \citenamefont {Endres}, \citenamefont
  {Greiner}, \citenamefont {Vuleti{\'{c}}},\ and\ \citenamefont
  {Lukin}}]{Bernien2017}%
  \BibitemOpen
  \bibfield  {author} {\bibinfo {author} {\bibfnamefont {H.}~\bibnamefont
  {Bernien}}, \bibinfo {author} {\bibfnamefont {S.}~\bibnamefont {Schwartz}},
  \bibinfo {author} {\bibfnamefont {A.}~\bibnamefont {Keesling}}, \bibinfo
  {author} {\bibfnamefont {H.}~\bibnamefont {Levine}}, \bibinfo {author}
  {\bibfnamefont {A.}~\bibnamefont {Omran}}, \bibinfo {author} {\bibfnamefont
  {H.}~\bibnamefont {Pichler}}, \bibinfo {author} {\bibfnamefont
  {S.}~\bibnamefont {Choi}}, \bibinfo {author} {\bibfnamefont {A.~S.}\
  \bibnamefont {Zibrov}}, \bibinfo {author} {\bibfnamefont {M.}~\bibnamefont
  {Endres}}, \bibinfo {author} {\bibfnamefont {M.}~\bibnamefont {Greiner}},
  \bibinfo {author} {\bibfnamefont {V.}~\bibnamefont {Vuleti{\'{c}}}},\ and\
  \bibinfo {author} {\bibfnamefont {M.~D.}\ \bibnamefont {Lukin}},\ }\bibfield
  {title} {\bibinfo {title} {Probing many-body dynamics on a 51-atom quantum
  simulator},\ }\href {https://doi.org/10.1038/nature24622} {\bibfield
  {journal} {\bibinfo  {journal} {Nature}\ }\textbf {\bibinfo {volume} {551}},\
  \bibinfo {pages} {579} (\bibinfo {year} {2017})}\BibitemShut {NoStop}%
\bibitem [{\citenamefont {Turner}\ \emph
  {et~al.}(2018{\natexlab{a}})\citenamefont {Turner}, \citenamefont
  {Michailidis}, \citenamefont {Abanin}, \citenamefont {Serbyn},\ and\
  \citenamefont {Papi{\'c}}}]{Turner2018}%
  \BibitemOpen
  \bibfield  {author} {\bibinfo {author} {\bibfnamefont {C.~J.}\ \bibnamefont
  {Turner}}, \bibinfo {author} {\bibfnamefont {A.~A.}\ \bibnamefont
  {Michailidis}}, \bibinfo {author} {\bibfnamefont {D.~A.}\ \bibnamefont
  {Abanin}}, \bibinfo {author} {\bibfnamefont {M.}~\bibnamefont {Serbyn}},\
  and\ \bibinfo {author} {\bibfnamefont {Z.}~\bibnamefont {Papi{\'c}}},\
  }\bibfield  {title} {\bibinfo {title} {Weak ergodicity breaking from quantum
  many-body scars},\ }\href {https://doi.org/10.1038/s41567-018-0137-5}
  {\bibfield  {journal} {\bibinfo  {journal} {Nature Physics}\ }\textbf
  {\bibinfo {volume} {14}},\ \bibinfo {pages} {745} (\bibinfo {year}
  {2018}{\natexlab{a}})}\BibitemShut {NoStop}%
\bibitem [{\citenamefont {Choi}\ \emph {et~al.}(2019)\citenamefont {Choi},
  \citenamefont {Turner}, \citenamefont {Pichler}, \citenamefont {Ho},
  \citenamefont {Michailidis}, \citenamefont {Papi\ifmmode~\acute{c}\else
  \'{c}\fi{}}, \citenamefont {Serbyn}, \citenamefont {Lukin},\ and\
  \citenamefont {Abanin}}]{PhysRevLett.122.220603}%
  \BibitemOpen
  \bibfield  {author} {\bibinfo {author} {\bibfnamefont {S.}~\bibnamefont
  {Choi}}, \bibinfo {author} {\bibfnamefont {C.~J.}\ \bibnamefont {Turner}},
  \bibinfo {author} {\bibfnamefont {H.}~\bibnamefont {Pichler}}, \bibinfo
  {author} {\bibfnamefont {W.~W.}\ \bibnamefont {Ho}}, \bibinfo {author}
  {\bibfnamefont {A.~A.}\ \bibnamefont {Michailidis}}, \bibinfo {author}
  {\bibfnamefont {Z.}~\bibnamefont {Papi\ifmmode~\acute{c}\else \'{c}\fi{}}},
  \bibinfo {author} {\bibfnamefont {M.}~\bibnamefont {Serbyn}}, \bibinfo
  {author} {\bibfnamefont {M.~D.}\ \bibnamefont {Lukin}},\ and\ \bibinfo
  {author} {\bibfnamefont {D.~A.}\ \bibnamefont {Abanin}},\ }\bibfield  {title}
  {\bibinfo {title} {Emergent su(2) dynamics and perfect quantum many-body
  scars},\ }\href {https://doi.org/10.1103/PhysRevLett.122.220603} {\bibfield
  {journal} {\bibinfo  {journal} {Phys. Rev. Lett.}\ }\textbf {\bibinfo
  {volume} {122}},\ \bibinfo {pages} {220603} (\bibinfo {year}
  {2019})}\BibitemShut {NoStop}%
\bibitem [{\citenamefont {Serbyn}\ \emph {et~al.}(2021)\citenamefont {Serbyn},
  \citenamefont {Abanin},\ and\ \citenamefont {Papi{\'{c}}}}]{Serbyn2021}%
  \BibitemOpen
  \bibfield  {author} {\bibinfo {author} {\bibfnamefont {M.}~\bibnamefont
  {Serbyn}}, \bibinfo {author} {\bibfnamefont {D.~A.}\ \bibnamefont {Abanin}},\
  and\ \bibinfo {author} {\bibfnamefont {Z.}~\bibnamefont {Papi{\'{c}}}},\
  }\bibfield  {title} {\bibinfo {title} {Quantum many-body scars and weak
  breaking of ergodicity},\ }\href {https://doi.org/10.1038/s41567-021-01230-2}
  {\bibfield  {journal} {\bibinfo  {journal} {Nature Physics}\ }\textbf
  {\bibinfo {volume} {17}},\ \bibinfo {pages} {675} (\bibinfo {year}
  {2021})}\BibitemShut {NoStop}%
\bibitem [{\citenamefont {Turner}\ \emph
  {et~al.}(2018{\natexlab{b}})\citenamefont {Turner}, \citenamefont
  {Michailidis}, \citenamefont {Abanin}, \citenamefont {Serbyn},\ and\
  \citenamefont {Papi\ifmmode~\acute{c}\else \'{c}\fi{}}}]{PhysRevB.98.155134}%
  \BibitemOpen
  \bibfield  {author} {\bibinfo {author} {\bibfnamefont {C.~J.}\ \bibnamefont
  {Turner}}, \bibinfo {author} {\bibfnamefont {A.~A.}\ \bibnamefont
  {Michailidis}}, \bibinfo {author} {\bibfnamefont {D.~A.}\ \bibnamefont
  {Abanin}}, \bibinfo {author} {\bibfnamefont {M.}~\bibnamefont {Serbyn}},\
  and\ \bibinfo {author} {\bibfnamefont {Z.}~\bibnamefont
  {Papi\ifmmode~\acute{c}\else \'{c}\fi{}}},\ }\bibfield  {title} {\bibinfo
  {title} {Quantum scarred eigenstates in a rydberg atom chain: Entanglement,
  breakdown of thermalization, and stability to perturbations},\ }\href
  {https://doi.org/10.1103/PhysRevB.98.155134} {\bibfield  {journal} {\bibinfo
  {journal} {Phys. Rev. B}\ }\textbf {\bibinfo {volume} {98}},\ \bibinfo
  {pages} {155134} (\bibinfo {year} {2018}{\natexlab{b}})}\BibitemShut
  {NoStop}%
\bibitem [{\citenamefont {Khemani}\ \emph {et~al.}(2019)\citenamefont
  {Khemani}, \citenamefont {Laumann},\ and\ \citenamefont
  {Chandran}}]{PhysRevB.99.161101}%
  \BibitemOpen
  \bibfield  {author} {\bibinfo {author} {\bibfnamefont {V.}~\bibnamefont
  {Khemani}}, \bibinfo {author} {\bibfnamefont {C.~R.}\ \bibnamefont
  {Laumann}},\ and\ \bibinfo {author} {\bibfnamefont {A.}~\bibnamefont
  {Chandran}},\ }\bibfield  {title} {\bibinfo {title} {Signatures of
  integrability in the dynamics of rydberg-blockaded chains},\ }\href
  {https://doi.org/10.1103/PhysRevB.99.161101} {\bibfield  {journal} {\bibinfo
  {journal} {Phys. Rev. B}\ }\textbf {\bibinfo {volume} {99}},\ \bibinfo
  {pages} {161101(R)} (\bibinfo {year} {2019})}\BibitemShut {NoStop}%
\bibitem [{\citenamefont {Turner}\ \emph {et~al.}(2021)\citenamefont {Turner},
  \citenamefont {Desaules}, \citenamefont {Bull},\ and\ \citenamefont
  {Papi\ifmmode~\acute{c}\else \'{c}\fi{}}}]{PhysRevX.11.021021}%
  \BibitemOpen
  \bibfield  {author} {\bibinfo {author} {\bibfnamefont {C.~J.}\ \bibnamefont
  {Turner}}, \bibinfo {author} {\bibfnamefont {J.-Y.}\ \bibnamefont
  {Desaules}}, \bibinfo {author} {\bibfnamefont {K.}~\bibnamefont {Bull}},\
  and\ \bibinfo {author} {\bibfnamefont {Z.}~\bibnamefont
  {Papi\ifmmode~\acute{c}\else \'{c}\fi{}}},\ }\bibfield  {title} {\bibinfo
  {title} {Correspondence principle for many-body scars in ultracold rydberg
  atoms},\ }\href {https://doi.org/10.1103/PhysRevX.11.021021} {\bibfield
  {journal} {\bibinfo  {journal} {Phys. Rev. X}\ }\textbf {\bibinfo {volume}
  {11}},\ \bibinfo {pages} {021021} (\bibinfo {year} {2021})}\BibitemShut
  {NoStop}%
\bibitem [{\citenamefont {Moudgalya}\ \emph {et~al.}(2022)\citenamefont
  {Moudgalya}, \citenamefont {Bernevig},\ and\ \citenamefont
  {Regnault}}]{Moudgalya2022}%
  \BibitemOpen
  \bibfield  {author} {\bibinfo {author} {\bibfnamefont {S.}~\bibnamefont
  {Moudgalya}}, \bibinfo {author} {\bibfnamefont {B.~A.}\ \bibnamefont
  {Bernevig}},\ and\ \bibinfo {author} {\bibfnamefont {N.}~\bibnamefont
  {Regnault}},\ }\bibfield  {title} {\bibinfo {title} {Quantum many-body scars
  and hilbert space fragmentation: a review of exact results},\ }\href
  {https://doi.org/10.1088/1361-6633/ac73a0} {\bibfield  {journal} {\bibinfo
  {journal} {Reports on Progress in Physics}\ }\textbf {\bibinfo {volume}
  {85}},\ \bibinfo {pages} {086501} (\bibinfo {year} {2022})}\BibitemShut
  {NoStop}%
\bibitem [{\citenamefont {Chandran}\ \emph {et~al.}(2023)\citenamefont
  {Chandran}, \citenamefont {Iadecola}, \citenamefont {Khemani},\ and\
  \citenamefont {Moessner}}]{chandran2023quantum}%
  \BibitemOpen
  \bibfield  {author} {\bibinfo {author} {\bibfnamefont {A.}~\bibnamefont
  {Chandran}}, \bibinfo {author} {\bibfnamefont {T.}~\bibnamefont {Iadecola}},
  \bibinfo {author} {\bibfnamefont {V.}~\bibnamefont {Khemani}},\ and\ \bibinfo
  {author} {\bibfnamefont {R.}~\bibnamefont {Moessner}},\ }\bibfield  {title}
  {\bibinfo {title} {Quantum many-body scars: A quasiparticle perspective},\
  }\href@noop {} {\bibfield  {journal} {\bibinfo  {journal} {Annual Review of
  Condensed Matter Physics}\ }\textbf {\bibinfo {volume} {14}},\ \bibinfo
  {pages} {443} (\bibinfo {year} {2023})}\BibitemShut {NoStop}%
\bibitem [{\citenamefont {Bu\ifmmode~\check{c}\else
  \v{c}\fi{}a}(2023)}]{PhysRevX.13.031013}%
  \BibitemOpen
  \bibfield  {author} {\bibinfo {author} {\bibfnamefont {B.}~\bibnamefont
  {Bu\ifmmode~\check{c}\else \v{c}\fi{}a}},\ }\bibfield  {title} {\bibinfo
  {title} {Unified theory of local quantum many-body dynamics: Eigenoperator
  thermalization theorems},\ }\href
  {https://doi.org/10.1103/PhysRevX.13.031013} {\bibfield  {journal} {\bibinfo
  {journal} {Phys. Rev. X}\ }\textbf {\bibinfo {volume} {13}},\ \bibinfo
  {pages} {031013} (\bibinfo {year} {2023})}\BibitemShut {NoStop}%
\bibitem [{\citenamefont {Nandkishore}\ and\ \citenamefont
  {Huse}(2015)}]{doi:10.1146/annurev-conmatphys-031214-014726}%
  \BibitemOpen
  \bibfield  {author} {\bibinfo {author} {\bibfnamefont {R.}~\bibnamefont
  {Nandkishore}}\ and\ \bibinfo {author} {\bibfnamefont {D.~A.}\ \bibnamefont
  {Huse}},\ }\bibfield  {title} {\bibinfo {title} {Many-body localization and
  thermalization in quantum statistical mechanics},\ }\href
  {https://doi.org/10.1146/annurev-conmatphys-031214-014726} {\bibfield
  {journal} {\bibinfo  {journal} {Annual Review of Condensed Matter Physics}\
  }\textbf {\bibinfo {volume} {6}},\ \bibinfo {pages} {15} (\bibinfo {year}
  {2015})},\ \Eprint
  {https://arxiv.org/abs/https://doi.org/10.1146/annurev-conmatphys-031214-014726}
  {https://doi.org/10.1146/annurev-conmatphys-031214-014726} \BibitemShut
  {NoStop}%
\bibitem [{\citenamefont {Abanin}\ \emph {et~al.}(2019)\citenamefont {Abanin},
  \citenamefont {Altman}, \citenamefont {Bloch},\ and\ \citenamefont
  {Serbyn}}]{RevModPhys.91.021001}%
  \BibitemOpen
  \bibfield  {author} {\bibinfo {author} {\bibfnamefont {D.~A.}\ \bibnamefont
  {Abanin}}, \bibinfo {author} {\bibfnamefont {E.}~\bibnamefont {Altman}},
  \bibinfo {author} {\bibfnamefont {I.}~\bibnamefont {Bloch}},\ and\ \bibinfo
  {author} {\bibfnamefont {M.}~\bibnamefont {Serbyn}},\ }\bibfield  {title}
  {\bibinfo {title} {Colloquium: Many-body localization, thermalization, and
  entanglement},\ }\href {https://doi.org/10.1103/RevModPhys.91.021001}
  {\bibfield  {journal} {\bibinfo  {journal} {Rev. Mod. Phys.}\ }\textbf
  {\bibinfo {volume} {91}},\ \bibinfo {pages} {021001} (\bibinfo {year}
  {2019})}\BibitemShut {NoStop}%
\bibitem [{\citenamefont {Kormos}\ \emph {et~al.}(2017)\citenamefont {Kormos},
  \citenamefont {Collura}, \citenamefont {Tak{\'a}cs},\ and\ \citenamefont
  {Calabrese}}]{kormos2017real}%
  \BibitemOpen
  \bibfield  {author} {\bibinfo {author} {\bibfnamefont {M.}~\bibnamefont
  {Kormos}}, \bibinfo {author} {\bibfnamefont {M.}~\bibnamefont {Collura}},
  \bibinfo {author} {\bibfnamefont {G.}~\bibnamefont {Tak{\'a}cs}},\ and\
  \bibinfo {author} {\bibfnamefont {P.}~\bibnamefont {Calabrese}},\ }\bibfield
  {title} {\bibinfo {title} {Real-time confinement following a quantum quench
  to a non-integrable model},\ }\href@noop {} {\bibfield  {journal} {\bibinfo
  {journal} {Nature Physics}\ }\textbf {\bibinfo {volume} {13}},\ \bibinfo
  {pages} {246} (\bibinfo {year} {2017})}\BibitemShut {NoStop}%
\bibitem [{\citenamefont {Lerose}\ \emph {et~al.}(2020)\citenamefont {Lerose},
  \citenamefont {Surace}, \citenamefont {Mazza}, \citenamefont {Perfetto},
  \citenamefont {Collura},\ and\ \citenamefont
  {Gambassi}}]{PhysRevB.102.041118}%
  \BibitemOpen
  \bibfield  {author} {\bibinfo {author} {\bibfnamefont {A.}~\bibnamefont
  {Lerose}}, \bibinfo {author} {\bibfnamefont {F.~M.}\ \bibnamefont {Surace}},
  \bibinfo {author} {\bibfnamefont {P.~P.}\ \bibnamefont {Mazza}}, \bibinfo
  {author} {\bibfnamefont {G.}~\bibnamefont {Perfetto}}, \bibinfo {author}
  {\bibfnamefont {M.}~\bibnamefont {Collura}},\ and\ \bibinfo {author}
  {\bibfnamefont {A.}~\bibnamefont {Gambassi}},\ }\bibfield  {title} {\bibinfo
  {title} {Quasilocalized dynamics from confinement of quantum excitations},\
  }\href {https://doi.org/10.1103/PhysRevB.102.041118} {\bibfield  {journal}
  {\bibinfo  {journal} {Phys. Rev. B}\ }\textbf {\bibinfo {volume} {102}},\
  \bibinfo {pages} {041118(R)} (\bibinfo {year} {2020})}\BibitemShut {NoStop}%
\bibitem [{\citenamefont {Surace}\ \emph {et~al.}(2020)\citenamefont {Surace},
  \citenamefont {Mazza}, \citenamefont {Giudici}, \citenamefont {Lerose},
  \citenamefont {Gambassi},\ and\ \citenamefont
  {Dalmonte}}]{PhysRevX.10.021041}%
  \BibitemOpen
  \bibfield  {author} {\bibinfo {author} {\bibfnamefont {F.~M.}\ \bibnamefont
  {Surace}}, \bibinfo {author} {\bibfnamefont {P.~P.}\ \bibnamefont {Mazza}},
  \bibinfo {author} {\bibfnamefont {G.}~\bibnamefont {Giudici}}, \bibinfo
  {author} {\bibfnamefont {A.}~\bibnamefont {Lerose}}, \bibinfo {author}
  {\bibfnamefont {A.}~\bibnamefont {Gambassi}},\ and\ \bibinfo {author}
  {\bibfnamefont {M.}~\bibnamefont {Dalmonte}},\ }\bibfield  {title} {\bibinfo
  {title} {Lattice gauge theories and string dynamics in rydberg atom quantum
  simulators},\ }\href {https://doi.org/10.1103/PhysRevX.10.021041} {\bibfield
  {journal} {\bibinfo  {journal} {Phys. Rev. X}\ }\textbf {\bibinfo {volume}
  {10}},\ \bibinfo {pages} {021041} (\bibinfo {year} {2020})}\BibitemShut
  {NoStop}%
\bibitem [{\citenamefont {Liu}\ \emph {et~al.}(2019)\citenamefont {Liu},
  \citenamefont {Lundgren}, \citenamefont {Titum}, \citenamefont {Pagano},
  \citenamefont {Zhang}, \citenamefont {Monroe},\ and\ \citenamefont
  {Gorshkov}}]{PhysRevLett.122.150601}%
  \BibitemOpen
  \bibfield  {author} {\bibinfo {author} {\bibfnamefont {F.}~\bibnamefont
  {Liu}}, \bibinfo {author} {\bibfnamefont {R.}~\bibnamefont {Lundgren}},
  \bibinfo {author} {\bibfnamefont {P.}~\bibnamefont {Titum}}, \bibinfo
  {author} {\bibfnamefont {G.}~\bibnamefont {Pagano}}, \bibinfo {author}
  {\bibfnamefont {J.}~\bibnamefont {Zhang}}, \bibinfo {author} {\bibfnamefont
  {C.}~\bibnamefont {Monroe}},\ and\ \bibinfo {author} {\bibfnamefont {A.~V.}\
  \bibnamefont {Gorshkov}},\ }\bibfield  {title} {\bibinfo {title} {Confined
  quasiparticle dynamics in long-range interacting quantum spin chains},\
  }\href {https://doi.org/10.1103/PhysRevLett.122.150601} {\bibfield  {journal}
  {\bibinfo  {journal} {Phys. Rev. Lett.}\ }\textbf {\bibinfo {volume} {122}},\
  \bibinfo {pages} {150601} (\bibinfo {year} {2019})}\BibitemShut {NoStop}%
\bibitem [{\citenamefont {Javier Valencia~Tortora}\ \emph
  {et~al.}(2020)\citenamefont {Javier Valencia~Tortora}, \citenamefont
  {Calabrese},\ and\ \citenamefont {Collura}}]{JavierValenciaTortora2020}%
  \BibitemOpen
  \bibfield  {author} {\bibinfo {author} {\bibfnamefont {R.}~\bibnamefont
  {Javier Valencia~Tortora}}, \bibinfo {author} {\bibfnamefont
  {P.}~\bibnamefont {Calabrese}},\ and\ \bibinfo {author} {\bibfnamefont
  {M.}~\bibnamefont {Collura}},\ }\bibfield  {title} {\bibinfo {title}
  {Relaxation of the order-parameter statistics and dynamical confinement},\
  }\href {https://doi.org/10.1209/0295-5075/132/50001} {\bibfield  {journal}
  {\bibinfo  {journal} {Europhysics Letters}\ }\textbf {\bibinfo {volume}
  {132}},\ \bibinfo {pages} {50001} (\bibinfo {year} {2020})}\BibitemShut
  {NoStop}%
\bibitem [{\citenamefont {Pomponio}\ \emph {et~al.}(2022)\citenamefont
  {Pomponio}, \citenamefont {Werner}, \citenamefont {Zarand},\ and\
  \citenamefont {Takacs}}]{10.21468/SciPostPhys.12.2.061}%
  \BibitemOpen
  \bibfield  {author} {\bibinfo {author} {\bibfnamefont {O.}~\bibnamefont
  {Pomponio}}, \bibinfo {author} {\bibfnamefont {M.~A.}\ \bibnamefont
  {Werner}}, \bibinfo {author} {\bibfnamefont {G.}~\bibnamefont {Zarand}},\
  and\ \bibinfo {author} {\bibfnamefont {G.}~\bibnamefont {Takacs}},\
  }\bibfield  {title} {\bibinfo {title} {{Bloch oscillations and the lack of
  the decay of the false vacuum in a one-dimensional quantum spin chain}},\
  }\href {https://doi.org/10.21468/SciPostPhys.12.2.061} {\bibfield  {journal}
  {\bibinfo  {journal} {SciPost Phys.}\ }\textbf {\bibinfo {volume} {12}},\
  \bibinfo {pages} {061} (\bibinfo {year} {2022})}\BibitemShut {NoStop}%
\bibitem [{\citenamefont {Bardarson}\ \emph {et~al.}(2012)\citenamefont
  {Bardarson}, \citenamefont {Pollmann},\ and\ \citenamefont
  {Moore}}]{PhysRevLett.109.017202}%
  \BibitemOpen
  \bibfield  {author} {\bibinfo {author} {\bibfnamefont {J.~H.}\ \bibnamefont
  {Bardarson}}, \bibinfo {author} {\bibfnamefont {F.}~\bibnamefont
  {Pollmann}},\ and\ \bibinfo {author} {\bibfnamefont {J.~E.}\ \bibnamefont
  {Moore}},\ }\bibfield  {title} {\bibinfo {title} {Unbounded growth of
  entanglement in models of many-body localization},\ }\href
  {https://doi.org/10.1103/PhysRevLett.109.017202} {\bibfield  {journal}
  {\bibinfo  {journal} {Phys. Rev. Lett.}\ }\textbf {\bibinfo {volume} {109}},\
  \bibinfo {pages} {017202} (\bibinfo {year} {2012})}\BibitemShut {NoStop}%
\bibitem [{\citenamefont {Serbyn}\ \emph {et~al.}(2013)\citenamefont {Serbyn},
  \citenamefont {Papi\ifmmode~\acute{c}\else \'{c}\fi{}},\ and\ \citenamefont
  {Abanin}}]{PhysRevLett.110.260601}%
  \BibitemOpen
  \bibfield  {author} {\bibinfo {author} {\bibfnamefont {M.}~\bibnamefont
  {Serbyn}}, \bibinfo {author} {\bibfnamefont {Z.}~\bibnamefont
  {Papi\ifmmode~\acute{c}\else \'{c}\fi{}}},\ and\ \bibinfo {author}
  {\bibfnamefont {D.~A.}\ \bibnamefont {Abanin}},\ }\bibfield  {title}
  {\bibinfo {title} {Universal slow growth of entanglement in interacting
  strongly disordered systems},\ }\href
  {https://doi.org/10.1103/PhysRevLett.110.260601} {\bibfield  {journal}
  {\bibinfo  {journal} {Phys. Rev. Lett.}\ }\textbf {\bibinfo {volume} {110}},\
  \bibinfo {pages} {260601} (\bibinfo {year} {2013})}\BibitemShut {NoStop}%
\bibitem [{\citenamefont {Ehrenberg}\ \emph {et~al.}(2022)\citenamefont
  {Ehrenberg}, \citenamefont {Deshpande}, \citenamefont {Baldwin},
  \citenamefont {Abanin},\ and\ \citenamefont
  {Gorshkov}}]{ehrenberg2022simulation}%
  \BibitemOpen
  \bibfield  {author} {\bibinfo {author} {\bibfnamefont {A.}~\bibnamefont
  {Ehrenberg}}, \bibinfo {author} {\bibfnamefont {A.}~\bibnamefont
  {Deshpande}}, \bibinfo {author} {\bibfnamefont {C.~L.}\ \bibnamefont
  {Baldwin}}, \bibinfo {author} {\bibfnamefont {D.~A.}\ \bibnamefont
  {Abanin}},\ and\ \bibinfo {author} {\bibfnamefont {A.~V.}\ \bibnamefont
  {Gorshkov}},\ }\bibfield  {title} {\bibinfo {title} {Simulation complexity of
  many-body localized systems},\ }\href@noop {} {\bibfield  {journal} {\bibinfo
   {journal} {arXiv preprint arXiv:2205.12967}\ } (\bibinfo {year}
  {2022})}\BibitemShut {NoStop}%
\bibitem [{\citenamefont {Sierant}\ and\ \citenamefont
  {Zakrzewski}(2022)}]{PhysRevB.105.224203}%
  \BibitemOpen
  \bibfield  {author} {\bibinfo {author} {\bibfnamefont {P.}~\bibnamefont
  {Sierant}}\ and\ \bibinfo {author} {\bibfnamefont {J.}~\bibnamefont
  {Zakrzewski}},\ }\bibfield  {title} {\bibinfo {title} {Challenges to
  observation of many-body localization},\ }\href
  {https://doi.org/10.1103/PhysRevB.105.224203} {\bibfield  {journal} {\bibinfo
   {journal} {Phys. Rev. B}\ }\textbf {\bibinfo {volume} {105}},\ \bibinfo
  {pages} {224203} (\bibinfo {year} {2022})}\BibitemShut {NoStop}%
\bibitem [{\citenamefont {Sierant}\ \emph {et~al.}(2024)\citenamefont
  {Sierant}, \citenamefont {Lewenstein}, \citenamefont {Scardicchio},
  \citenamefont {Vidmar},\ and\ \citenamefont {Zakrzewski}}]{sierant2024many}%
  \BibitemOpen
  \bibfield  {author} {\bibinfo {author} {\bibfnamefont {P.}~\bibnamefont
  {Sierant}}, \bibinfo {author} {\bibfnamefont {M.}~\bibnamefont {Lewenstein}},
  \bibinfo {author} {\bibfnamefont {A.}~\bibnamefont {Scardicchio}}, \bibinfo
  {author} {\bibfnamefont {L.}~\bibnamefont {Vidmar}},\ and\ \bibinfo {author}
  {\bibfnamefont {J.}~\bibnamefont {Zakrzewski}},\ }\bibfield  {title}
  {\bibinfo {title} {Many-body localization in the age of classical
  computing},\ }\href@noop {} {\bibfield  {journal} {\bibinfo  {journal} {arXiv
  preprint arXiv:2403.07111}\ } (\bibinfo {year} {2024})}\BibitemShut {NoStop}%
\bibitem [{\citenamefont {Kim}\ and\ \citenamefont
  {Huse}(2013)}]{PhysRevLett.111.127205}%
  \BibitemOpen
  \bibfield  {author} {\bibinfo {author} {\bibfnamefont {H.}~\bibnamefont
  {Kim}}\ and\ \bibinfo {author} {\bibfnamefont {D.~A.}\ \bibnamefont {Huse}},\
  }\bibfield  {title} {\bibinfo {title} {Ballistic spreading of entanglement in
  a diffusive nonintegrable system},\ }\href
  {https://doi.org/10.1103/PhysRevLett.111.127205} {\bibfield  {journal}
  {\bibinfo  {journal} {Phys. Rev. Lett.}\ }\textbf {\bibinfo {volume} {111}},\
  \bibinfo {pages} {127205} (\bibinfo {year} {2013})}\BibitemShut {NoStop}%
\bibitem [{\citenamefont {van Horssen}\ \emph {et~al.}(2015)\citenamefont {van
  Horssen}, \citenamefont {Levi},\ and\ \citenamefont
  {Garrahan}}]{PhysRevB.92.100305}%
  \BibitemOpen
  \bibfield  {author} {\bibinfo {author} {\bibfnamefont {M.}~\bibnamefont {van
  Horssen}}, \bibinfo {author} {\bibfnamefont {E.}~\bibnamefont {Levi}},\ and\
  \bibinfo {author} {\bibfnamefont {J.~P.}\ \bibnamefont {Garrahan}},\
  }\bibfield  {title} {\bibinfo {title} {Dynamics of many-body localization in
  a translation-invariant quantum glass model},\ }\href
  {https://doi.org/10.1103/PhysRevB.92.100305} {\bibfield  {journal} {\bibinfo
  {journal} {Phys. Rev. B}\ }\textbf {\bibinfo {volume} {92}},\ \bibinfo
  {pages} {100305(R)} (\bibinfo {year} {2015})}\BibitemShut {NoStop}%
\bibitem [{\citenamefont {Pancotti}\ \emph {et~al.}(2020)\citenamefont
  {Pancotti}, \citenamefont {Giudice}, \citenamefont {Cirac}, \citenamefont
  {Garrahan},\ and\ \citenamefont {Ba\~nuls}}]{PhysRevX.10.021051}%
  \BibitemOpen
  \bibfield  {author} {\bibinfo {author} {\bibfnamefont {N.}~\bibnamefont
  {Pancotti}}, \bibinfo {author} {\bibfnamefont {G.}~\bibnamefont {Giudice}},
  \bibinfo {author} {\bibfnamefont {J.~I.}\ \bibnamefont {Cirac}}, \bibinfo
  {author} {\bibfnamefont {J.~P.}\ \bibnamefont {Garrahan}},\ and\ \bibinfo
  {author} {\bibfnamefont {M.~C.}\ \bibnamefont {Ba\~nuls}},\ }\bibfield
  {title} {\bibinfo {title} {Quantum east model: Localization, nonthermal
  eigenstates, and slow dynamics},\ }\href
  {https://doi.org/10.1103/PhysRevX.10.021051} {\bibfield  {journal} {\bibinfo
  {journal} {Phys. Rev. X}\ }\textbf {\bibinfo {volume} {10}},\ \bibinfo
  {pages} {021051} (\bibinfo {year} {2020})}\BibitemShut {NoStop}%
\bibitem [{\citenamefont {Valencia-Tortora}\ \emph {et~al.}(2024)\citenamefont
  {Valencia-Tortora}, \citenamefont {Pancotti}, \citenamefont {Fleischhauer},
  \citenamefont {Bernien},\ and\ \citenamefont
  {Marino}}]{PhysRevLett.132.223201}%
  \BibitemOpen
  \bibfield  {author} {\bibinfo {author} {\bibfnamefont {R.~J.}\ \bibnamefont
  {Valencia-Tortora}}, \bibinfo {author} {\bibfnamefont {N.}~\bibnamefont
  {Pancotti}}, \bibinfo {author} {\bibfnamefont {M.}~\bibnamefont
  {Fleischhauer}}, \bibinfo {author} {\bibfnamefont {H.}~\bibnamefont
  {Bernien}},\ and\ \bibinfo {author} {\bibfnamefont {J.}~\bibnamefont
  {Marino}},\ }\bibfield  {title} {\bibinfo {title} {Rydberg platform for
  nonergodic chiral quantum dynamics},\ }\href
  {https://doi.org/10.1103/PhysRevLett.132.223201} {\bibfield  {journal}
  {\bibinfo  {journal} {Phys. Rev. Lett.}\ }\textbf {\bibinfo {volume} {132}},\
  \bibinfo {pages} {223201} (\bibinfo {year} {2024})}\BibitemShut {NoStop}%
\bibitem [{\citenamefont {Bhore}\ \emph {et~al.}(2023)\citenamefont {Bhore},
  \citenamefont {Desaules},\ and\ \citenamefont {Papi\ifmmode~\acute{c}\else
  \'{c}\fi{}}}]{PhysRevB.108.104317}%
  \BibitemOpen
  \bibfield  {author} {\bibinfo {author} {\bibfnamefont {T.}~\bibnamefont
  {Bhore}}, \bibinfo {author} {\bibfnamefont {J.-Y.}\ \bibnamefont
  {Desaules}},\ and\ \bibinfo {author} {\bibfnamefont {Z.}~\bibnamefont
  {Papi\ifmmode~\acute{c}\else \'{c}\fi{}}},\ }\bibfield  {title} {\bibinfo
  {title} {Deep thermalization in constrained quantum systems},\ }\href
  {https://doi.org/10.1103/PhysRevB.108.104317} {\bibfield  {journal} {\bibinfo
   {journal} {Phys. Rev. B}\ }\textbf {\bibinfo {volume} {108}},\ \bibinfo
  {pages} {104317} (\bibinfo {year} {2023})}\BibitemShut {NoStop}%
\bibitem [{\citenamefont {Bertini}\ \emph
  {et~al.}(2024{\natexlab{a}})\citenamefont {Bertini}, \citenamefont
  {De~Fazio}, \citenamefont {Garrahan},\ and\ \citenamefont
  {Klobas}}]{PhysRevLett.132.120402}%
  \BibitemOpen
  \bibfield  {author} {\bibinfo {author} {\bibfnamefont {B.}~\bibnamefont
  {Bertini}}, \bibinfo {author} {\bibfnamefont {C.}~\bibnamefont {De~Fazio}},
  \bibinfo {author} {\bibfnamefont {J.~P.}\ \bibnamefont {Garrahan}},\ and\
  \bibinfo {author} {\bibfnamefont {K.}~\bibnamefont {Klobas}},\ }\bibfield
  {title} {\bibinfo {title} {Exact quench dynamics of the floquet quantum east
  model at the deterministic point},\ }\href
  {https://doi.org/10.1103/PhysRevLett.132.120402} {\bibfield  {journal}
  {\bibinfo  {journal} {Phys. Rev. Lett.}\ }\textbf {\bibinfo {volume} {132}},\
  \bibinfo {pages} {120402} (\bibinfo {year} {2024}{\natexlab{a}})}\BibitemShut
  {NoStop}%
\bibitem [{\citenamefont {Bertini}\ \emph
  {et~al.}(2024{\natexlab{b}})\citenamefont {Bertini}, \citenamefont {Kos},\
  and\ \citenamefont {Prosen}}]{PhysRevLett.132.080401}%
  \BibitemOpen
  \bibfield  {author} {\bibinfo {author} {\bibfnamefont {B.}~\bibnamefont
  {Bertini}}, \bibinfo {author} {\bibfnamefont {P.}~\bibnamefont {Kos}},\ and\
  \bibinfo {author} {\bibfnamefont {T.~c.~v.}\ \bibnamefont {Prosen}},\
  }\bibfield  {title} {\bibinfo {title} {Localized dynamics in the floquet
  quantum east model},\ }\href {https://doi.org/10.1103/PhysRevLett.132.080401}
  {\bibfield  {journal} {\bibinfo  {journal} {Phys. Rev. Lett.}\ }\textbf
  {\bibinfo {volume} {132}},\ \bibinfo {pages} {080401} (\bibinfo {year}
  {2024}{\natexlab{b}})}\BibitemShut {NoStop}%
\bibitem [{\citenamefont {Valencia-Tortora}\ \emph {et~al.}(2022)\citenamefont
  {Valencia-Tortora}, \citenamefont {Pancotti},\ and\ \citenamefont
  {Marino}}]{PRXQuantum.3.020346}%
  \BibitemOpen
  \bibfield  {author} {\bibinfo {author} {\bibfnamefont {R.~J.}\ \bibnamefont
  {Valencia-Tortora}}, \bibinfo {author} {\bibfnamefont {N.}~\bibnamefont
  {Pancotti}},\ and\ \bibinfo {author} {\bibfnamefont {J.}~\bibnamefont
  {Marino}},\ }\bibfield  {title} {\bibinfo {title} {Kinetically constrained
  quantum dynamics in superconducting circuits},\ }\href
  {https://doi.org/10.1103/PRXQuantum.3.020346} {\bibfield  {journal} {\bibinfo
   {journal} {PRX Quantum}\ }\textbf {\bibinfo {volume} {3}},\ \bibinfo {pages}
  {020346} (\bibinfo {year} {2022})}\BibitemShut {NoStop}%
\bibitem [{\citenamefont {Gei{\ss}ler}\ and\ \citenamefont
  {Garrahan}(2023)}]{geissler2023slow}%
  \BibitemOpen
  \bibfield  {author} {\bibinfo {author} {\bibfnamefont {A.}~\bibnamefont
  {Gei{\ss}ler}}\ and\ \bibinfo {author} {\bibfnamefont {J.~P.}\ \bibnamefont
  {Garrahan}},\ }\bibfield  {title} {\bibinfo {title} {Slow dynamics and
  nonergodicity of the bosonic quantum east model in the semiclassical limit},\
  }\href@noop {} {\bibfield  {journal} {\bibinfo  {journal} {Physical Review
  E}\ }\textbf {\bibinfo {volume} {108}},\ \bibinfo {pages} {034207} (\bibinfo
  {year} {2023})}\BibitemShut {NoStop}%
\bibitem [{\citenamefont {Brighi}\ \emph {et~al.}(2023)\citenamefont {Brighi},
  \citenamefont {Ljubotina},\ and\ \citenamefont {Serbyn}}]{brighi2023hilbert}%
  \BibitemOpen
  \bibfield  {author} {\bibinfo {author} {\bibfnamefont {P.}~\bibnamefont
  {Brighi}}, \bibinfo {author} {\bibfnamefont {M.}~\bibnamefont {Ljubotina}},\
  and\ \bibinfo {author} {\bibfnamefont {M.}~\bibnamefont {Serbyn}},\
  }\bibfield  {title} {\bibinfo {title} {Hilbert space fragmentation and slow
  dynamics in particle-conserving quantum east models},\ }\href@noop {}
  {\bibfield  {journal} {\bibinfo  {journal} {SciPost Physics}\ }\textbf
  {\bibinfo {volume} {15}},\ \bibinfo {pages} {093} (\bibinfo {year}
  {2023})}\BibitemShut {NoStop}%
\bibitem [{\citenamefont {Maity}\ and\ \citenamefont
  {Hamazaki}(2024)}]{PhysRevB.110.014301}%
  \BibitemOpen
  \bibfield  {author} {\bibinfo {author} {\bibfnamefont {S.}~\bibnamefont
  {Maity}}\ and\ \bibinfo {author} {\bibfnamefont {R.}~\bibnamefont
  {Hamazaki}},\ }\bibfield  {title} {\bibinfo {title} {Kinetically constrained
  models constructed from dissipative quantum dynamics},\ }\href
  {https://doi.org/10.1103/PhysRevB.110.014301} {\bibfield  {journal} {\bibinfo
   {journal} {Phys. Rev. B}\ }\textbf {\bibinfo {volume} {110}},\ \bibinfo
  {pages} {014301} (\bibinfo {year} {2024})}\BibitemShut {NoStop}%
\bibitem [{\citenamefont {Klobas}\ \emph {et~al.}(2023)\citenamefont {Klobas},
  \citenamefont {De~Fazio},\ and\ \citenamefont {Garrahan}}]{klobas2023exact}%
  \BibitemOpen
  \bibfield  {author} {\bibinfo {author} {\bibfnamefont {K.}~\bibnamefont
  {Klobas}}, \bibinfo {author} {\bibfnamefont {C.}~\bibnamefont {De~Fazio}},\
  and\ \bibinfo {author} {\bibfnamefont {J.~P.}\ \bibnamefont {Garrahan}},\
  }\bibfield  {title} {\bibinfo {title} {Exact" hydrophobicity" in
  deterministic circuits: dynamical fluctuations in the floquet-east model},\
  }\href@noop {} {\bibfield  {journal} {\bibinfo  {journal} {arXiv preprint
  arXiv:2305.07423}\ } (\bibinfo {year} {2023})}\BibitemShut {NoStop}%
\bibitem [{\citenamefont {Bertini}\ \emph
  {et~al.}(2024{\natexlab{c}})\citenamefont {Bertini}, \citenamefont
  {De~Fazio}, \citenamefont {Garrahan},\ and\ \citenamefont
  {Klobas}}]{bertini2024exact}%
  \BibitemOpen
  \bibfield  {author} {\bibinfo {author} {\bibfnamefont {B.}~\bibnamefont
  {Bertini}}, \bibinfo {author} {\bibfnamefont {C.}~\bibnamefont {De~Fazio}},
  \bibinfo {author} {\bibfnamefont {J.~P.}\ \bibnamefont {Garrahan}},\ and\
  \bibinfo {author} {\bibfnamefont {K.}~\bibnamefont {Klobas}},\ }\bibfield
  {title} {\bibinfo {title} {Exact quench dynamics of the floquet quantum east
  model at the deterministic point},\ }\href@noop {} {\bibfield  {journal}
  {\bibinfo  {journal} {Physical Review Letters}\ }\textbf {\bibinfo {volume}
  {132}},\ \bibinfo {pages} {120402} (\bibinfo {year}
  {2024}{\natexlab{c}})}\BibitemShut {NoStop}%
\bibitem [{\citenamefont {De~Fazio}\ \emph {et~al.}(2024)\citenamefont
  {De~Fazio}, \citenamefont {Garrahan},\ and\ \citenamefont
  {Klobas}}]{de2024exact}%
  \BibitemOpen
  \bibfield  {author} {\bibinfo {author} {\bibfnamefont {C.}~\bibnamefont
  {De~Fazio}}, \bibinfo {author} {\bibfnamefont {J.~P.}\ \bibnamefont
  {Garrahan}},\ and\ \bibinfo {author} {\bibfnamefont {K.}~\bibnamefont
  {Klobas}},\ }\bibfield  {title} {\bibinfo {title} {Exact results on the
  dynamics of the stochastic floquet-east model},\ }\href@noop {} {\bibfield
  {journal} {\bibinfo  {journal} {arXiv preprint arXiv:2406.17464}\ } (\bibinfo
  {year} {2024})}\BibitemShut {NoStop}%
\bibitem [{\citenamefont {Causer}\ \emph {et~al.}(2024)\citenamefont {Causer},
  \citenamefont {Ba{\~n}uls},\ and\ \citenamefont
  {Garrahan}}]{causer2024quantum}%
  \BibitemOpen
  \bibfield  {author} {\bibinfo {author} {\bibfnamefont {L.}~\bibnamefont
  {Causer}}, \bibinfo {author} {\bibfnamefont {M.~C.}\ \bibnamefont
  {Ba{\~n}uls}},\ and\ \bibinfo {author} {\bibfnamefont {J.~P.}\ \bibnamefont
  {Garrahan}},\ }\bibfield  {title} {\bibinfo {title} {Quantum many-body scars
  and non-thermal behaviour in fredkin spin chains},\ }\href@noop {} {\bibfield
   {journal} {\bibinfo  {journal} {arXiv preprint arXiv:2403.03986}\ }
  (\bibinfo {year} {2024})}\BibitemShut {NoStop}%
\bibitem [{\citenamefont {Lan}\ \emph {et~al.}(2018{\natexlab{a}})\citenamefont
  {Lan}, \citenamefont {van Horssen}, \citenamefont {Powell},\ and\
  \citenamefont {Garrahan}}]{PhysRevLett.121.040603}%
  \BibitemOpen
  \bibfield  {author} {\bibinfo {author} {\bibfnamefont {Z.}~\bibnamefont
  {Lan}}, \bibinfo {author} {\bibfnamefont {M.}~\bibnamefont {van Horssen}},
  \bibinfo {author} {\bibfnamefont {S.}~\bibnamefont {Powell}},\ and\ \bibinfo
  {author} {\bibfnamefont {J.~P.}\ \bibnamefont {Garrahan}},\ }\bibfield
  {title} {\bibinfo {title} {Quantum slow relaxation and metastability due to
  dynamical constraints},\ }\href
  {https://doi.org/10.1103/PhysRevLett.121.040603} {\bibfield  {journal}
  {\bibinfo  {journal} {Phys. Rev. Lett.}\ }\textbf {\bibinfo {volume} {121}},\
  \bibinfo {pages} {040603} (\bibinfo {year} {2018}{\natexlab{a}})}\BibitemShut
  {NoStop}%
\bibitem [{\citenamefont {Garrahan}\ \emph {et~al.}(2009)\citenamefont
  {Garrahan}, \citenamefont {Jack}, \citenamefont {Lecomte}, \citenamefont
  {Pitard}, \citenamefont {van Duijvendijk},\ and\ \citenamefont {van
  Wijland}}]{Garrahan2009}%
  \BibitemOpen
  \bibfield  {author} {\bibinfo {author} {\bibfnamefont {J.~P.}\ \bibnamefont
  {Garrahan}}, \bibinfo {author} {\bibfnamefont {R.~L.}\ \bibnamefont {Jack}},
  \bibinfo {author} {\bibfnamefont {V.}~\bibnamefont {Lecomte}}, \bibinfo
  {author} {\bibfnamefont {E.}~\bibnamefont {Pitard}}, \bibinfo {author}
  {\bibfnamefont {K.}~\bibnamefont {van Duijvendijk}},\ and\ \bibinfo {author}
  {\bibfnamefont {F.}~\bibnamefont {van Wijland}},\ }\bibfield  {title}
  {\bibinfo {title} {First-order dynamical phase transition in models of
  glasses: an approach based on ensembles of histories},\ }\href
  {https://doi.org/10.1088/1751-8113/42/7/075007} {\bibfield  {journal}
  {\bibinfo  {journal} {Journal of Physics A: Mathematical and Theoretical}\
  }\textbf {\bibinfo {volume} {42}},\ \bibinfo {pages} {075007} (\bibinfo
  {year} {2009})}\BibitemShut {NoStop}%
\bibitem [{\citenamefont {Chleboun}\ \emph {et~al.}(2013)\citenamefont
  {Chleboun}, \citenamefont {Faggionato},\ and\ \citenamefont
  {Martinelli}}]{Chleboun2013}%
  \BibitemOpen
  \bibfield  {author} {\bibinfo {author} {\bibfnamefont {P.}~\bibnamefont
  {Chleboun}}, \bibinfo {author} {\bibfnamefont {A.}~\bibnamefont
  {Faggionato}},\ and\ \bibinfo {author} {\bibfnamefont {F.}~\bibnamefont
  {Martinelli}},\ }\bibfield  {title} {\bibinfo {title} {Time scale separation
  in the low temperature east model: rigorous results},\ }\href
  {https://doi.org/10.1088/1742-5468/2013/04/l04001} {\bibfield  {journal}
  {\bibinfo  {journal} {J. Stat. Mech.}\ }\textbf {\bibinfo {volume} {2013}},\
  \bibinfo {pages} {L04001} (\bibinfo {year} {2013})}\BibitemShut {NoStop}%
\bibitem [{\citenamefont {Garrahan}(2018)}]{Garrahan2018}%
  \BibitemOpen
  \bibfield  {author} {\bibinfo {author} {\bibfnamefont {J.~P.}\ \bibnamefont
  {Garrahan}},\ }\bibfield  {title} {\bibinfo {title} {Aspects of
  non-equilibrium in classical and quantum systems: Slow relaxation and
  glasses, dynamical large deviations, quantum non-ergodicity, and open quantum
  dynamics},\ }\href {https://doi.org/10.1016/j.physa.2017.12.149} {\bibfield
  {journal} {\bibinfo  {journal} {Physica A: Statistical Mechanics and its
  Applications}\ }\textbf {\bibinfo {volume} {504}},\ \bibinfo {pages} {130}
  (\bibinfo {year} {2018})}\BibitemShut {NoStop}%
\bibitem [{\citenamefont {Causer}\ \emph {et~al.}(2020)\citenamefont {Causer},
  \citenamefont {Lesanovsky}, \citenamefont {Ba\~nuls},\ and\ \citenamefont
  {Garrahan}}]{PhysRevE.102.052132}%
  \BibitemOpen
  \bibfield  {author} {\bibinfo {author} {\bibfnamefont {L.}~\bibnamefont
  {Causer}}, \bibinfo {author} {\bibfnamefont {I.}~\bibnamefont {Lesanovsky}},
  \bibinfo {author} {\bibfnamefont {M.~C.}\ \bibnamefont {Ba\~nuls}},\ and\
  \bibinfo {author} {\bibfnamefont {J.~P.}\ \bibnamefont {Garrahan}},\
  }\bibfield  {title} {\bibinfo {title} {Dynamics and large deviation
  transitions of the xor-fredrickson-andersen kinetically constrained model},\
  }\href {https://doi.org/10.1103/PhysRevE.102.052132} {\bibfield  {journal}
  {\bibinfo  {journal} {Phys. Rev. E}\ }\textbf {\bibinfo {volume} {102}},\
  \bibinfo {pages} {052132} (\bibinfo {year} {2020})}\BibitemShut {NoStop}%
\bibitem [{\citenamefont {Sala}\ \emph {et~al.}(2020)\citenamefont {Sala},
  \citenamefont {Rakovszky}, \citenamefont {Verresen}, \citenamefont {Knap},\
  and\ \citenamefont {Pollmann}}]{PhysRevX.10.011047}%
  \BibitemOpen
  \bibfield  {author} {\bibinfo {author} {\bibfnamefont {P.}~\bibnamefont
  {Sala}}, \bibinfo {author} {\bibfnamefont {T.}~\bibnamefont {Rakovszky}},
  \bibinfo {author} {\bibfnamefont {R.}~\bibnamefont {Verresen}}, \bibinfo
  {author} {\bibfnamefont {M.}~\bibnamefont {Knap}},\ and\ \bibinfo {author}
  {\bibfnamefont {F.}~\bibnamefont {Pollmann}},\ }\bibfield  {title} {\bibinfo
  {title} {Ergodicity breaking arising from hilbert space fragmentation in
  dipole-conserving hamiltonians},\ }\href
  {https://doi.org/10.1103/PhysRevX.10.011047} {\bibfield  {journal} {\bibinfo
  {journal} {Phys. Rev. X}\ }\textbf {\bibinfo {volume} {10}},\ \bibinfo
  {pages} {011047} (\bibinfo {year} {2020})}\BibitemShut {NoStop}%
\bibitem [{\citenamefont {Moudgalya}\ and\ \citenamefont
  {Motrunich}(2022)}]{PhysRevX.12.011050}%
  \BibitemOpen
  \bibfield  {author} {\bibinfo {author} {\bibfnamefont {S.}~\bibnamefont
  {Moudgalya}}\ and\ \bibinfo {author} {\bibfnamefont {O.~I.}\ \bibnamefont
  {Motrunich}},\ }\bibfield  {title} {\bibinfo {title} {Hilbert space
  fragmentation and commutant algebras},\ }\href
  {https://doi.org/10.1103/PhysRevX.12.011050} {\bibfield  {journal} {\bibinfo
  {journal} {Phys. Rev. X}\ }\textbf {\bibinfo {volume} {12}},\ \bibinfo
  {pages} {011050} (\bibinfo {year} {2022})}\BibitemShut {NoStop}%
\bibitem [{\citenamefont {Khemani}\ \emph {et~al.}(2016)\citenamefont
  {Khemani}, \citenamefont {Pollmann},\ and\ \citenamefont
  {Sondhi}}]{PhysRevLett.116.247204}%
  \BibitemOpen
  \bibfield  {author} {\bibinfo {author} {\bibfnamefont {V.}~\bibnamefont
  {Khemani}}, \bibinfo {author} {\bibfnamefont {F.}~\bibnamefont {Pollmann}},\
  and\ \bibinfo {author} {\bibfnamefont {S.~L.}\ \bibnamefont {Sondhi}},\
  }\bibfield  {title} {\bibinfo {title} {Obtaining highly excited eigenstates
  of many-body localized hamiltonians by the density matrix renormalization
  group approach},\ }\href {https://doi.org/10.1103/PhysRevLett.116.247204}
  {\bibfield  {journal} {\bibinfo  {journal} {Phys. Rev. Lett.}\ }\textbf
  {\bibinfo {volume} {116}},\ \bibinfo {pages} {247204} (\bibinfo {year}
  {2016})}\BibitemShut {NoStop}%
\bibitem [{\citenamefont {Mondragon-Shem}\ \emph {et~al.}(2015)\citenamefont
  {Mondragon-Shem}, \citenamefont {Pal}, \citenamefont {Hughes},\ and\
  \citenamefont {Laumann}}]{PhysRevB.92.064203}%
  \BibitemOpen
  \bibfield  {author} {\bibinfo {author} {\bibfnamefont {I.}~\bibnamefont
  {Mondragon-Shem}}, \bibinfo {author} {\bibfnamefont {A.}~\bibnamefont {Pal}},
  \bibinfo {author} {\bibfnamefont {T.~L.}\ \bibnamefont {Hughes}},\ and\
  \bibinfo {author} {\bibfnamefont {C.~R.}\ \bibnamefont {Laumann}},\
  }\bibfield  {title} {\bibinfo {title} {Many-body mobility edge due to
  symmetry-constrained dynamics and strong interactions},\ }\href
  {https://doi.org/10.1103/PhysRevB.92.064203} {\bibfield  {journal} {\bibinfo
  {journal} {Phys. Rev. B}\ }\textbf {\bibinfo {volume} {92}},\ \bibinfo
  {pages} {064203} (\bibinfo {year} {2015})}\BibitemShut {NoStop}%
\bibitem [{\citenamefont {Luitz}\ \emph {et~al.}(2015)\citenamefont {Luitz},
  \citenamefont {Laflorencie},\ and\ \citenamefont
  {Alet}}]{PhysRevB.91.081103}%
  \BibitemOpen
  \bibfield  {author} {\bibinfo {author} {\bibfnamefont {D.~J.}\ \bibnamefont
  {Luitz}}, \bibinfo {author} {\bibfnamefont {N.}~\bibnamefont {Laflorencie}},\
  and\ \bibinfo {author} {\bibfnamefont {F.}~\bibnamefont {Alet}},\ }\bibfield
  {title} {\bibinfo {title} {Many-body localization edge in the random-field
  heisenberg chain},\ }\href {https://doi.org/10.1103/PhysRevB.91.081103}
  {\bibfield  {journal} {\bibinfo  {journal} {Phys. Rev. B}\ }\textbf {\bibinfo
  {volume} {91}},\ \bibinfo {pages} {081103(R)} (\bibinfo {year}
  {2015})}\BibitemShut {NoStop}%
\bibitem [{\citenamefont {Naldesi}\ \emph {et~al.}(2016)\citenamefont
  {Naldesi}, \citenamefont {Ercolessi},\ and\ \citenamefont
  {Roscilde}}]{Naldesi2016}%
  \BibitemOpen
  \bibfield  {author} {\bibinfo {author} {\bibfnamefont {P.}~\bibnamefont
  {Naldesi}}, \bibinfo {author} {\bibfnamefont {E.}~\bibnamefont {Ercolessi}},\
  and\ \bibinfo {author} {\bibfnamefont {T.}~\bibnamefont {Roscilde}},\
  }\bibfield  {title} {\bibinfo {title} {Detecting a many-body mobility edge
  with quantum quenches},\ }\bibfield  {journal} {\bibinfo  {journal} {SciPost
  Physics}\ }\textbf {\bibinfo {volume} {1}},\ \href
  {https://doi.org/10.21468/scipostphys.1.1.010} {10.21468/scipostphys.1.1.010}
  (\bibinfo {year} {2016})\BibitemShut {NoStop}%
\bibitem [{\citenamefont {S\"underhauf}\ \emph {et~al.}(2018)\citenamefont
  {S\"underhauf}, \citenamefont {P\'erez-Garc\'{\i}a}, \citenamefont {Huse},
  \citenamefont {Schuch},\ and\ \citenamefont {Cirac}}]{PhysRevB.98.134204}%
  \BibitemOpen
  \bibfield  {author} {\bibinfo {author} {\bibfnamefont {C.}~\bibnamefont
  {S\"underhauf}}, \bibinfo {author} {\bibfnamefont {D.}~\bibnamefont
  {P\'erez-Garc\'{\i}a}}, \bibinfo {author} {\bibfnamefont {D.~A.}\
  \bibnamefont {Huse}}, \bibinfo {author} {\bibfnamefont {N.}~\bibnamefont
  {Schuch}},\ and\ \bibinfo {author} {\bibfnamefont {J.~I.}\ \bibnamefont
  {Cirac}},\ }\bibfield  {title} {\bibinfo {title} {Localization with random
  time-periodic quantum circuits},\ }\href
  {https://doi.org/10.1103/PhysRevB.98.134204} {\bibfield  {journal} {\bibinfo
  {journal} {Phys. Rev. B}\ }\textbf {\bibinfo {volume} {98}},\ \bibinfo
  {pages} {134204} (\bibinfo {year} {2018})}\BibitemShut {NoStop}%
\bibitem [{\citenamefont {Pai}\ \emph {et~al.}(2019)\citenamefont {Pai},
  \citenamefont {Pretko},\ and\ \citenamefont
  {Nandkishore}}]{PhysRevX.9.021003}%
  \BibitemOpen
  \bibfield  {author} {\bibinfo {author} {\bibfnamefont {S.}~\bibnamefont
  {Pai}}, \bibinfo {author} {\bibfnamefont {M.}~\bibnamefont {Pretko}},\ and\
  \bibinfo {author} {\bibfnamefont {R.~M.}\ \bibnamefont {Nandkishore}},\
  }\bibfield  {title} {\bibinfo {title} {Localization in fractonic random
  circuits},\ }\href {https://doi.org/10.1103/PhysRevX.9.021003} {\bibfield
  {journal} {\bibinfo  {journal} {Phys. Rev. X}\ }\textbf {\bibinfo {volume}
  {9}},\ \bibinfo {pages} {021003} (\bibinfo {year} {2019})}\BibitemShut
  {NoStop}%
\bibitem [{\citenamefont {Lesanovsky}\ and\ \citenamefont
  {Garrahan}(2013)}]{PhysRevLett.111.215305}%
  \BibitemOpen
  \bibfield  {author} {\bibinfo {author} {\bibfnamefont {I.}~\bibnamefont
  {Lesanovsky}}\ and\ \bibinfo {author} {\bibfnamefont {J.~P.}\ \bibnamefont
  {Garrahan}},\ }\bibfield  {title} {\bibinfo {title} {Kinetic constraints,
  hierarchical relaxation, and onset of glassiness in strongly interacting and
  dissipative rydberg gases},\ }\href
  {https://doi.org/10.1103/PhysRevLett.111.215305} {\bibfield  {journal}
  {\bibinfo  {journal} {Phys. Rev. Lett.}\ }\textbf {\bibinfo {volume} {111}},\
  \bibinfo {pages} {215305} (\bibinfo {year} {2013})}\BibitemShut {NoStop}%
\bibitem [{\citenamefont {P\'erez-Espigares}\ \emph {et~al.}(2018)\citenamefont
  {P\'erez-Espigares}, \citenamefont {Lesanovsky}, \citenamefont {Garrahan},\
  and\ \citenamefont {Guti\'errez}}]{PhysRevA.98.021804}%
  \BibitemOpen
  \bibfield  {author} {\bibinfo {author} {\bibfnamefont {C.}~\bibnamefont
  {P\'erez-Espigares}}, \bibinfo {author} {\bibfnamefont {I.}~\bibnamefont
  {Lesanovsky}}, \bibinfo {author} {\bibfnamefont {J.~P.}\ \bibnamefont
  {Garrahan}},\ and\ \bibinfo {author} {\bibfnamefont {R.}~\bibnamefont
  {Guti\'errez}},\ }\bibfield  {title} {\bibinfo {title} {Glassy dynamics due
  to a trajectory phase transition in dissipative rydberg gases},\ }\href
  {https://doi.org/10.1103/PhysRevA.98.021804} {\bibfield  {journal} {\bibinfo
  {journal} {Phys. Rev. A}\ }\textbf {\bibinfo {volume} {98}},\ \bibinfo
  {pages} {021804(R)} (\bibinfo {year} {2018})}\BibitemShut {NoStop}%
\bibitem [{\citenamefont {Valado}\ \emph {et~al.}(2016)\citenamefont {Valado},
  \citenamefont {Simonelli}, \citenamefont {Hoogerland}, \citenamefont
  {Lesanovsky}, \citenamefont {Garrahan}, \citenamefont {Arimondo},
  \citenamefont {Ciampini},\ and\ \citenamefont {Morsch}}]{PhysRevA.93.040701}%
  \BibitemOpen
  \bibfield  {author} {\bibinfo {author} {\bibfnamefont {M.~M.}\ \bibnamefont
  {Valado}}, \bibinfo {author} {\bibfnamefont {C.}~\bibnamefont {Simonelli}},
  \bibinfo {author} {\bibfnamefont {M.~D.}\ \bibnamefont {Hoogerland}},
  \bibinfo {author} {\bibfnamefont {I.}~\bibnamefont {Lesanovsky}}, \bibinfo
  {author} {\bibfnamefont {J.~P.}\ \bibnamefont {Garrahan}}, \bibinfo {author}
  {\bibfnamefont {E.}~\bibnamefont {Arimondo}}, \bibinfo {author}
  {\bibfnamefont {D.}~\bibnamefont {Ciampini}},\ and\ \bibinfo {author}
  {\bibfnamefont {O.}~\bibnamefont {Morsch}},\ }\bibfield  {title} {\bibinfo
  {title} {Experimental observation of controllable kinetic constraints in a
  cold atomic gas},\ }\href {https://doi.org/10.1103/PhysRevA.93.040701}
  {\bibfield  {journal} {\bibinfo  {journal} {Phys. Rev. A}\ }\textbf {\bibinfo
  {volume} {93}},\ \bibinfo {pages} {040701(R)} (\bibinfo {year}
  {2016})}\BibitemShut {NoStop}%
\bibitem [{\citenamefont {Lesanovsky}\ and\ \citenamefont
  {Garrahan}(2014)}]{PhysRevA.90.011603}%
  \BibitemOpen
  \bibfield  {author} {\bibinfo {author} {\bibfnamefont {I.}~\bibnamefont
  {Lesanovsky}}\ and\ \bibinfo {author} {\bibfnamefont {J.~P.}\ \bibnamefont
  {Garrahan}},\ }\bibfield  {title} {\bibinfo {title} {Out-of-equilibrium
  structures in strongly interacting rydberg gases with dissipation},\ }\href
  {https://doi.org/10.1103/PhysRevA.90.011603} {\bibfield  {journal} {\bibinfo
  {journal} {Phys. Rev. A}\ }\textbf {\bibinfo {volume} {90}},\ \bibinfo
  {pages} {011603(R)} (\bibinfo {year} {2014})}\BibitemShut {NoStop}%
\bibitem [{\citenamefont {Gribben}\ \emph {et~al.}(2018)\citenamefont
  {Gribben}, \citenamefont {Lesanovsky},\ and\ \citenamefont
  {Guti\'errez}}]{PhysRevA.97.011603}%
  \BibitemOpen
  \bibfield  {author} {\bibinfo {author} {\bibfnamefont {D.}~\bibnamefont
  {Gribben}}, \bibinfo {author} {\bibfnamefont {I.}~\bibnamefont
  {Lesanovsky}},\ and\ \bibinfo {author} {\bibfnamefont {R.}~\bibnamefont
  {Guti\'errez}},\ }\bibfield  {title} {\bibinfo {title} {Quench dynamics of a
  dissipative rydberg gas in the classical and quantum regimes},\ }\href
  {https://doi.org/10.1103/PhysRevA.97.011603} {\bibfield  {journal} {\bibinfo
  {journal} {Phys. Rev. A}\ }\textbf {\bibinfo {volume} {97}},\ \bibinfo
  {pages} {011603(R)} (\bibinfo {year} {2018})}\BibitemShut {NoStop}%
\bibitem [{\citenamefont {Ostmann}\ \emph {et~al.}(2019)\citenamefont
  {Ostmann}, \citenamefont {Marcuzzi}, \citenamefont {Garrahan},\ and\
  \citenamefont {Lesanovsky}}]{PhysRevA.99.060101}%
  \BibitemOpen
  \bibfield  {author} {\bibinfo {author} {\bibfnamefont {M.}~\bibnamefont
  {Ostmann}}, \bibinfo {author} {\bibfnamefont {M.}~\bibnamefont {Marcuzzi}},
  \bibinfo {author} {\bibfnamefont {J.~P.}\ \bibnamefont {Garrahan}},\ and\
  \bibinfo {author} {\bibfnamefont {I.}~\bibnamefont {Lesanovsky}},\ }\bibfield
   {title} {\bibinfo {title} {Localization in spin chains with facilitation
  constraints and disordered interactions},\ }\href
  {https://doi.org/10.1103/PhysRevA.99.060101} {\bibfield  {journal} {\bibinfo
  {journal} {Phys. Rev. A}\ }\textbf {\bibinfo {volume} {99}},\ \bibinfo
  {pages} {060101(R)} (\bibinfo {year} {2019})}\BibitemShut {NoStop}%
\bibitem [{\citenamefont {Marcuzzi}\ \emph {et~al.}(2017)\citenamefont
  {Marcuzzi}, \citenamefont {Min\'a\ifmmode~\check{r}\else \v{r}\fi{}},
  \citenamefont {Barredo}, \citenamefont {de~L\'es\'eleuc}, \citenamefont
  {Labuhn}, \citenamefont {Lahaye}, \citenamefont {Browaeys}, \citenamefont
  {Levi},\ and\ \citenamefont {Lesanovsky}}]{PhysRevLett.118.063606}%
  \BibitemOpen
  \bibfield  {author} {\bibinfo {author} {\bibfnamefont {M.}~\bibnamefont
  {Marcuzzi}}, \bibinfo {author} {\bibfnamefont {J.~c.~v.}\ \bibnamefont
  {Min\'a\ifmmode~\check{r}\else \v{r}\fi{}}}, \bibinfo {author} {\bibfnamefont
  {D.}~\bibnamefont {Barredo}}, \bibinfo {author} {\bibfnamefont
  {S.}~\bibnamefont {de~L\'es\'eleuc}}, \bibinfo {author} {\bibfnamefont
  {H.}~\bibnamefont {Labuhn}}, \bibinfo {author} {\bibfnamefont
  {T.}~\bibnamefont {Lahaye}}, \bibinfo {author} {\bibfnamefont
  {A.}~\bibnamefont {Browaeys}}, \bibinfo {author} {\bibfnamefont
  {E.}~\bibnamefont {Levi}},\ and\ \bibinfo {author} {\bibfnamefont
  {I.}~\bibnamefont {Lesanovsky}},\ }\bibfield  {title} {\bibinfo {title}
  {Facilitation dynamics and localization phenomena in rydberg lattice gases
  with position disorder},\ }\href
  {https://doi.org/10.1103/PhysRevLett.118.063606} {\bibfield  {journal}
  {\bibinfo  {journal} {Phys. Rev. Lett.}\ }\textbf {\bibinfo {volume} {118}},\
  \bibinfo {pages} {063606} (\bibinfo {year} {2017})}\BibitemShut {NoStop}%
\bibitem [{\citenamefont {Lorenzo}\ \emph {et~al.}(2017)\citenamefont
  {Lorenzo}, \citenamefont {Marino}, \citenamefont {Plastina}, \citenamefont
  {Palma},\ and\ \citenamefont {Apollaro}}]{Lorenzo2017}%
  \BibitemOpen
  \bibfield  {author} {\bibinfo {author} {\bibfnamefont {S.}~\bibnamefont
  {Lorenzo}}, \bibinfo {author} {\bibfnamefont {J.}~\bibnamefont {Marino}},
  \bibinfo {author} {\bibfnamefont {F.}~\bibnamefont {Plastina}}, \bibinfo
  {author} {\bibfnamefont {G.~M.}\ \bibnamefont {Palma}},\ and\ \bibinfo
  {author} {\bibfnamefont {T.~J.~G.}\ \bibnamefont {Apollaro}},\ }\bibfield
  {title} {\bibinfo {title} {Quantum critical scaling under periodic driving},\
  }\bibfield  {journal} {\bibinfo  {journal} {Scientific Reports}\ }\textbf
  {\bibinfo {volume} {7}},\ \href {https://doi.org/10.1038/s41598-017-06025-1}
  {10.1038/s41598-017-06025-1} (\bibinfo {year} {2017})\BibitemShut {NoStop}%
\bibitem [{\citenamefont {Lan}\ \emph {et~al.}(2018{\natexlab{b}})\citenamefont
  {Lan}, \citenamefont {van Horssen}, \citenamefont {Powell},\ and\
  \citenamefont {Garrahan}}]{lan2018quantum}%
  \BibitemOpen
  \bibfield  {author} {\bibinfo {author} {\bibfnamefont {Z.}~\bibnamefont
  {Lan}}, \bibinfo {author} {\bibfnamefont {M.}~\bibnamefont {van Horssen}},
  \bibinfo {author} {\bibfnamefont {S.}~\bibnamefont {Powell}},\ and\ \bibinfo
  {author} {\bibfnamefont {J.~P.}\ \bibnamefont {Garrahan}},\ }\bibfield
  {title} {\bibinfo {title} {Quantum slow relaxation and metastability due to
  dynamical constraints},\ }\href@noop {} {\bibfield  {journal} {\bibinfo
  {journal} {Physical review letters}\ }\textbf {\bibinfo {volume} {121}},\
  \bibinfo {pages} {040603} (\bibinfo {year} {2018}{\natexlab{b}})}\BibitemShut
  {NoStop}%
\bibitem [{\citenamefont {Zadnik}\ and\ \citenamefont
  {Garrahan}(2023)}]{zadnik2023slow}%
  \BibitemOpen
  \bibfield  {author} {\bibinfo {author} {\bibfnamefont {L.}~\bibnamefont
  {Zadnik}}\ and\ \bibinfo {author} {\bibfnamefont {J.~P.}\ \bibnamefont
  {Garrahan}},\ }\bibfield  {title} {\bibinfo {title} {Slow heterogeneous
  relaxation due to constraints in dual xxz models},\ }\href@noop {} {\bibfield
   {journal} {\bibinfo  {journal} {Physical Review B}\ }\textbf {\bibinfo
  {volume} {108}},\ \bibinfo {pages} {L100304} (\bibinfo {year}
  {2023})}\BibitemShut {NoStop}%
\bibitem [{\citenamefont {{\v{S}}untajs}\ \emph {et~al.}(2020)\citenamefont
  {{\v{S}}untajs}, \citenamefont {Bon{\v{c}}a}, \citenamefont {Prosen},\ and\
  \citenamefont {Vidmar}}]{vsuntajs2020quantum}%
  \BibitemOpen
  \bibfield  {author} {\bibinfo {author} {\bibfnamefont {J.}~\bibnamefont
  {{\v{S}}untajs}}, \bibinfo {author} {\bibfnamefont {J.}~\bibnamefont
  {Bon{\v{c}}a}}, \bibinfo {author} {\bibfnamefont {T.}~\bibnamefont
  {Prosen}},\ and\ \bibinfo {author} {\bibfnamefont {L.}~\bibnamefont
  {Vidmar}},\ }\bibfield  {title} {\bibinfo {title} {Quantum chaos challenges
  many-body localization},\ }\href@noop {} {\bibfield  {journal} {\bibinfo
  {journal} {Physical Review E}\ }\textbf {\bibinfo {volume} {102}},\ \bibinfo
  {pages} {062144} (\bibinfo {year} {2020})}\BibitemShut {NoStop}%
\bibitem [{\citenamefont {Abanin}\ \emph {et~al.}(2021)\citenamefont {Abanin},
  \citenamefont {Bardarson}, \citenamefont {De~Tomasi}, \citenamefont
  {Gopalakrishnan}, \citenamefont {Khemani}, \citenamefont {Parameswaran},
  \citenamefont {Pollmann}, \citenamefont {Potter}, \citenamefont {Serbyn},\
  and\ \citenamefont {Vasseur}}]{Abanin2021}%
  \BibitemOpen
  \bibfield  {author} {\bibinfo {author} {\bibfnamefont {D.}~\bibnamefont
  {Abanin}}, \bibinfo {author} {\bibfnamefont {J.}~\bibnamefont {Bardarson}},
  \bibinfo {author} {\bibfnamefont {G.}~\bibnamefont {De~Tomasi}}, \bibinfo
  {author} {\bibfnamefont {S.}~\bibnamefont {Gopalakrishnan}}, \bibinfo
  {author} {\bibfnamefont {V.}~\bibnamefont {Khemani}}, \bibinfo {author}
  {\bibfnamefont {S.}~\bibnamefont {Parameswaran}}, \bibinfo {author}
  {\bibfnamefont {F.}~\bibnamefont {Pollmann}}, \bibinfo {author}
  {\bibfnamefont {A.}~\bibnamefont {Potter}}, \bibinfo {author} {\bibfnamefont
  {M.}~\bibnamefont {Serbyn}},\ and\ \bibinfo {author} {\bibfnamefont
  {R.}~\bibnamefont {Vasseur}},\ }\bibfield  {title} {\bibinfo {title}
  {Distinguishing localization from chaos: Challenges in finite-size systems},\
  }\href {https://doi.org/10.1016/j.aop.2021.168415} {\bibfield  {journal}
  {\bibinfo  {journal} {Annals of Physics}\ }\textbf {\bibinfo {volume}
  {427}},\ \bibinfo {pages} {168415} (\bibinfo {year} {2021})}\BibitemShut
  {NoStop}%
\bibitem [{\citenamefont {De~Roeck}\ and\ \citenamefont
  {Huveneers}(2017)}]{PhysRevB.95.155129}%
  \BibitemOpen
  \bibfield  {author} {\bibinfo {author} {\bibfnamefont {W.}~\bibnamefont
  {De~Roeck}}\ and\ \bibinfo {author} {\bibfnamefont {F.~m.~c.}\ \bibnamefont
  {Huveneers}},\ }\bibfield  {title} {\bibinfo {title} {Stability and
  instability towards delocalization in many-body localization systems},\
  }\href {https://doi.org/10.1103/PhysRevB.95.155129} {\bibfield  {journal}
  {\bibinfo  {journal} {Phys. Rev. B}\ }\textbf {\bibinfo {volume} {95}},\
  \bibinfo {pages} {155129} (\bibinfo {year} {2017})}\BibitemShut {NoStop}%
\bibitem [{\citenamefont {Fisher}\ \emph {et~al.}(2023)\citenamefont {Fisher},
  \citenamefont {Khemani}, \citenamefont {Nahum},\ and\ \citenamefont
  {Vijay}}]{Fisher2023}%
  \BibitemOpen
  \bibfield  {author} {\bibinfo {author} {\bibfnamefont {M.~P.}\ \bibnamefont
  {Fisher}}, \bibinfo {author} {\bibfnamefont {V.}~\bibnamefont {Khemani}},
  \bibinfo {author} {\bibfnamefont {A.}~\bibnamefont {Nahum}},\ and\ \bibinfo
  {author} {\bibfnamefont {S.}~\bibnamefont {Vijay}},\ }\bibfield  {title}
  {\bibinfo {title} {Random quantum circuits},\ }\href
  {https://doi.org/10.1146/annurev-conmatphys-031720-030658} {\bibfield
  {journal} {\bibinfo  {journal} {Annual Review of Condensed Matter Physics}\
  }\textbf {\bibinfo {volume} {14}},\ \bibinfo {pages} {335–379} (\bibinfo
  {year} {2023})}\BibitemShut {NoStop}%
\bibitem [{\citenamefont {Parker}\ \emph {et~al.}(2019)\citenamefont {Parker},
  \citenamefont {Cao}, \citenamefont {Avdoshkin}, \citenamefont {Scaffidi},\
  and\ \citenamefont {Altman}}]{PhysRevX.9.041017}%
  \BibitemOpen
  \bibfield  {author} {\bibinfo {author} {\bibfnamefont {D.~E.}\ \bibnamefont
  {Parker}}, \bibinfo {author} {\bibfnamefont {X.}~\bibnamefont {Cao}},
  \bibinfo {author} {\bibfnamefont {A.}~\bibnamefont {Avdoshkin}}, \bibinfo
  {author} {\bibfnamefont {T.}~\bibnamefont {Scaffidi}},\ and\ \bibinfo
  {author} {\bibfnamefont {E.}~\bibnamefont {Altman}},\ }\bibfield  {title}
  {\bibinfo {title} {A universal operator growth hypothesis},\ }\href
  {https://doi.org/10.1103/PhysRevX.9.041017} {\bibfield  {journal} {\bibinfo
  {journal} {Phys. Rev. X}\ }\textbf {\bibinfo {volume} {9}},\ \bibinfo {pages}
  {041017} (\bibinfo {year} {2019})}\BibitemShut {NoStop}%
\bibitem [{\citenamefont {Rabinovici}\ \emph {et~al.}(2022)\citenamefont
  {Rabinovici}, \citenamefont {S{\'a}nchez-Garrido}, \citenamefont {Shir},\
  and\ \citenamefont {Sonner}}]{rabinovici2022krylov}%
  \BibitemOpen
  \bibfield  {author} {\bibinfo {author} {\bibfnamefont {E.}~\bibnamefont
  {Rabinovici}}, \bibinfo {author} {\bibfnamefont {A.}~\bibnamefont
  {S{\'a}nchez-Garrido}}, \bibinfo {author} {\bibfnamefont {R.}~\bibnamefont
  {Shir}},\ and\ \bibinfo {author} {\bibfnamefont {J.}~\bibnamefont {Sonner}},\
  }\bibfield  {title} {\bibinfo {title} {Krylov complexity from integrability
  to chaos},\ }\href@noop {} {\bibfield  {journal} {\bibinfo  {journal}
  {Journal of High Energy Physics}\ }\textbf {\bibinfo {volume} {2022}},\
  \bibinfo {pages} {1} (\bibinfo {year} {2022})}\BibitemShut {NoStop}%
\bibitem [{\citenamefont {Menzler}\ and\ \citenamefont
  {Jha}(2024)}]{menzler2024krylov}%
  \BibitemOpen
  \bibfield  {author} {\bibinfo {author} {\bibfnamefont {H.~G.}\ \bibnamefont
  {Menzler}}\ and\ \bibinfo {author} {\bibfnamefont {R.}~\bibnamefont {Jha}},\
  }\bibfield  {title} {\bibinfo {title} {Krylov localization as a probe for
  ergodicity breaking},\ }\href@noop {} {\bibfield  {journal} {\bibinfo
  {journal} {arXiv preprint arXiv:2403.14384}\ } (\bibinfo {year}
  {2024})}\BibitemShut {NoStop}%
\bibitem [{\citenamefont {Ba\~nuls}\ \emph {et~al.}(2017)\citenamefont
  {Ba\~nuls}, \citenamefont {Yao}, \citenamefont {Choi}, \citenamefont
  {Lukin},\ and\ \citenamefont {Cirac}}]{PhysRevB.96.174201}%
  \BibitemOpen
  \bibfield  {author} {\bibinfo {author} {\bibfnamefont {M.~C.}\ \bibnamefont
  {Ba\~nuls}}, \bibinfo {author} {\bibfnamefont {N.~Y.}\ \bibnamefont {Yao}},
  \bibinfo {author} {\bibfnamefont {S.}~\bibnamefont {Choi}}, \bibinfo {author}
  {\bibfnamefont {M.~D.}\ \bibnamefont {Lukin}},\ and\ \bibinfo {author}
  {\bibfnamefont {J.~I.}\ \bibnamefont {Cirac}},\ }\bibfield  {title} {\bibinfo
  {title} {Dynamics of quantum information in many-body localized systems},\
  }\href {https://doi.org/10.1103/PhysRevB.96.174201} {\bibfield  {journal}
  {\bibinfo  {journal} {Phys. Rev. B}\ }\textbf {\bibinfo {volume} {96}},\
  \bibinfo {pages} {174201} (\bibinfo {year} {2017})}\BibitemShut {NoStop}%
\bibitem [{\citenamefont {Li}\ \emph {et~al.}(2023)\citenamefont {Li},
  \citenamefont {Sala},\ and\ \citenamefont
  {Pollmann}}]{PhysRevResearch.5.043239}%
  \BibitemOpen
  \bibfield  {author} {\bibinfo {author} {\bibfnamefont {Y.}~\bibnamefont
  {Li}}, \bibinfo {author} {\bibfnamefont {P.}~\bibnamefont {Sala}},\ and\
  \bibinfo {author} {\bibfnamefont {F.}~\bibnamefont {Pollmann}},\ }\bibfield
  {title} {\bibinfo {title} {Hilbert space fragmentation in open quantum
  systems},\ }\href {https://doi.org/10.1103/PhysRevResearch.5.043239}
  {\bibfield  {journal} {\bibinfo  {journal} {Phys. Rev. Res.}\ }\textbf
  {\bibinfo {volume} {5}},\ \bibinfo {pages} {043239} (\bibinfo {year}
  {2023})}\BibitemShut {NoStop}%
\bibitem [{\citenamefont {De~Tomasi}\ \emph {et~al.}(2019)\citenamefont
  {De~Tomasi}, \citenamefont {Hetterich}, \citenamefont {Sala},\ and\
  \citenamefont {Pollmann}}]{PhysRevB.100.214313}%
  \BibitemOpen
  \bibfield  {author} {\bibinfo {author} {\bibfnamefont {G.}~\bibnamefont
  {De~Tomasi}}, \bibinfo {author} {\bibfnamefont {D.}~\bibnamefont
  {Hetterich}}, \bibinfo {author} {\bibfnamefont {P.}~\bibnamefont {Sala}},\
  and\ \bibinfo {author} {\bibfnamefont {F.}~\bibnamefont {Pollmann}},\
  }\bibfield  {title} {\bibinfo {title} {Dynamics of strongly interacting
  systems: From fock-space fragmentation to many-body localization},\ }\href
  {https://doi.org/10.1103/PhysRevB.100.214313} {\bibfield  {journal} {\bibinfo
   {journal} {Phys. Rev. B}\ }\textbf {\bibinfo {volume} {100}},\ \bibinfo
  {pages} {214313} (\bibinfo {year} {2019})}\BibitemShut {NoStop}%
\bibitem [{\citenamefont {Herviou}\ \emph {et~al.}(2021)\citenamefont
  {Herviou}, \citenamefont {Bardarson},\ and\ \citenamefont
  {Regnault}}]{PhysRevB.103.134207}%
  \BibitemOpen
  \bibfield  {author} {\bibinfo {author} {\bibfnamefont {L.}~\bibnamefont
  {Herviou}}, \bibinfo {author} {\bibfnamefont {J.~H.}\ \bibnamefont
  {Bardarson}},\ and\ \bibinfo {author} {\bibfnamefont {N.}~\bibnamefont
  {Regnault}},\ }\bibfield  {title} {\bibinfo {title} {Many-body localization
  in a fragmented hilbert space},\ }\href
  {https://doi.org/10.1103/PhysRevB.103.134207} {\bibfield  {journal} {\bibinfo
   {journal} {Phys. Rev. B}\ }\textbf {\bibinfo {volume} {103}},\ \bibinfo
  {pages} {134207} (\bibinfo {year} {2021})}\BibitemShut {NoStop}%
\bibitem [{\citenamefont {Moudgalya}\ \emph {et~al.}(2021)\citenamefont
  {Moudgalya}, \citenamefont {Prem}, \citenamefont {Nandkishore}, \citenamefont
  {Regnault},\ and\ \citenamefont {Bernevig}}]{Moudgalya2021}%
  \BibitemOpen
  \bibfield  {author} {\bibinfo {author} {\bibfnamefont {S.}~\bibnamefont
  {Moudgalya}}, \bibinfo {author} {\bibfnamefont {A.}~\bibnamefont {Prem}},
  \bibinfo {author} {\bibfnamefont {R.}~\bibnamefont {Nandkishore}}, \bibinfo
  {author} {\bibfnamefont {N.}~\bibnamefont {Regnault}},\ and\ \bibinfo
  {author} {\bibfnamefont {B.~A.}\ \bibnamefont {Bernevig}},\ }\bibinfo {title}
  {Thermalization and its absence within krylov subspaces of a constrained
  hamiltonian},\ in\ \href {https://doi.org/10.1142/9789811231711_0009} {\emph
  {\bibinfo {booktitle} {Memorial Volume for Shoucheng Zhang}}}\ (\bibinfo
  {publisher} {WORLD SCIENTIFIC},\ \bibinfo {year} {2021})\ p.\ \bibinfo
  {pages} {147–209}\BibitemShut {NoStop}%
\bibitem [{\citenamefont {Han}\ \emph {et~al.}(2024)\citenamefont {Han},
  \citenamefont {Chen},\ and\ \citenamefont {Lake}}]{han2024exponentially}%
  \BibitemOpen
  \bibfield  {author} {\bibinfo {author} {\bibfnamefont {Y.}~\bibnamefont
  {Han}}, \bibinfo {author} {\bibfnamefont {X.}~\bibnamefont {Chen}},\ and\
  \bibinfo {author} {\bibfnamefont {E.}~\bibnamefont {Lake}},\ }\bibfield
  {title} {\bibinfo {title} {Exponentially slow thermalization and the
  robustness of hilbert space fragmentation},\ }\href@noop {} {\bibfield
  {journal} {\bibinfo  {journal} {arXiv preprint arXiv:2401.11294}\ } (\bibinfo
  {year} {2024})}\BibitemShut {NoStop}%
\bibitem [{\citenamefont {Vidal}(2003)}]{PhysRevLett.91.147902}%
  \BibitemOpen
  \bibfield  {author} {\bibinfo {author} {\bibfnamefont {G.}~\bibnamefont
  {Vidal}},\ }\bibfield  {title} {\bibinfo {title} {Efficient classical
  simulation of slightly entangled quantum computations},\ }\href
  {https://doi.org/10.1103/PhysRevLett.91.147902} {\bibfield  {journal}
  {\bibinfo  {journal} {Phys. Rev. Lett.}\ }\textbf {\bibinfo {volume} {91}},\
  \bibinfo {pages} {147902} (\bibinfo {year} {2003})}\BibitemShut {NoStop}%
\bibitem [{\citenamefont {Gray}(2018)}]{Gray2018}%
  \BibitemOpen
  \bibfield  {author} {\bibinfo {author} {\bibfnamefont {J.}~\bibnamefont
  {Gray}},\ }\bibfield  {title} {\bibinfo {title} {quimb: A python package for
  quantum information and many-body calculations},\ }\href
  {https://doi.org/10.21105/joss.00819} {\bibfield  {journal} {\bibinfo
  {journal} {Journal of Open Source Software}\ }\textbf {\bibinfo {volume}
  {3}},\ \bibinfo {pages} {819} (\bibinfo {year} {2018})}\BibitemShut {NoStop}%
\bibitem [{\citenamefont {Lieb}\ and\ \citenamefont
  {Robinson}(1972)}]{lieb1972finite}%
  \BibitemOpen
  \bibfield  {author} {\bibinfo {author} {\bibfnamefont {E.~H.}\ \bibnamefont
  {Lieb}}\ and\ \bibinfo {author} {\bibfnamefont {D.~W.}\ \bibnamefont
  {Robinson}},\ }\bibfield  {title} {\bibinfo {title} {The finite group
  velocity of quantum spin systems},\ }\href@noop {} {\bibfield  {journal}
  {\bibinfo  {journal} {Communications in mathematical physics}\ }\textbf
  {\bibinfo {volume} {28}},\ \bibinfo {pages} {251} (\bibinfo {year}
  {1972})}\BibitemShut {NoStop}%
\bibitem [{\citenamefont {Haah}\ \emph {et~al.}(2021)\citenamefont {Haah},
  \citenamefont {Hastings}, \citenamefont {Kothari},\ and\ \citenamefont
  {Low}}]{Haah2021}%
  \BibitemOpen
  \bibfield  {author} {\bibinfo {author} {\bibfnamefont {J.}~\bibnamefont
  {Haah}}, \bibinfo {author} {\bibfnamefont {M.~B.}\ \bibnamefont {Hastings}},
  \bibinfo {author} {\bibfnamefont {R.}~\bibnamefont {Kothari}},\ and\ \bibinfo
  {author} {\bibfnamefont {G.~H.}\ \bibnamefont {Low}},\ }\bibfield  {title}
  {\bibinfo {title} {Quantum algorithm for simulating real time evolution of
  lattice hamiltonians},\ }\href {https://doi.org/10.1137/18m1231511}
  {\bibfield  {journal} {\bibinfo  {journal} {SIAM Journal on Computing}\
  }\textbf {\bibinfo {volume} {52}},\ \bibinfo {pages} {FOCS18} (\bibinfo
  {year} {2021})}\BibitemShut {NoStop}%
\bibitem [{\citenamefont {(Anthony)~Chen}\ \emph {et~al.}(2023)\citenamefont
  {(Anthony)~Chen}, \citenamefont {Lucas},\ and\ \citenamefont
  {Yin}}]{AnthonyChen2023}%
  \BibitemOpen
  \bibfield  {author} {\bibinfo {author} {\bibfnamefont {C.-F.}\ \bibnamefont
  {(Anthony)~Chen}}, \bibinfo {author} {\bibfnamefont {A.}~\bibnamefont
  {Lucas}},\ and\ \bibinfo {author} {\bibfnamefont {C.}~\bibnamefont {Yin}},\
  }\bibfield  {title} {\bibinfo {title} {Speed limits and locality in many-body
  quantum dynamics},\ }\href {https://doi.org/10.1088/1361-6633/acfaae}
  {\bibfield  {journal} {\bibinfo  {journal} {Reports on Progress in Physics}\
  }\textbf {\bibinfo {volume} {86}},\ \bibinfo {pages} {116001} (\bibinfo
  {year} {2023})}\BibitemShut {NoStop}%
\bibitem [{\citenamefont {Aaronson}\ and\ \citenamefont
  {Arkhipov}(2010)}]{aaronson2010computational}%
  \BibitemOpen
  \bibfield  {author} {\bibinfo {author} {\bibfnamefont {S.}~\bibnamefont
  {Aaronson}}\ and\ \bibinfo {author} {\bibfnamefont {A.}~\bibnamefont
  {Arkhipov}},\ }\href@noop {} {\bibinfo {title} {The computational complexity
  of linear optics}} (\bibinfo {year} {2010}),\ \Eprint
  {https://arxiv.org/abs/1011.3245} {arXiv:1011.3245 [quant-ph]} \BibitemShut
  {NoStop}%
\bibitem [{\citenamefont {Lund}\ \emph {et~al.}(2017)\citenamefont {Lund},
  \citenamefont {Bremner},\ and\ \citenamefont {Ralph}}]{Lund2017}%
  \BibitemOpen
  \bibfield  {author} {\bibinfo {author} {\bibfnamefont {A.~P.}\ \bibnamefont
  {Lund}}, \bibinfo {author} {\bibfnamefont {M.~J.}\ \bibnamefont {Bremner}},\
  and\ \bibinfo {author} {\bibfnamefont {T.~C.}\ \bibnamefont {Ralph}},\
  }\bibfield  {title} {\bibinfo {title} {Quantum sampling problems,
  bosonsampling and quantum supremacy},\ }\bibfield  {journal} {\bibinfo
  {journal} {npj Quantum Information}\ }\textbf {\bibinfo {volume} {3}},\ \href
  {https://doi.org/10.1038/s41534-017-0018-2} {10.1038/s41534-017-0018-2}
  (\bibinfo {year} {2017})\BibitemShut {NoStop}%
\bibitem [{\citenamefont {Deshpande}\ \emph {et~al.}(2018)\citenamefont
  {Deshpande}, \citenamefont {Fefferman}, \citenamefont {Tran}, \citenamefont
  {Foss-Feig},\ and\ \citenamefont {Gorshkov}}]{PhysRevLett.121.030501}%
  \BibitemOpen
  \bibfield  {author} {\bibinfo {author} {\bibfnamefont {A.}~\bibnamefont
  {Deshpande}}, \bibinfo {author} {\bibfnamefont {B.}~\bibnamefont
  {Fefferman}}, \bibinfo {author} {\bibfnamefont {M.~C.}\ \bibnamefont {Tran}},
  \bibinfo {author} {\bibfnamefont {M.}~\bibnamefont {Foss-Feig}},\ and\
  \bibinfo {author} {\bibfnamefont {A.~V.}\ \bibnamefont {Gorshkov}},\
  }\bibfield  {title} {\bibinfo {title} {Dynamical phase transitions in
  sampling complexity},\ }\href
  {https://doi.org/10.1103/PhysRevLett.121.030501} {\bibfield  {journal}
  {\bibinfo  {journal} {Phys. Rev. Lett.}\ }\textbf {\bibinfo {volume} {121}},\
  \bibinfo {pages} {030501} (\bibinfo {year} {2018})}\BibitemShut {NoStop}%
\bibitem [{\citenamefont {Tindall}\ and\ \citenamefont
  {Fishman}(2023)}]{10.21468/SciPostPhys.15.6.222}%
  \BibitemOpen
  \bibfield  {author} {\bibinfo {author} {\bibfnamefont {J.}~\bibnamefont
  {Tindall}}\ and\ \bibinfo {author} {\bibfnamefont {M.}~\bibnamefont
  {Fishman}},\ }\bibfield  {title} {\bibinfo {title} {{Gauging tensor networks
  with belief propagation}},\ }\href
  {https://doi.org/10.21468/SciPostPhys.15.6.222} {\bibfield  {journal}
  {\bibinfo  {journal} {SciPost Phys.}\ }\textbf {\bibinfo {volume} {15}},\
  \bibinfo {pages} {222} (\bibinfo {year} {2023})}\BibitemShut {NoStop}%
\bibitem [{\citenamefont {Tindall}\ and\ \citenamefont
  {Sels}(2024)}]{tindall2024confinement}%
  \BibitemOpen
  \bibfield  {author} {\bibinfo {author} {\bibfnamefont {J.}~\bibnamefont
  {Tindall}}\ and\ \bibinfo {author} {\bibfnamefont {D.}~\bibnamefont {Sels}},\
  }\bibfield  {title} {\bibinfo {title} {Confinement in the transverse field
  ising model on the heavy hex lattice},\ }\href
  {https://doi.org/10.1103/PhysRevLett.133.180402} {\bibfield  {journal}
  {\bibinfo  {journal} {Phys. Rev. Lett.}\ }\textbf {\bibinfo {volume} {133}},\
  \bibinfo {pages} {180402} (\bibinfo {year} {2024})}\BibitemShut {NoStop}%
\bibitem [{\citenamefont {Pave{\v{s}}i{\'c}}\ \emph {et~al.}(2024)\citenamefont
  {Pave{\v{s}}i{\'c}}, \citenamefont {Jaschke},\ and\ \citenamefont
  {Montangero}}]{pavevsic2024constrained}%
  \BibitemOpen
  \bibfield  {author} {\bibinfo {author} {\bibfnamefont {L.}~\bibnamefont
  {Pave{\v{s}}i{\'c}}}, \bibinfo {author} {\bibfnamefont {D.}~\bibnamefont
  {Jaschke}},\ and\ \bibinfo {author} {\bibfnamefont {S.}~\bibnamefont
  {Montangero}},\ }\bibfield  {title} {\bibinfo {title} {Constrained dynamics
  and confinement in the two-dimensional quantum ising model},\ }\href@noop {}
  {\bibfield  {journal} {\bibinfo  {journal} {arXiv preprint arXiv:2406.11979}\
  } (\bibinfo {year} {2024})}\BibitemShut {NoStop}%
\bibitem [{\citenamefont {Thiery}\ \emph {et~al.}(2018)\citenamefont {Thiery},
  \citenamefont {Huveneers}, \citenamefont {M\"uller},\ and\ \citenamefont
  {De~Roeck}}]{PhysRevLett.121.140601}%
  \BibitemOpen
  \bibfield  {author} {\bibinfo {author} {\bibfnamefont {T.}~\bibnamefont
  {Thiery}}, \bibinfo {author} {\bibfnamefont {F.~m.~c.}\ \bibnamefont
  {Huveneers}}, \bibinfo {author} {\bibfnamefont {M.}~\bibnamefont
  {M\"uller}},\ and\ \bibinfo {author} {\bibfnamefont {W.}~\bibnamefont
  {De~Roeck}},\ }\bibfield  {title} {\bibinfo {title} {Many-body delocalization
  as a quantum avalanche},\ }\href
  {https://doi.org/10.1103/PhysRevLett.121.140601} {\bibfield  {journal}
  {\bibinfo  {journal} {Phys. Rev. Lett.}\ }\textbf {\bibinfo {volume} {121}},\
  \bibinfo {pages} {140601} (\bibinfo {year} {2018})}\BibitemShut {NoStop}%
\bibitem [{\citenamefont {Luitz}\ \emph {et~al.}(2017)\citenamefont {Luitz},
  \citenamefont {Huveneers},\ and\ \citenamefont
  {De~Roeck}}]{PhysRevLett.119.150602}%
  \BibitemOpen
  \bibfield  {author} {\bibinfo {author} {\bibfnamefont {D.~J.}\ \bibnamefont
  {Luitz}}, \bibinfo {author} {\bibfnamefont {F.~m.~c.}\ \bibnamefont
  {Huveneers}},\ and\ \bibinfo {author} {\bibfnamefont {W.}~\bibnamefont
  {De~Roeck}},\ }\bibfield  {title} {\bibinfo {title} {How a small quantum bath
  can thermalize long localized chains},\ }\href
  {https://doi.org/10.1103/PhysRevLett.119.150602} {\bibfield  {journal}
  {\bibinfo  {journal} {Phys. Rev. Lett.}\ }\textbf {\bibinfo {volume} {119}},\
  \bibinfo {pages} {150602} (\bibinfo {year} {2017})}\BibitemShut {NoStop}%
\bibitem [{\citenamefont {Morningstar}\ \emph {et~al.}(2022)\citenamefont
  {Morningstar}, \citenamefont {Colmenarez}, \citenamefont {Khemani},
  \citenamefont {Luitz},\ and\ \citenamefont {Huse}}]{PhysRevB.105.174205}%
  \BibitemOpen
  \bibfield  {author} {\bibinfo {author} {\bibfnamefont {A.}~\bibnamefont
  {Morningstar}}, \bibinfo {author} {\bibfnamefont {L.}~\bibnamefont
  {Colmenarez}}, \bibinfo {author} {\bibfnamefont {V.}~\bibnamefont {Khemani}},
  \bibinfo {author} {\bibfnamefont {D.~J.}\ \bibnamefont {Luitz}},\ and\
  \bibinfo {author} {\bibfnamefont {D.~A.}\ \bibnamefont {Huse}},\ }\bibfield
  {title} {\bibinfo {title} {Avalanches and many-body resonances in many-body
  localized systems},\ }\href {https://doi.org/10.1103/PhysRevB.105.174205}
  {\bibfield  {journal} {\bibinfo  {journal} {Phys. Rev. B}\ }\textbf {\bibinfo
  {volume} {105}},\ \bibinfo {pages} {174205} (\bibinfo {year}
  {2022})}\BibitemShut {NoStop}%
\bibitem [{\citenamefont {Sels}(2022)}]{PhysRevB.106.L020202}%
  \BibitemOpen
  \bibfield  {author} {\bibinfo {author} {\bibfnamefont {D.}~\bibnamefont
  {Sels}},\ }\bibfield  {title} {\bibinfo {title} {Bath-induced delocalization
  in interacting disordered spin chains},\ }\href
  {https://doi.org/10.1103/PhysRevB.106.L020202} {\bibfield  {journal}
  {\bibinfo  {journal} {Phys. Rev. B}\ }\textbf {\bibinfo {volume} {106}},\
  \bibinfo {pages} {L020202} (\bibinfo {year} {2022})}\BibitemShut {NoStop}%
\end{thebibliography}%

\end{document}